\documentclass[fleqn,usenatbib]{mnras}

\usepackage{newtxtext,newtxmath}

\usepackage[T1]{fontenc}

\usepackage{graphicx}	
\usepackage{amsmath}	
\usepackage{upgreek}
\usepackage{bm}
\usepackage{makecell}
\usepackage{multirow}
\usepackage{booktabs}
\usepackage{hyperref}
\usepackage{ulem}  

\title[Multicolour Validation of Two Mini-Neptunes]{Multicolour Validation of Two Temperate Mini-Neptunes Around M-dwarf Habitable Zones}

\author[Jiang et al.]{
Chengzi Jiang,$^{1,2}$\thanks{E-mail: cjiang@iac.es}
Aleksandra Selezneva,$^{1,2}$
Hannu Parviainen,$^{1,2}$ 
Felipe Murgas,$^{1,2}$
Enric Pall\'e,$^{1,2}$
\newauthor
~Gareb Fernández-Rodríguez,$^{1,2}$
Samuel Geraldía-González,$^{1,2}$
Jaume Orell-Miquel,$^{3,1,2}$
Norio Narita,$^{4,5,1}$
\newauthor
~Akihiko Fukui,$^{4,1}$
Jerome de Leon,$^{4}$
Izuru Fukuda,$^{6}$
Kai Ikuta,$^{7}$
Kiyoe Kawauchi,$^{8}$
Steve B. Howell,$^{9}$
\newauthor
~Colin Littlefield,$^{10,9}$
Sarah J. Deveny,$^{10,9}$
Joseph D. Twicken,$^{11,9}$
Richard P. Schwarz,$^{12}$
and Avi Shporer $^{13}$ 
\\
$^{1}$Instituto de Astrof\'isica de Canarias, V\'ia L\'actea s/n, 38205 La Laguna, Tenerife, Spain\\
$^{2}$Departamento de Astrof\'isica, Universidad de La Laguna, C/ Padre Herrera, 38206 La Laguna, Tenerife, Spain\\
$^{3}$Department of Astronomy, University of Texas at Austin, 2515 Speedway, Austin, TX 78712, USA \\
$^{4}$Komaba Institute for Science, The University of Tokyo, 3-8-1 Komaba, Meguro, Tokyo 153-8902, Japan \\
$^{5}$Astrobiology Center, 2-21-1 Osawa, Mitaka, Tokyo 181-8588, Japan \\
$^{6}$Department of Multi-Disciplinary Sciences, Graduate School of Arts and Sciences, The University of Tokyo, 3-8-1 Komaba, Meguro, Tokyo 153-8902, Japan \\
$^{7}$Graduate School of Social Data Science, Hitotsubashi University, 2-1 Naka, Kunitachi, Tokyo 186-8601, Japan \\
$^{8}$Department of Physical Sciences, Ritsumeikan University, Kusatsu, Shiga 525-8577, Japan\\
$^{9}$NASA Ames Research Center, Moffett Field, CA 94035, USA \\
$^{10}$Bay Area Environmental Research Institute, Moffett Field, CA 94035, USA \\
$^{11}$SETI Institute, Mountain View, CA 94043 USA \\
$^{12}$Center for Astrophysics \textbar \ Harvard \& Smithsonian, 60 Garden Street, Cambridge, MA 02138, USA \\
$^{13}$Department of Physics and Kavli Institute for Astrophysics and Space Research, Massachusetts Institute of Technology, Cambridge, MA 02139, USA  
}

\date{Accepted XXX. Received YYY; in original form ZZZ}

\pubyear{2025}

\begin{document}
\label{firstpage}
\pagerange{\pageref{firstpage}--\pageref{lastpage}}
\maketitle

\begin{abstract}
For small planets orbiting within the habitable zones of their host stars, multicolour validation via photometric transit observations offers an efficient alternative to prioritize targets before intensive radial-velocity follow-up, thereby expanding the sample of habitable-zone exoplanets amenable for atmospheric characterisation. In this study, we validate two exceptional habitable-zone TESS candidates, orbiting around M-dwarfs, as genuine planets, precisely determining their transit and physical parameters. We perform Bayesian model comparison by jointly fitting multicolour light curves from TESS and ground-based follow-up, including observations with HiPERCAM at the 10.4-m GTC. Our approach uses wavelength-dependent transit depth variations and precise transit geometry to reject false positives. We validate TOI-2094 b and TOI-7166 b as two new benchmark temperate mini-Neptunes. TOI-2094 b (1.90 $R_{\oplus}$) orbits its M3V star (V=14.4, d=50.22~pc) with a period of $\sim$18.79 days, well within the habitable zone ($\sim$0.98 Earth insolation). TOI-7166 b (2.39 $R_{\oplus}$) orbits its M4.5V host star (V=15.8, d=35.24~pc) with a period of $\sim$12.92 days, placing it near the inner edge of the habitable zone ($\sim$1.93 Earth insolation). Statistical mass and density estimates suggest that TOI-2094 b may be a volatile-rich planet, such as a water world or a gaseous planet, and is less likely to be rocky, while TOI-7166 b is likely to be volatile-rich. Both planets are of great interest for detailed atmospheric characterisation with the JWST and future ELTs, which requires further precise mass measurements. 
\end{abstract}




\section{Introduction}

The discovery of an Earth analogue---a rocky planet orbiting within the habitable zone of a Sun-like star---remains a fundamental goal in exoplanet research. Although such a planet has not been detected yet, recent transit surveys have revealed a growing population of small planets, from Earth-size to mini-Neptunes, orbiting within the habitable zone of M dwarfs (e.g. LHS 1140 b, \citealt{2017Natur.544..333D}; TOI-700 d, \citealt{2020AJ....160..116G, 2020AJ....160..117R}; and TOI-715 b, \citealt{2024MNRAS.527...35D}). M dwarfs offer favourable conditions for detecting transiting small planets: their habitable zones are closer in, yielding shorter orbital periods (typically less than 100 days) and deeper transits ($\sim$10$^3$~ppm), significantly enhancing detectability. However, confirming these planets is challenging due to the faintness of the host stars, their very small radii and masses and the potentially strong activity of M dwarfs. Notably, precise radial velocity (RV) confirmation of small habitable-zone planets usually requires large amounts of telescope time, even for favourable M-dwarf systems. Consequently, many high-value candidates remain unconfirmed, which limits our understanding of the demographics and potential habitability of small exoplanets. 

To address this bottleneck, we use multicolour transit photometry as an efficient validation approach (hereafter multicolour validation) prior to intensive RV campaigns. This method leverages wavelength-dependent features of broadband transit depths to distinguish true planets from false positives, such as blended eclipsing binaries or hierarchical triple systems. The general idea has been illustrated in previous studies using the MuSCAT2 instrument at the Carlos Sánchez Telescope \citep[TCS; e.g.][]{2019A&A...630A..89P, 2020A&A...633A..28P}. Multicolour validation has been applied to a series of successful validations of planetary candidates \citep[e.g. ][]{2021AJ....162...54H, 2022A&A...666A..10E, 2024A&A...683A.170P, 2024A&A...690A.263G, 2024A&A...690A..62P}. However, the photometric precisions of the 1--2-meter telescopes in previous studies limit their targets of interest down to either Neptune-sized candidates or bright sources. This study demonstrates, for the first time, the power of the 10.4-m Gran Telescopio Canarias (GTC) equipped with the quintuple-beam, high-speed optical imager HiPERCAM \citep{2021MNRAS.507..350D} to validate small planetary candidates orbiting faint stars. By combining multi-epoch light curves from the Transiting Exoplanet Survey Satellite \citep[TESS;][]{2015JATIS...1a4003R} with multi-colour light curves from the GTC/HiPERCAM and other instruments, we can (1) reject false positives through Bayesian model comparison and (2) precisely measure the true planetary radius and orbital parameters.

In this work, we validate two high-priority TESS candidates, TOI-2094.01 and TOI-7166.01. They are located inside and around the habitable zones of their M-dwarf hosts according to the ExoFOP archive\footnote{\url{https://exofop.ipac.caltech.edu/tess/}}. Both targets exemplify the challenges of traditional confirmation methods: their small sizes, long orbital periods, and faint host stars render RV confirmation inefficient, and the TESS light curves lack the colour features to rule out astrophysical false positives. Our analysis combines TESS data and ground-based follow-up observations including new multicolour photometry from GTC/HiPERCAM to confirm the planetary nature of these candidates and refine their physical properties in a Bayesian framework. This approach not only expands the sample of validated terrestrial exoplanets but also establishes a pathway for future studies to prioritize terrestrial planets for atmospheric characterisation with the James Webb Space Telescope (JWST), the Atmospheric Remote-sensing Infrared Exoplanet Large-survey (Ariel), and upcoming extremely large telescopes.

The paper is structured as follows: Section \ref{sect:method} details the methodology of multicolour validation; Section \ref{sect:stars} presents the analysis on host stars; Section \ref{sect:observations} describes the transit observations and data reduction; Section \ref{sect:results} presents the validation results of target candidates; Section \ref{sect:discussion} discusses the statistical properties of the planets and their implications for future atmospheric studies, and Section \ref{sect:conclusion} draws our conclusion.

\section{Methods}
\label{sect:method}

\subsection{Framework of multicolour validation}
\label{sect:framework}

\begin{figure}
    \centering
    \includegraphics[width=\linewidth]{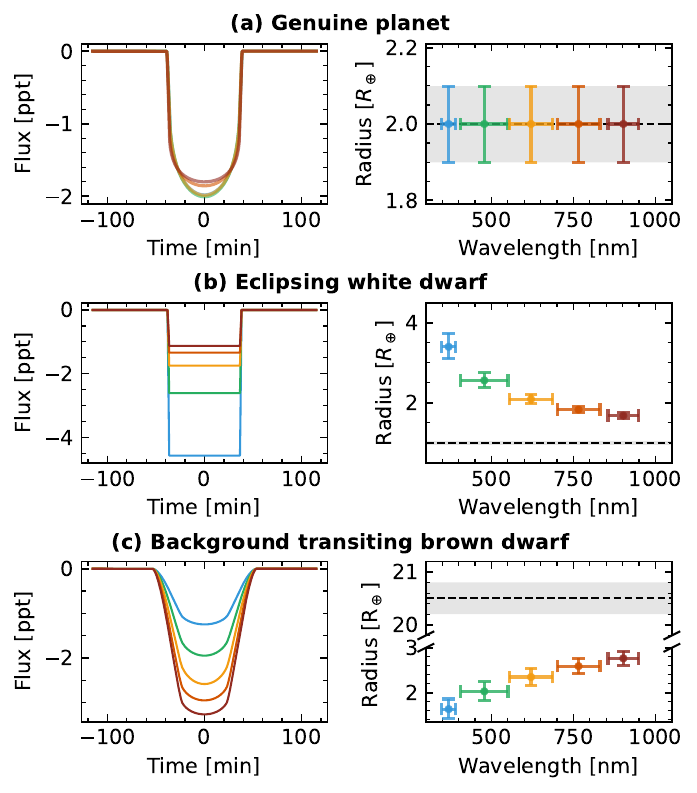} 
    \caption{ Toy models of transit light curves (\textit{left}) and corresponding apparent radii in the HiPERCAM passbands (\textit{right}), using different model hypotheses ($\mathcal{H}_1$ to $\mathcal{H}_3$ from top to bottom). The dashed line and gray shaded area are the input true radius and 1-$\sigma$ uncertainty. In all cases, we assume an edge-on circular orbit with a period of 20 days and a semi-major axis of 120~$R_\star$, and the primary star has $T_{\rm eff}=3500$~K, $\lg g=5.0$, and $\rm [Fe/H]=0$. Panel (a) assumes a transiting planet with 2 Earth radii. Panel (b) assumes an eclipsing white dwarf of Earth radius with $T_{\rm eff}=4500$~K. Panel (c) assumes a brown dwarf transiting on an unresolved, faint, background star with $T_{\rm eff}=3000$~K.}
    \label{fig:toymodels}
\end{figure}

We start with the quick validation combining TESS light curves and ground-based high-resolution imaging using the \texttt{TRICERATOPS} code \citep{2021AJ....161...24G}. This helps to exclude 17 different false positive scenarios defined in the Table 1 of \citet{2021AJ....161...24G} by calculating the false positive probability (FPP) and the nearby false positive probability (NFPP). Since there are slight scatters in the calculated FPPs and NFPPs across different runs, we repeat the calculations using \texttt{TRICERATOPS} by 20 times to estimate the mean and standard deviation of the FPPs and NFPPs for robust results, following \citet{2021AJ....161...24G}. After that we perform the multicolour validation using our own Bayesian framework. By combining TESS light curves and ground-based multi-colour transit light curves, we further compare the genuine transiting planet scenario ($\mathcal{H}_1$) with two important false positive scenarios ($\mathcal{H}_2$, $\mathcal{H}_3$) for small candidates around faint sources, thereby contributing to the reliability of TESS data validation. The three scenarios for Bayesian model comparison are illustrated as follows:
 
$\mathcal{H}_1$: a genuine transiting planet. For small planets around habitable zones, the atmospheric signals in optical broad bands are negligible compared to photometric uncertainties of current instruments. For example, for typical temperate sub-Neptunes TOI-270 d and K2-18 b, the amplitudes of optical transmission signals induced by Rayleigh scattering are several tens of ppm \citep[e.g.,][]{2024arXiv240303325B, 2025AJ....170..298S}, while the broadband transit depth uncertainties from 10.4-m GTC/HiPERCAM are $10^2$~ppm for target sources with V-magnitudes around 15. Thus, a genuine planet would exhibit a constant transit depth ($R_{\rm p}^2/R_\star^2$) across wavelengths within uncertainties (Fig. \ref{fig:toymodels} a). Significant deviations from this behaviour indicate a false positive in planetary detection. 

$\mathcal{H}_2$: a transiting white dwarf exhibiting secondary eclipse signals. Since transit signals are very shallow ($R_{\rm p}^2/R_\star^2<0.01$) in cases of small candidates around late-type stars, the eclipsing component other than a white dwarf can be naturally excluded. White dwarfs can usually be excluded if both their primary and secondary eclipse are detectable in TESS light curve validation. However, for white dwarfs around faint sources, their secondary eclipses may mimic planetary transits in low-cadence TESS light curves, while the weaker primary eclipse signals can be buried in noise. A precise constraints on the transit geometry and limb-darkening features with very high-cadence transit light curves would help distinguish this scenario. In addition, the colour difference between the white dwarf and host star can produce chromatic flux modifications and thus a slope in the apparent radii (Fig. \ref{fig:toymodels} b). 

$\mathcal{H}_3$: a brown dwarf transiting an unresolved faint star. The signals are produced by a substellar object transiting an unresolved star much fainter than the primary target star. Due to flux dilution, the true radius of the transiting object can be underestimated by one order of magnitude, making a brown dwarf appear to be a sub-Neptune in TESS light curves. A large, true transit depth can produce V-shaped transit signals, which can be constrained by high-cadence, high-precision transit follow-ups. In addition, the chromatic flux modifications due to different colours of the background binaries and host star can also produce a slope in the apparent radii (Fig. \ref{fig:toymodels} c). 

Figure \ref{fig:toymodels} shows the toy models corresponding to the above hypotheses, where the light curves were calculated assuming an edge-on circular orbit with a period of 20 days and a semi-major axis of 120~$R_\star$, and the primary star were assumed to have $T_{\rm eff}=3500$~K, $\log g=5.0$, and $\rm [Fe/H]=0$. These hypotheses can be distinguished through precisely measuring the transit geometry and chromatic transit depths. This would be challenging without high-cadence, high-precision transit photometry, especially for those weak detections from TESS observations. The GTC/HiPERCAM observations address these limitations by (1) simultaneous five-band transit photometry detecting chromatic features with amplitudes of $10^{-4}$, and (2) sub-second cadence constraining precise ingress/egress geometry to exclude V-shaped false positives.

We perform Bayesian model comparison through joint light curve analysis of TESS and ground-based multicolour follow-ups. For each hypothesis $\mathcal{H}_i$, we compute the Bayesian evidence $\mathcal{Z}_i$ and adopt the logarithmic Bayes factor $\Delta\ln\mathcal{Z}$ as the criterion of model comparison. If the hypothesis $\mathcal{H}_1$ exhibit the highest Bayesian evidence, we validate the candidate as a transiting exoplanet. Following \cite{2008ConPh..49...71T} and \cite{2013ApJ...778..153B}, we consider $\Delta\ln\mathcal{Z}\ge5$ as strong evidence, $2.5\le\Delta\ln\mathcal{Z}<5$ as moderate, $1\le\Delta\ln\mathcal{Z}<2.5$ as weak, and $\Delta\ln\mathcal{Z}<1$ as inconclusive.

\subsection{Light curve modelling and fitting} 
 
The transit light curves were modelled using the Python package \texttt{PyTransit} \citep{2015MNRAS.450.3233P, 2020MNRAS.499.1633P}. We assumed circular orbits in transit models due to the lack of additional constraints on planetary orbits from the RV data. Under the circular orbit assumption, the transit parameters for light curve modelling are radius ratio ($R_{\rm p}/R_\star$), orbital period ($P$), transit zero-epoch ($T_0$), orbital semimajor axis scaled by stellar radius ($a/R_\star$), orbital impact parameter ($b$), and limb-darkening coefficients. \texttt{PyTransit} allows super-sampling when calculating long-cadence light curve. Thus, we used a super-sampling factor of 12 to model the two-minute cadence TESS light curves with better accuracy.
We used the \texttt{LDTK} \citep{2015MNRAS.453.3821P} module integrated in \texttt{PyTransit} to account for the stellar limb-darkening effect. This module performs direct modelling of limb darkening profiles based on interpolation of PHOENIX stellar models \citep{2013A&A...553A...6H} without assuming any parameterized profiles. The ``limb-darkening coefficients'' thus become the effective temperature ($T_{\rm eff}$), surface gravity ($\lg g$), and metallicity ([Fe/H]) of the host star, and are applied to the light curves across all wavebands. This method greatly reduces the amount of free parameters when performing a joint fit of multicolour light curves.  

For the false positive scenario $\mathcal{H}_2$, we assume the observed signals were produced during the secondary eclipse of a white dwarf. We estimate the flux ratio between the white dwarf and the host star using the blackbody assumption, with an additional free parameter: the white dwarf's effective temperature ($T_{\rm eff, WD}$). The eclipse models were directly calculated using \texttt{PyTransit}. The primary transits were also modelled for the TESS light curves in this case, where the transit model is modified by the flux dilution effect as follows:   
\begin{align}\label{eq:dilution} 
    &f^*(t;F_\lambda)=\frac{f(t)+F_\lambda}{1+F_\lambda}, \\
    &F_\lambda = \left(\frac{R_{\rm p}}{R_\star}\right)^2\frac{\int_\lambda w_\lambda B_\lambda(T_{\rm eff, WD})\rm d\lambda}{\int_\lambda w_\lambda B_\lambda(T_{\rm eff})\rm d\lambda}, 
\end{align}
where $f(t)$ is the original transit model, $F_\lambda$ is the blackbody flux ratio of the two objects averaged by the broadband instrumental responses $w (\lambda)$, and $B_\lambda(T)$ is the Planck function.  

For the false positive scenario $\mathcal{H}_3$, we assume a brown dwarf transiting an unresolved faint star.  
This can be characterised by adding two parameters: an effective temperature $T_{\rm eff, EB}$ of the faint host and a luminosity rescaling factor $\gamma$ accounting for the radius and distance degeneracy, so that the target-to-binary flux ratio is 
\begin{equation}\label{eq:dilution2}  
    F_\lambda = \gamma \frac{\int_\lambda w_\lambda B_\lambda(T_{\rm eff})\rm d\lambda}{\int_\lambda w_\lambda B_\lambda(T_{\rm eff, EB})\rm d\lambda},
\end{equation}
which can be used in Eq. \ref{eq:dilution} to correct the flux dilution during transits.

Ground-based light curves are usually affected by significant systematics, such as variations in seeing and airmass. Although differential photometry mitigates the majority of this systematic noise, we accounted for the residual time-correlated components using Gaussian process (GP) regression. The kernel function of GPs was defined as the combination of a 3/2-order Matern kernel and a jitter term modelled with \texttt{celerite} \citep{2017AJ....154..220F}:
\begin{equation}
    k(\tau_{i,j}; \rho, \sigma^2, \sigma^2_{\rm n}) = \sigma^2\left(1+\frac{\sqrt{3}\tau_{i,j}}{\rho}\right)\exp\left(-\frac{\sqrt{3}\tau_{i,j}}{\rho}\right) + \sigma_{\rm n}^2\delta_{i,j},
\end{equation}
where $k$ is the covariance, $\tau_{i,j}$ is the time difference between two data points, $\rho$ and $\sigma^2$ are the length scale and the variance of the Mat\'ern-3/2 kernel, $\sigma^2_{\rm n}$ is the jitter variance to compensate for additional white noise, and $\delta_{i,j}$ is the Kronecker delta. The GP parameters $\rho$, $\sigma^2$, and $\sigma^2_{\rm n}$ are free parameters with non-informative priors in light curve fitting. 

We assigned each light curve a separate GP to compute its correlated noise. The TESS light curves were separated by each sector, while the ground-based light curves were separated by each observation and each passbands.  
We performed joint light curve fitting so that the transit parameters are consistent across all transits. The likelihood function was defined as the sum of logarithmic likelihood of each GP. 

We use the nested sampling algorithm implemented with \texttt{PyMultiNest} \citep{2009MNRAS.398.1601F, 2014A&A...564A.125B} to estimate Bayesian evidence and parameter posteriors. We used 5000 live points in \texttt{PyMultiNest} to generate $\sim$$4\times10^5$ accepted likelihood evaluations and $\sim$$7\times10^4$ weighted posterior samples for each run, reaching a $\ln \mathcal{Z}$ uncertainty of $\sim$0.1.

\section{Host star characterisation}
\label{sect:stars}

\subsection{High-resolution imaging}

\begin{figure}
    \centering
    \includegraphics[width=\linewidth]{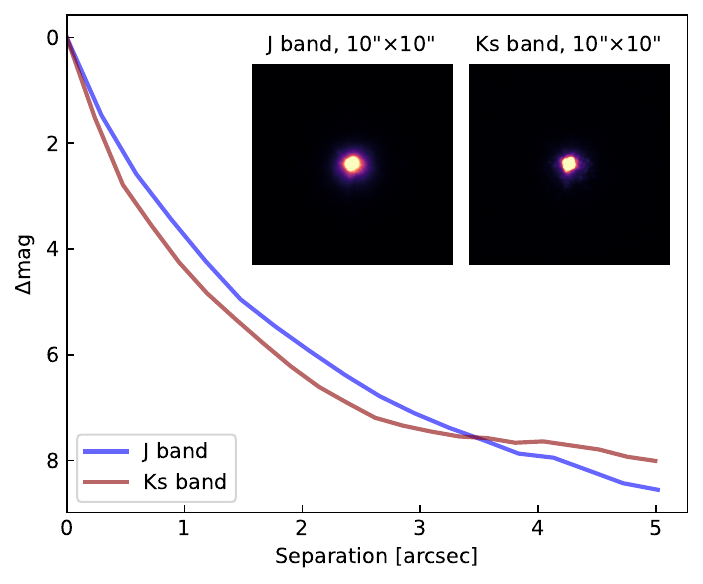} 
    \caption{Infrared AO imaging of TOI-2094 and corresponding magnitude sensitivity curves in the $J$ and $K_s$ bands observed using ShaneAO/ShARCS instrument, obtained from ExoFOP archival data (PI: C. Dressing). No close stellar companion is noted.}
    \label{fig:speckle_2094}
\end{figure}

\begin{figure} 
    \centering
    \includegraphics[width=\linewidth]{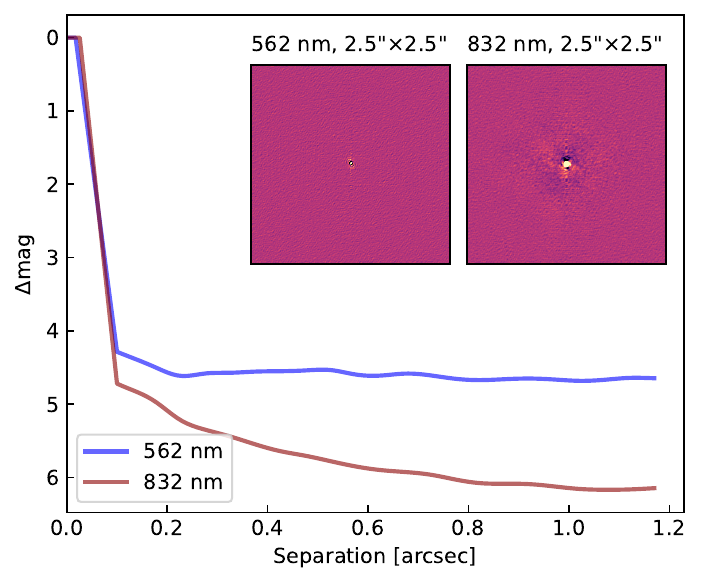}
    \caption{Optical speckle imaging of TOI-7166 and corresponding 5-$\sigma$ magnitude sensitivity curves observed with Gemini/Zorro instrument using filters EO 562 and EO 832, obtained from ExoFOP archival data (PI: S. B. Howell). No close stellar companion is noted.}
    \label{fig:speckle_7166}
\end{figure}

Both host stars were observed using high-resolution imaging to rule out very close contamination sources. 
TOI-2094 was observed on 27 March 2021 (PI: C. Dressing) using the Shane AO infraRed Camera and Spectrograph (ShARCS) equipped on the Shane 3-m telescope at Lick Observatory \citep{2014SPIE.9148E..05G,2014SPIE.9148E..3AM}. The infrared adaptive optics (IR/AO) high-resolution images were taken with a field of view of $19.8''\times19.8''$ at a pixel scale of $0.033''$ in the $J$ and $K_{\rm s}$ bands (Fig. \ref{fig:speckle_2094}). There is no nearby source detected brighter than $\Delta m_{K_{\rm s}}=4.0$ at a separation of 1$''$ or $\Delta m_{K_{\rm s}}=7.5$ at 5$''$. 
TOI-7166 was observed on 3 July 2025 (PI: S. B. Howell) using the Zorro optical imager mounted on the 8.1-m Gemini South telescope at Gemini Observatory in Cerro Pachón \citep{2021FrASS...8..138S}. The images were obtained with a field of view of $2.5''\times2.5''$ having pixel scales of $0.0095''$ at 562 nm and $0.010''$ at 832 nm (Fig. \ref{fig:speckle_7166}). There is no nearby source detected brighter than $\Delta{\rm mag}=5$ at separations from $0.2''$ to $1.2''$ through the 832-nm filter.
In addition, we note that according to the renormalized unit weight error (RUWE) of the astrometric measurements from \textit{Gaia} DR3 \citep{2016A&A...595A...1G, 2023A&A...674A...1G}, both TOI-2094 ($\rm RUWE=1.32$) and TOI-7166 ($\rm RUWE=1.01$) satisfy the generally correct single-star criterion of 1.4\footnote{\url{https://dms.cosmos.esa.int/COSMOS/doc_fetch.php?id=3757412}}. 

\subsection{Spectroscopic analysis}
The determination of stellar properties through spectral energy distribution (SED) analysis relies on the spectroscopic measurements of stellar effective temperature ($T_{\rm eff}$), gravity ($\lg g$), and metallicity ([Fe/H]) as prior inputs. 
For TOI-2094, we referenced these parameters from \cite{2022AJ....163..152S} through APOGEE spectroscopic measurements ($T_{\rm eff}=3464\pm5$~K, $\lg g=4.925\pm0.037$ and $\rm [Fe/H]=-0.071\pm0.023$).
For TOI-7166, a low resolution spectrum was observed on 29 July 2025 using the Alhambra Faint Object Spectrograph and Camera\footnote{\url{https://www.not.iac.es/instruments/alfosc/}} (ALFOSC) at the Nordic Optical Telescope \citep[NOT;][]{2010ASSP...14..211D}. We used the grism \#5 and a slit of $0.5''$ to obtain the stellar spectra with a resolution of $\sim$820 covering 510--1000 nm. The spectral image was reduced using standard \texttt{IRAF} routines \citep{1986SPIE..627..733T} including bias subtraction, flat correction, wavelength calibrations with arc lamps of HeNe and ThAr, and flux calibration with a spectroscopic flux standard star BD+17 4708. Before spectroscopic analysis, strong telluric oxygen absorption bands (685--695 nm and 759--771 nm) were masked, together with the band redder than 920 nm with second order spectral contamination and strong fringing patterns. Using PHOENIX stellar flux templates, we performed three dimensional linear interpolation on model grids ($T_{\rm eff}$, $\lg g$, and [Fe/H]) to fit the normalized spectra. Using nested sampling of \texttt{PyMultiNest}, we obtained $T_{\rm eff}=3084\pm7$~K, $\lg g=5.00\pm0.06$ and $\rm [Fe/H]=0.24\pm0.04$ for TOI-7166.  
We note that the uncertainties of $T_{\rm eff}$ from spectroscopic analysis are usually underestimated due to flux scales, extinctions, and bolometric corrections as addressed in \cite{2022ApJ...927...31T}, for which a systematic uncertainty floor of 2\% in $T_{\rm eff}$ was added to the prior estimates before SED analysis. 

\subsection{HiPERCAM multicolour photometry}

On the night of transit observation for TOI-7166.01, we also observed a standard star HZ~44 for flux calibration. Its magnitudes across HiPERCAM $u_{\rm s}g_{\rm s}r_{\rm s}i_{\rm s}z_{\rm s}$ bands are 10.999, 11.389, 11.885, 12.310, and 12.653. We observed a total of 158 images at an exposure time of 0.2 seconds with slow readout and $1\times1$ pixel binning. The data were reduced using the standard HiPERCAM pipeline. In each waveband, we correct the atmospheric extinction for both standard star and the target star based on the extinction curve at the Roque de los Muchachos Observatory provided by the La Palma Technical Note 31\footnote{
\url{https://www.ing.iac.es/Astronomy/observing/manuals/ps/tech_notes/tn031.pdf}}. 
Only the out-of-transit flux measurements were used for flux calibration of TOI-7166. After extinction correction, we used the 3$\sigma$ clipping method to filter out flux outliers before averaging the measurements. The magnitude uncertainties were estimated statistically by the standard deviation of flux measurements to account for non-photon noise from the telluric and instrumental origins. An additional 1\% noise floor was added to account for uncertainties of the flux standard. The measured magnitudes are presented in Table~\ref{tab:stellar_parameters}.

\begin{table} 
    \caption{Stellar magnitudes and parameters of TOI-2094 and TOI-7166.}
    \begin{tabular}{lccc}
    \hline
        Parameter & TOI-2094 & TOI-7166 & Ref.\\
    \hline
        TIC ID & 356016119 & 288421619 & [1] \\
        RA [J2000] & 16:56:34.37 & 21:22:43.35 & [2]\\
        Dec [J2000] & +70:01:38.34 & +08:53:21.83 & [2]\\
        \hline
        $m_{\rm B}$ & $15.980\pm0.089$ & $-$ & [1]\\
        $m_{\rm V}$ & $14.409\pm0.050$ & $15.790\pm0.200$ & [1]\\
        $m_{\rm TESS}$ & $12.267\pm0.007$ & $13.123 \pm 0.008$ & [1]\\
        $m_{\rm Gaia\_B_P}$ & $14.753\pm0.003$ & $16.2398\pm0.0053$ & [2]\\
        $m_{\rm Gaia\_G}$ & $13.433\pm0.003$ & $14.4703\pm0.0029$ & [2]\\
        $m_{\rm Gaia\_R_P}$ & $12.299\pm0.004$ & $13.2068\pm0.0044$ & [2]\\
        $m_{\mathrm{PS}\_g}$ & $15.1126\pm0.0031$ & $16.5602\pm0.0046$ & [3]\\
        $m_{\mathrm{PS}\_r}$ & $13.9425\pm0.0015$ & $15.3372\pm0.0018$ & [3]\\
        $m_{\mathrm{PS}\_i}$ & $12.8850\pm0.0020$ & $13.7610\pm0.0086$ & [3]\\
        $m_{\mathrm{PS}\_z}$ & $12.3680\pm0.0020$ & $13.0323\pm0.0037$ & [3]\\
        $m_{\mathrm{PS}\_y}$ & $11.9977\pm0.0022$ & $12.6860\pm0.0046$ & [3]\\
        $m_{\rm 2MASS\_J}$ & $10.801\pm0.022$ & $11.407\pm0.022$ & [4]\\
        $m_{\rm 2MASS\_H}$ & $10.205\pm0.026$ & $10.862\pm0.023$ & [4]\\
        $m_{\rm 2MASS\_Ks}$ & $9.970\pm0.020$ & $10.598\pm0.023$ & [4]\\
        $m_{\rm WISE\_3.4}$ & $9.828\pm0.022$ & $10.421 \pm 0.023$ & [5]\\
        $m_{\rm WISE\_4.6}$ & $9.684\pm0.020$ & $10.211 \pm 0.022$ & [5]\\
        $m_{\mathrm{HCAM}\_u_s}$ & $-$ & $19.642\pm0.022$ & $\dagger$\\
        $m_{\mathrm{HCAM}\_g_s}$ & $-$ & $16.832\pm0.019$ & $\dagger$\\
        $m_{\mathrm{HCAM}\_r_s}$ & $-$ & $15.412\pm0.020$ & $\dagger$\\
        $m_{\mathrm{HCAM}\_i_s}$ & $-$ & $13.842\pm0.021$ & $\dagger$\\
        $m_{\mathrm{HCAM}\_z_s}$ & $-$ & $13.013\pm0.021$ & $\dagger$\\
        \hline
        Sp. type & M3V & M4.5V & $\dagger$\\
        $T_{\rm eff}$ [K] & $3403\pm{70}$ & $3092\pm63$ & $\dagger$\\
        $\lg g$ [cgs] & $4.89\pm{0.04}$ & $5.01\pm{0.06}$ & $\dagger$\\
        $\rm [Fe/H]$ & $-0.06\pm0.02$ & $0.24 \pm{0.04}$ & $\dagger$\\
        $L_\star$ [$L_\odot$] & $0.0180\pm{0.0014}$ & $0.0049\pm{0.0003}$ & $\dagger$\\
        $R_\star$ [$R_\odot$] & $0.384\pm{0.018}$ & $0.245\pm{0.010}$ & $\dagger$\\
        $M_\star$ [$M_\odot$] & $0.425\pm{0.059}$ & $0.221\pm{0.051}$ & $\dagger$\\  
        Dist. [pc] & $50.22 \pm 0.03 $ & $35.24\pm 0.03$ & [2]\\ 
        PM [mas/yr] & $55.30 \pm 0.02$ & $375.43 \pm 0.02$ & [2]\\  
        RV [km/s] & $-24.98 \pm 0.88$ & $-50.14 \pm 2.20$ & [2]\\
    \hline
    \end{tabular} 
    
    \medskip
    \begin{minipage}{\linewidth}
    \textbf{Notes.} PS refers to PanSTARRS. HCAM refers to HiPERCAM.\\
    \textbf{References:} 
        [1] TESS Input Catalogue \citep{2018AJ....156..102S, 2019AJ....158..138S}
        [2] Gaia DR3 \citep{2016A&A...595A...1G, 2023A&A...674A...1G} 
        [3] PanSTARRS Catalogue \citep{2016arXiv161205560C, 2020ApJS..251....7F}
        [4] 2MASS All-Sky Catalogue \citep{2003tmc..book.....C}
        [5] AllWISE Catalogue \citep{2013wise.rept....1C}
        $\dagger$ This work.
    \end{minipage}
    
    \label{tab:stellar_parameters}
\end{table}

\subsection{SED analysis}
\begin{figure*} 
    \centering
    \includegraphics[width=\linewidth]{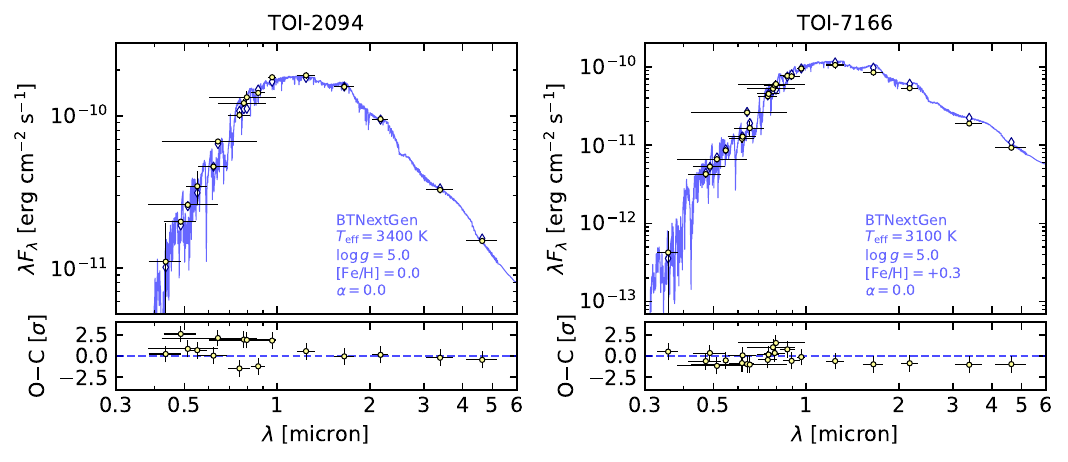} 
    \caption{Best SED models for TOI-2094 and TOI-7166. The blue solid lines are the grid models of BT-NextGen AGSS2009 \citep{2011ASPC..448...91A, 2012RSPTA.370.2765A}. The lower panels show the normalized residuals. }
    \label{fig:stellar_spectra}
\end{figure*}

The photometric measurements listed in Table \ref{tab:stellar_parameters} were used for fitting the stellar SEDs to determine stellar luminosity $L_\star$, radius $R_\star$, and mass $M_\star$. The SED fitting was performed using \texttt{astroARIADNE} developed by \cite{2022MNRAS.513.2719V}. The Bayesian model averaging method was used to consider multiple stellar models of low-mass stars: PHOENIX \citep{2013A&A...553A...6H}, BT-COND, BT-NextGen, and BT-Settl \citep{2011ASPC..448...91A,2012RSPTA.370.2765A}. We used the nested sampling algorithm to estimate model parameters. Similar to the estimates of $T_{\rm eff}$, we set systematic uncertainty floors of 2\% in luminosity, 4\% in radius, and 5\% in mass when summarizing the posterior estimates, suggested by \cite{2022ApJ...927...31T}.  
Table \ref{tab:stellar_parameters} lists the stellar spectral types, effective temperatures $T_{\rm eff}$, gravities $\lg g$, metallicities [Fe/H], luminosities $L_\star$, radii $R_\star$, and masses $M_\star$ derived from the SED fitting. Figure \ref{fig:stellar_spectra} shows the corresponding stellar SED models.
The host stars TOI-2094 and TOI-7166 are found to be M3 and M4.5 dwarf stars. The measured temperature and gravity are consistent with the reference values in the TESS Input Catalogue \citep[TIC;][]{2018AJ....156..102S, 2019AJ....158..138S} within $1\sigma$ uncertainties ($T_{\rm eff}=3457\pm157~{\rm K}$ and $\lg g=4.848\pm0.002$ for TOI-2094; $T_{\rm eff}=3166\pm157~{\rm K}$ and $\lg g=5.025\pm0.019$ for TOI-7166).

\section{Transit observations} 
\label{sect:observations}

\subsection{TESS light curves} 

TOI 2094 (TIC 356016119) was observed by the TESS in 39 sectors (14-17, 19-26, 40, 47, 49-60, 73-74, 76-86), covering 44 transits. The data were processed in the TESS Science Processing Operations Center \citep[SPOC;][]{2016SPIE.9913E..3EJ} at NASA Ames Research Center. The SPOC conducted a transit search of the light curve with all available data through sector 23 on 6 May 2020 with an adaptive, noise-compensating matched filter \citep{2002ApJ...575..493J, 2010SPIE.7740E..0DJ, 2020ksci.rept....9J}, producing a Threshold Crossing Event (TCE) with 18.79-d period. A limb-darkened transit model was fitted \citep{2019PASP..131b4506L} and a suite of diagnostic tests were conducted to help assess the planetary nature of the signal \citep{2018PASP..130f4502T}. The transit signature passed all diagnostic tests presented in the SPOC Data Validation \citep[DV;][]{2019PASP..131b4506L} reports, and the source of the transit signal was localized within $4.71 \pm 3.42$ arcsec. The TESS Science Office (TSO) reviewed the vetting information and issued an alert for TOI 2094.01 on 15 July 2020 \citep{2021ApJS..254...39G}. 

TOI 7166 (TIC 288421619) was observed by TESS in sector 82 only, covering two transits. The SPOC transit search on 30 September 2024 produced a TCE with 12.92-d period. A limb-darkened transit model was fitted, and the transit signature passed all DV diagnostic tests. The source of the transit signal was localized within $2.08 \pm 2.97$ arcsec. An alert for TOI 7166.01 was issued by the TSO on 14 November 2024.

\begin{figure}
    \centering
    \includegraphics[width=\linewidth]{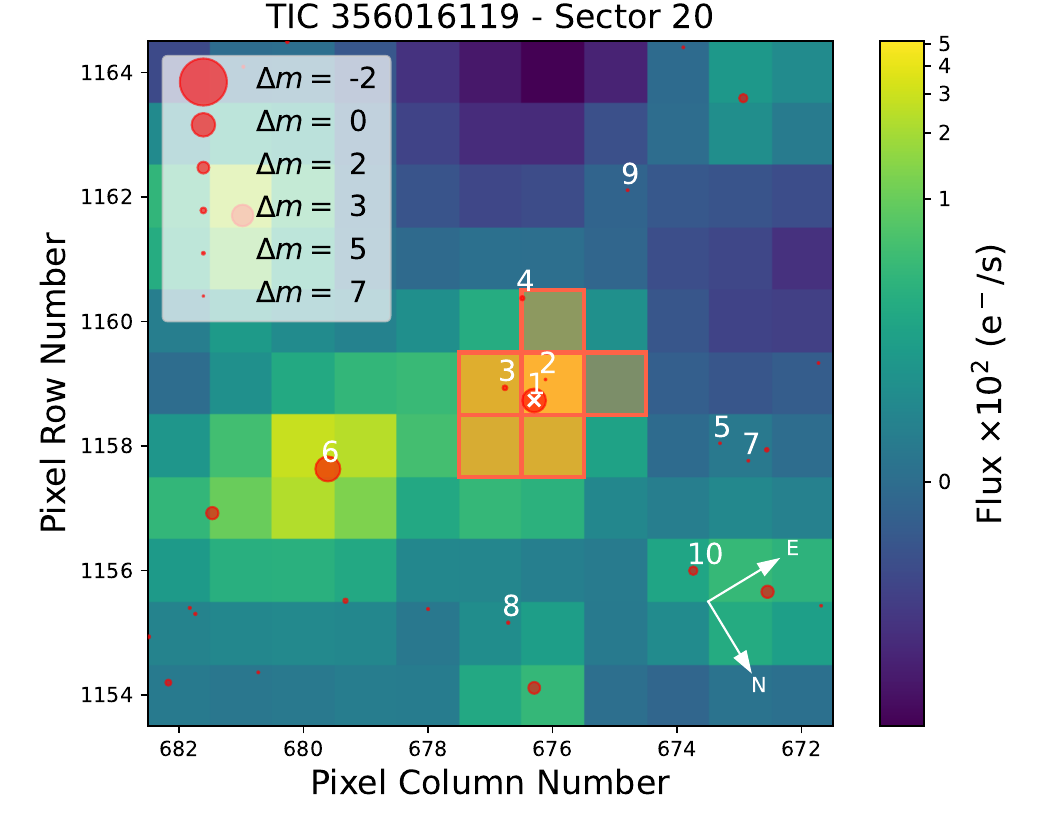} \\
    \includegraphics[width=\linewidth]{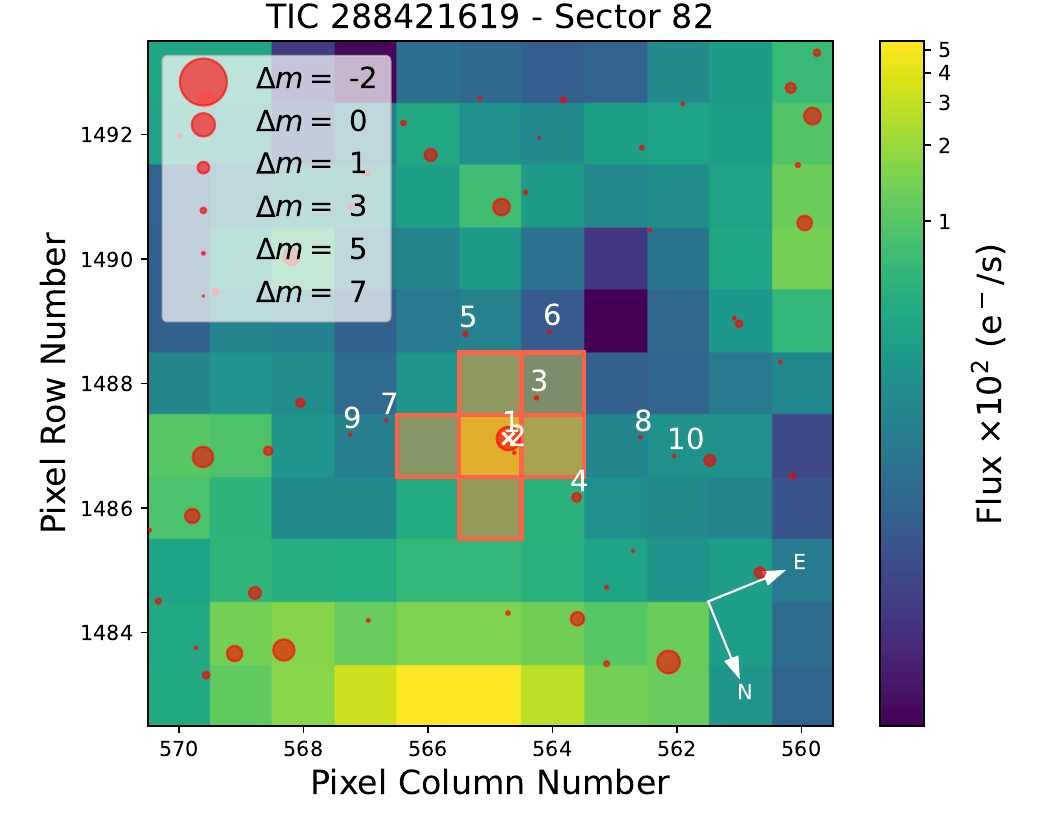}
    \caption{TESS target pixel file images of TOI-2094 ({\it top}) and TOI-7166 ({\it bottom}). The red circles are sources identified by \textit{Gaia} DR3 catalogue \citep{2016A&A...595A...1G, 2023A&A...674A...1G}, labelled with numbers sorted by angular distance to the target source. The size of the circles indicates the  magnitude contrast to the target source. The mosaic of orange pixels shows the aperture of TESS photometry, where the pixel scale is 20$''$. The plots were produced using \texttt{tpfplotter} \citep{2020A&A...635A.128A}.}
    \label{fig:tpf}
\end{figure}

The TESS data were retrieved from the Mikulski Archive for Space Telescopes\footnote{\url{https://mast.stsci.edu/portal/Mashup/Clients/Mast/Portal.html}} (MAST). We employed the SPOC Presearch Data Conditioning Simple Aperture Photometry \citep[PDCSAP;][]{2012PASP..124..985S, 2014PASP..126..100S, 2012PASP..124.1000S}. The PDCSAP light curves of both candidates are in two-minute cadence with systematic errors and long-term trends removed.  
The SAP aperture for TOI-2094 contains up to three contamination sources across different sectors (Fig. \ref{fig:tpf}). According to Gaia DR3 catalogue, their IDs are 1649377156604091520 ($\Delta m_{\rm G}=7.44$ at a distance of 7.10$''$), 1649377152308845952 ($\Delta m_{\rm G}=4.57$ at 11.25$''$), and 1649376946150348800 ($\Delta m_{\rm G}=4.96$ at 34.03$''$). The SAP aperture for TOI-7166 contains two contamination sources in section 82 (Fig. \ref{fig:tpf}): Gaia DR3 1740534087955142912 ($\Delta m_{\rm G}=6.31$ at 3.75$''$) and 1740531133017635072 ($\Delta m_{\rm G}=5.45$ at 19.96$''$). For the contamination sources that were included in TIC, their dilution effects have already been accounted for in the PDCSAP light curves. The remaining one contamination source of TOI-7166.01 would only cause a a negligible underestimation of transit depth by $\sim$26 ppm, compared to the 871-ppm uncertainty of TESS transit depths of TOI-7166.01. This can be accounted for by modelling systematic noise through a joint fit with ground-based transit light curves.

\subsection{MuSCAT2/3 light curves}

One transit of TOI-2094.01 was observed in the $g$, $r$, $i$, $z$ wavebands on 5 June 2023 using the 2m Las Cumbres Observatory Global Telescope (LCOGT) at Haleakal\=a Observatory on Maui, Hawaii. The telescope is equipped with the MuSCAT3 multi-band imager \citep{2020SPIE11447E..5KN}. The instrument MuSCAT3 has a field of view of $\sim$$9.1'\times9.1'$ and a pixel scale of $0.27''$, covering multiple comparison stars for differential photometry. The images were calibrated using the standard LCOGT \texttt{BANZAI} pipeline \citep{2018SPIE10707E..0KM}, and photometric data were extracted using \texttt{AstroImageJ} \citep{2017AJ....153...77C} with an aperture radius of 5.3$''$ in all four wavebands without contamination sources. 

Two transits of TOI-7166.01 were observed in the $g$, $r$, $i$, and $z$ wavebands on 16 and 29 July 2025 using the TCS/MuSCAT2 instrument \citep{Narita2019} at the Teide Observatory. MuSCAT2 has a field of view of $\sim$$7.4'\times7.4'$ and a pixel scale of $0.44''$. The raw data were reduced using the MuSCAT2 pipeline\footnote{
\url{https://github.com/hpparvi/MuSCAT2_transit_pipeline}}  \citep{Parviainen2019}, which performs dark and flat-field calibrations, aperture photometry, and a transit model fit including instrumental systematics. Differential photometry was carried out with fluxes from four comparison stars in the field of view. The photometric aperture size was 4.35$''$ for the target star and 2.61$''$ for the comparison stars on the first night, and 5.65$''$ and 3.04$''$ respectively on the second night, the same across all passbands. On both nights, technical issues with the guiding caused a jump in the field of view, though these drifts occurred outside the predicted transit events. Despite this, we were able to model the induced systematics with time-correlated GPs, leading to clear detections of the transits on both nights. 

\subsection{GTC/HiPERCAM light curves}

HiPERCAM is a five-channel photometric instrument covering the $u_{\rm s}$, $g_{\rm s}$, $r_{\rm s}$, $i_{\rm s}$, and $z_{\rm s}$ (300--1000 nm) bands simultaneously. It has a field of view of 1.4$'$ $\times$ 2.8$'$ with a pixel scale of 0.081$''$. It also has a COMParison star Pick-Off system (COMPO) to observe a comparison star as far as $\sim$6.5$'$. The light from COMPO was redirected to one corner of HiPERCAM CCDs. The exposure time in each passband can be stacked before readout so that fainter bands can accumulate higher flux counts and increase signal-to-noise ratios (SNRs). The 10.4-m aperture of GTC allows very fast exposures down to the order of one second for both target stars. In general, GTC/HiPERCAM can achieve very high temporal sampling rates while maintaining a high duty cycle (>97\%), ensuring the total integration time for a transit observation and precisely constraining the transit parameters especially the transit duration.

One transit of TOI-2094 b was observed using HiPERCAM on the night of 14 October 2023 (PI: J. Orell-Miquel). The observation was conducted in the full-frame slow readout mode with $1\times1$ pixel binning and lasted 1.6 hours. The exposure time from $u_{\rm s}$ to $z_{\rm s}$ bands are 24, 12, 6, 3, and 3 seconds. One comparison star at a separation of $1.2'$ (Gaia DR3 1649377190964268416, $m_{\rm G}=13.28$) was used for differential photometry.
One transit of TOI-7166 b was observed on 29 July 2025 (PI: C. Jiang). The observation wad conducted in the full-frame fast readout mode with $2\times2$ pixel binning and lasted 4.3 hours. The exposure time from $u_{\rm s}$ to $z_{\rm s}$ bands are 42.55, 1.70, 0.85, 0.43, and 0.43 seconds. One comparison star at a separation of $1.5'$ (Gaia DR3 1740534264049447936, $m_{\rm G}=14.56$) was used for differential photometry. 

We reduced the images and extracted the light curves using the HiPERCAM pipeline\footnote{
\url{https://cygnus.astro.warwick.ac.uk/phsaap/hipercam/docs/html/index.html}}. The optimal photometric aperture radius was determined as 1.8 times the measured full width at half maximum (FWHM) of stellar point source function (PSF) for each image by default. The PSF FWHMs were 0.60--0.79$''$ during the observations of TOI-2094 b and 0.52--0.91$''$ during that of TOI-7166 b. The corresponding aperture radii were 1.08--1.42$''$ for the former and 0.94--1.64$''$ for the latter. Therefore, the photometric apertures did not include any known contamination sources.
The extracted light curve of the target star was divided by that of the comparison star to reduce telluric variations, then normalized by the median values of out-of-transit part. We applied the sigma clipping method to the first differences of normalized light curves to remove outliers at the $5\sigma$ confidence level due to cosmic ray impacts.

\section{Results}
\label{sect:results}

\begin{figure}
    \centering
    \includegraphics[width=\linewidth]{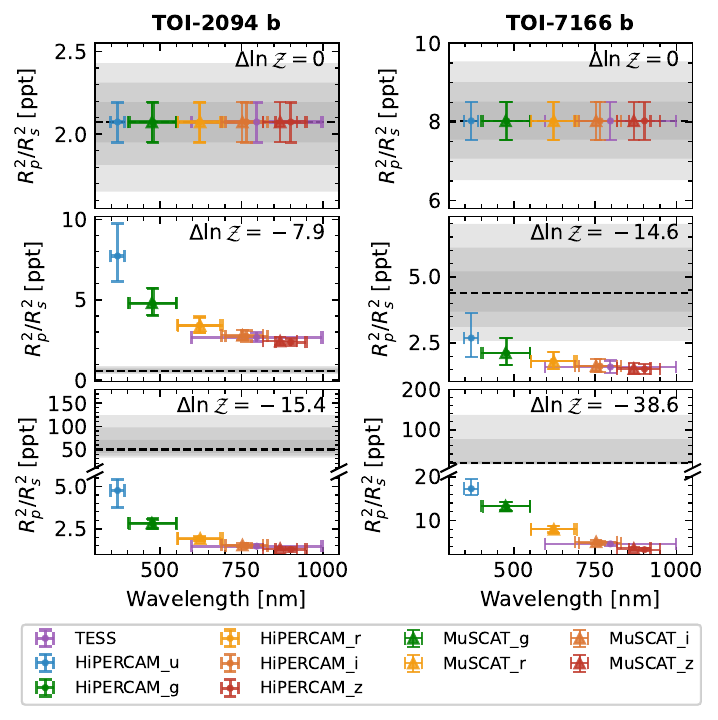}
    \caption{Apparent transit depths of TOI-2094 b ({\it left}) and TOI-7166 b ({\it right}) under different hypotheses (from top to bottom: $\mathcal{H}_1$ to $\mathcal{H}_3$). In each panel, the dashed line and gray shaded area indicate the median and 3-$\sigma$ posteriors of the true transit depth, while the error bars show the apparent transit depths.}
    \label{fig:apparent_radii}
\end{figure}

According to the false positive probabilities computed by \texttt{TRICERATOPS} combining TESS light curves and high-resolution imaging, the total FPP and NFPP are $0.133 \pm 0.086$ and $0.0050 \pm 0.0046$ for TOI-2094.01, and $0.0012 \pm 0.0005$ and $<1\times10^{-10}$ for TOI-7166.01. Considering that the photometric apertures in our ground-based transit follow-ups contains no resolved Gaia sources in addition to the target sources, we can conclude that the observed transit signals are unlikely to be contributed by contaminant sources included in the TESS photometric apertures. The false positive cases can be rejected at high confidence for TOI-7166.01, but the FPP for TOI-2094.01 is higher than the suggested 1\% FPP criterion with a large uncertainty. The major false positive case for TOI-2094.01 raised by \texttt{TRICERATOPS} is the ``STP'' scenario: there is an unresolved bound companion while the planet is transiting the secondary star. We note that this is equivalent to our $\mathcal{H}_3$ scenario, which will be further validated through our multi-colour data. In addition, as mentioned in \citet{2021AJ....161...24G}, \texttt{TRICERATOPS} performs best when $\rm S/N>15$ in TESS transit detections. Therefore, for TOI-2094.01 with a low $\rm S/N$ of 11.6, it is difficult to rule out false positives with a high confidence without high-precision follow-ups. 

\begin{figure}
    \centering
    \includegraphics[width=\linewidth]{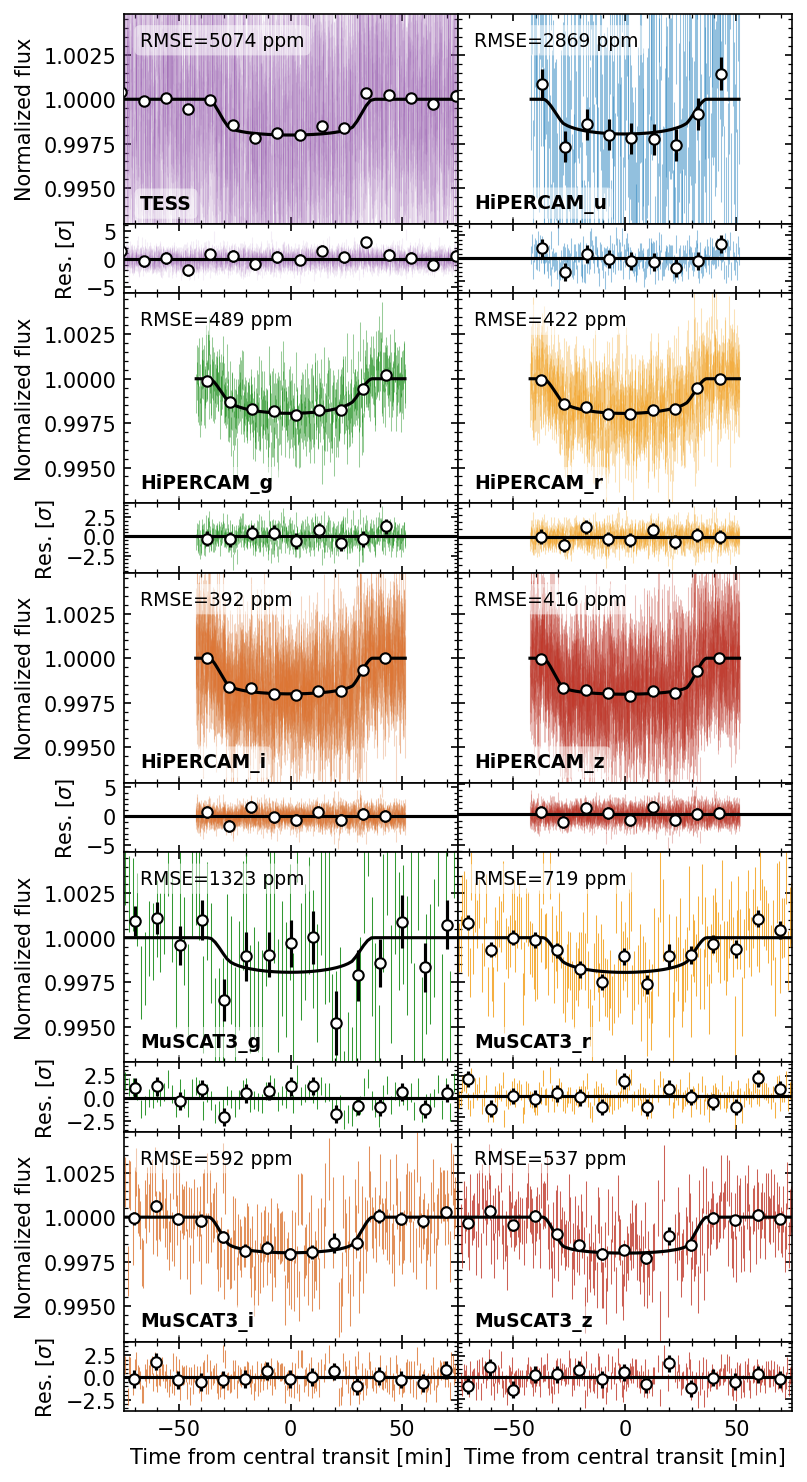}
    \caption{Best-fit light curves for TOI-2094 b. The coloured error bars are the normalized and detrended flux curves, while the black error bars are the detrended fluxes after 10-min binning. The black solid lines are the best-fit model. The small panels below the light curves are the residuals normalized by flux uncertainties. The TESS light curves have been phase-folded before binning. The RMSE in each panel has been normalized to one-minute integration time for comparison.}
    \label{fig:lc-2094}
\end{figure}

\begin{figure}
    \centering
    \includegraphics[width=\linewidth]{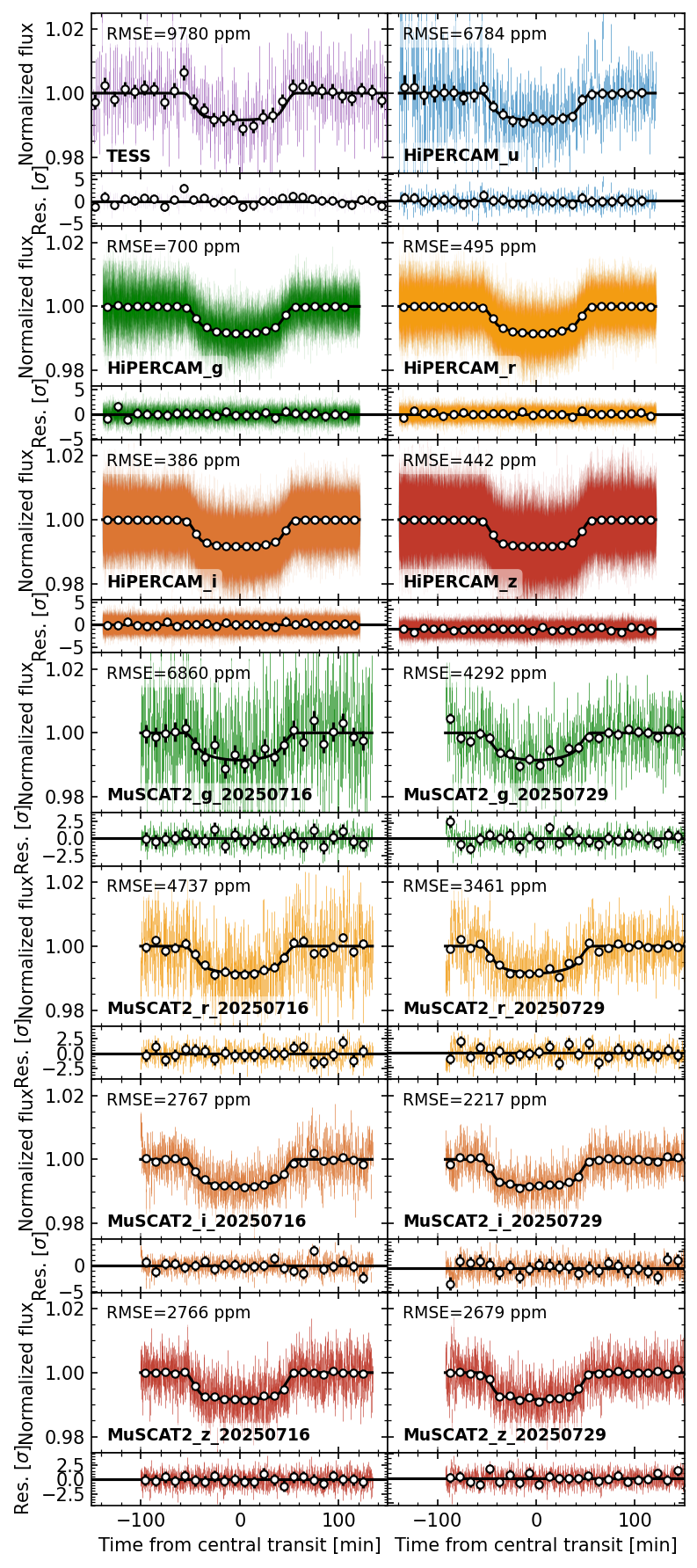}
    \caption{Best-fit light curves for TOI-7166 b. The coloured error bars are the normalized and detrended flux curves, while the black error bars are the detrended fluxes after 10-min binning. The black solid lines are the best-fit model. The small panels below the light curves are the residuals normalized by flux uncertainties. The TESS light curves have been phase-folded before binning. The RMSE in each panel has been normalized to one-minute integration time for comparison. }
    \label{fig:lc-7166}
\end{figure}

Through the joint fit of TESS, HiPERCAM, and MuSCAT2/3 light curves, we obtained the Bayesian evidence of the three model hypotheses illustrated in Section \ref{sect:framework} for TOI-2094.01 and TOI-7166.01. Figure \ref{fig:apparent_radii} shows the resulting chromatic transit depths corresponding to each hypothesis. The logarithmic Bayesian evidence ($\ln\mathcal{Z}$) of being a genuine planet transiting the target star is $2\,227\,715.09\pm0.11$ for TOI-2094.01 and $472\,537.20\pm0.11$ for TOI-7166.01, which serves as the baseline for model comparison. For TOI-2094.01, the false positive scenarios $\mathcal{H}_2$ and $\mathcal{H}_3$ are lower than $\mathcal{H}_1$ by 7.9 and 15.4; for TOI-7166.01, the false positive scenarios $\mathcal{H}_2$ and $\mathcal{H}_3$ are lower than $\mathcal{H}_1$ by 14.6 and 38.6, respectively. According to the model comparison criteria in Section \ref{sect:framework}, $\mathcal{H}_2$ and $\mathcal{H}_3$ can be rejected with strong evidence for both candidates. Combined with previous diagnostic tests by the TESS data validation and high-resolution imaging, we conclude that both candidates can be validated as transiting exoplanets (hereafter TOI-2094 b and TOI-7166 b) of their respective host stars. 
Table \ref{tab:planet_parameters} summarizes the fitted and derived parameters for TOI-2094 b and TOI-7166 b. The corresponding best-fit light curves are shown in Figs. \ref{fig:lc-2094} and \ref{fig:lc-7166}. For comparison purpose, we show the transit parameters and best-fit curves from the false positive hypotheses in Table \ref{tab:planet_params_other_scenarios} and Figs. \ref{fig:lc-2094-s23} and \ref{fig:lc-7166-s23}.

\begin{table*} 
    \caption{Posterior estimates of the planetary parameters of TOI-2094 b and TOI-7166 b assuming circular orbits.}
    \begin{tabular}{llcc}
    \hline
        Parameter & Prior & TOI-2094 b & TOI-7166 b \\
    \hline
    {\it Fitted parameters} \\ 
        Radius ratio $R_{\rm p}/R_\star$ & $\mathcal{U}(0, 0.2)$ & $0.0455^{+0.0013}_{-0.0014}$ & $0.0896^{+0.0027}_{-0.0027}$\\
        Transit epoch $T_0$ [$\rm BJD_{TDB}$] & $\mathcal{U}(-0.1, 0.1)+{\rm Const.}$ $^a$ & $2\,460\,232.36095^{+0.00032}_{-0.00032}$ & $2\,460\,886.50494^{+0.00037}_{-0.00037}$\\
        Orbital period $P$ [d] & $\mathcal{U}(-0.1, 0.1)+{\rm Const.}$ $^b$ & $18.793193^{+0.000018}_{-0.000018}$ & $12.920616^{+0.000059}_{-0.000061}$\\
        Scaled semimajor axis $a/R_\star$ & $\mathcal{U}(0.3, 1)\times {\rm Const.}$ $^c$  & $76.5^{+5.6}_{-4.4}$ & $44.2^{+1.5}_{-1.5}$\\
        Orbital impact parameter $b$ & $\mathcal{U}(0, 1+R_{\rm p}/R_\star)$ & $0.818^{+0.023}_{-0.033}$ & $0.702^{+0.025}_{-0.027}$\\
        Stellar temperature $T_{\rm eff}$ [K] & $\mathcal{N}(\mu,\sigma)$ from Table \ref{tab:stellar_parameters} & $3435^{+55}_{-65}$ & $3093^{+60}_{-63}$ \\
        Stellar gravity $\lg g$ [c.g.s] & $\mathcal{N}(\mu,\sigma)$ from Table \ref{tab:stellar_parameters} & $4.89^{+0.04}_{-0.04}$ & $5.01^{+0.06}_{-0.06}$\\
        Stellar metallicity [Fe/H] & $\mathcal{N}(\mu,\sigma)$ from Table \ref{tab:stellar_parameters} & $-0.06^{+0.02}_{-0.02}$ & $0.24^{+0.04}_{-0.04}$\\ 
    \hline
    {\it Derived parameters} \\ 
        Transit depth [ppm] & ... & $2073\pm124$ & $8027\pm488$\\
        Transit duration $T_{14}$ [h] & ... & $1.221\pm0.023$ & $1.863\pm{0.025}$\\
        Ingress duration $T_{12}$ [h] & $T_{12}=T_{34}$ & $0.150\pm0.023$ & $0.283\pm{0.025}$\\
        Semimajor axis $a$ [au] & ... & $0.136\pm0.011$ & $0.0503\pm{0.0027}$\\
        Orbital inclination $i$ [deg] & ... & $89.387\pm0.064$ & $89.089\pm{0.067}$\\
        Equilibrium temperature $T_{\rm eq}$~[K]$^d$ & null albedo & $276\pm{11}$ & $329\pm{9}$\\
        Insolation $S$ [$S_\oplus$]$^e$ & ... & $0.98\pm{0.18}$ & $1.93\pm{0.25}$\\
        Radius $R_{\rm p}$ [$R_\oplus$] & ... & $1.90\pm0.10$ & $2.39\pm{0.12}$\\
        Mass $M_{\rm p}$ [$R_\oplus$] & {\tt SPRIGHT} model & $4.2^{+0.9}_{-0.9}$ | $8.2^{+2.3}_{-1.9}$ & $6.4^{+1.9}_{-1.5}$\\
        Density $\rho_{\rm p}$ [g/cm$^3$] & \texttt{SPRIGHT} model & $3.3^{+0.6}_{-0.6}$ | $7.7^{+0.8}_{-0.7}$ & $2.6^{+0.9}_{-0.6}$\\
        RV semi-amplitude $K_\star$ [m/s] & {\tt SPRIGHT} model & $1.8^{+0.4}_{-0.4}$ | $3.5^{+1.0}_{-0.8}$ & $4.8^{+1.4}_{-1.0}$ \\
        TSM$^f$ & {\tt SPRIGHT} model & $20\pm6$ | $10\pm2$ & $78\pm5$ \\
    \hline
    \end{tabular}
    
    \medskip
    \begin{minipage}{\linewidth}
    \textbf{Notes.}
    $^a$ Constants are 2\,460\,232.35970 for TOI-2094 b and 2\,460\,886.50520 for TOI-7166 b, corresponding to the mid-transit epoch predicted by the ephemeris from SPOC DV for the latest HiPERCAM observation.
    $^b$ Constants are 18.793175 for TOI-2094 b and 12.920670 for TOI-7166 b.
    $^c$ Using Kepler's third law, the constants are calculated to be 120 for TOI-2094 b and 100 for TOI-7166 b, given the prior knowledge of the stellar bulk density and planetary orbital period. 
    $^d$ Equilibrium temperature $T_{\rm eq}=T_{\rm eff}\sqrt{R_\star/(2a)}(1-A_B)^{1/4}$, assuming a zero Bond albedo ($A_B=0$). 
    $^e$ Insolation $S=(L_\star/L_\odot) (a/{\rm au})^{-2}$.
    $^f$ Transmission spectroscopy metric in $J$ band proposed by \cite{2018PASP..130k4401K}, assuming a scale factor of 1.26 according to their Table 1.
    \end{minipage}
    \label{tab:planet_parameters}
\end{table*}

As listed in Table \ref{tab:planet_parameters}, TOI-2094 b has a radius of $1.90\pm0.1~R_\oplus$ and a transit depth of $2073\pm 124$~ppm, consistent with the SPOC DV estimates ($1.72\pm0.69~R_\oplus$, $1999\pm 173$~ppm). It orbits its M3V host star with a period of $\sim$18.793193 days at a separation of $0.136\pm 0.011$ au. With a null-albedo equilibrium temperature of $276\pm 11$~K and an Earth-like insolation of $0.98\pm0.18~S_\oplus$, it resides within the habitable zone as a temperate mini-Neptune. 
TOI-7166 b is also a temperate mini-Neptune but larger ($2.39\pm0.12~R_\oplus$) and warmer ($T_{\rm eq}=329\pm9$ K). It orbits its M4.5V host star every $\sim$12.920616 days at a separation of $\sim$0.0503 au, exhibiting a deeper transit depth of $8027\pm 488$~ppm. While our radius and transit depth measurements agree with SPOC DV estimates ($2.10\pm0.32~R_\oplus$, $8723\pm871$~ppm) within $1\sigma$, we found a significantly higher equilibrium temperature ($329\pm9$~K) and planetary insolation ($1.93\pm 0.25~S_\oplus$) than the SPOC DV estimate of $290\pm14$~K (null albedo) and $1.17\pm 0.33~S_\oplus$.  
This discrepancy comes from our tighter constraint on the semimajor axis ($44.2\pm1.5~R_\star$) compared to the $\sim$$60~R_\star$ predicted from Kepler's third law using the stellar density and planetary orbital period from the SPOC DV. 

To verify that this discrepancy arises from improved data rather than methodological differences, we re-analysed the TESS-only light curve of TOI-7166 b using the same model and fitting method presented in this work. The resulting value of $62.5^{+4.8}_{-8.4}~R_\star$ for the semi-major axis of TOI-7166 b is consistent with the SPOC DV, indicating that methodology is not the cause. We also tested whether the assumption of circular orbits could introduce bias. By setting the orbital eccentricity ($e$) and the argument of periastron ($\omega$) as free parameters in the genuine-planet hypothesis $\mathcal{H}_0$, we found that these parameters were poorly constrained for both TOI-2094 b and TOI-7166 b (see Table \ref{tab:planet_parameters_eccentric}). While this leads to larger uncertainties in the scaled semi-major axis ($a/R_\star$) and impact parameter ($b$), it did not produce significantly biased estimates. We therefore conclude that the revised orbital parameters of TOI-7166 b result from better constraints on transit geometry provided by the ground-based follow-up observations. However, the model preferences for eccentric orbits were still inconclusive based on Bayesian evidence ($\Delta \ln \mathcal{Z}<1$ for both planets). Precise constraints on eccentric orbits would require sufficient RV measurements.

Figures \ref{fig:lc-2094} and \ref{fig:lc-7166} show the detrended light curves of TOI-2094 b and TOI-7166 b. The transits of both planets can be clearly detected by  GTC/HiPERCAM and MuSCAT2/3. The SNRs of transit detections, as defined by the measured transit depth, increased from 11.6 (TESS only, $1999\pm173$ ppm) to 16.7 ($2073\pm124$ ppm) for TOI-2094 b and from 10.5 ($8723\pm871$ ppm) to 16.4 ($8027\pm488$ ppm) for TOI-7166 b.
The photometric precision achieved by different instruments and passbands can be compared by normalizing the root mean square errors (RMSEs) of the light curve residuals to a common integration time. After normalization to 1-min cadence, we found that GTC/HiPERCAM photometry exhibits RMSEs of $\sim$400 ppm in the $i$ and $z$ bands for M dwarfs of $\sim$14 G-magnitudes (Figs. \ref{fig:lc-2094} and \ref{fig:lc-7166}). When normalized by transit duration, the RMSEs in the $griz$ bands range from 46--57 ppm for TOI-2094 and 36--66 ppm for TOI-7166. This precision enables GTC/HiPERCAM to conduct multicolour validation of Earth-sized planets with transit depths of 200--300 ppm around faint K dwarfs in their habitable zones, creating valuable synergies with ongoing missions like CHEOPS \citep[Characterising Exoplanet Satellite;][]{2021ExA....51..109B} and future transit surveys such as PLATO \citep[PLAnetary Transits and Oscillations of stars;][]{2025ExA....59...26R} and ET \citep[Earth 2.0;][]{2024ChJSS..44..400G} satellites.

\section{Discussion}
\label{sect:discussion}

\subsection{Planet properties from statistical perspectives}
\label{sect:statistical}

While the bulk composition of exoplanets is complex, statistical analyses of planets with well-measured masses and radii have revealed certain mass--radius relationships \citep[e.g.][]{2014ApJ...783L...6W, 2019PNAS..116.9723Z, 2020A&A...634A..43O, 2022Sci...377.1211L, 2024A&A...688A..59P}. For small transiting planets around M dwarfs (STPMs; $R_{\rm p}<4R_\oplus$), \cite{2022Sci...377.1211L} proposed a classification into three populations: rocky, water-rich, and gas-rich, based on bulk density. However, \cite{2024A&A...688A..59P} found no clear compositional or radius gap between super-Earths and sub-Neptunes orbiting M dwarfs. Thus, the nature of ``water worlds'' remains particularly degenerate, as their observed properties can be explained by several distinct internal structures and compositions. To better characterise TOI-2094 b and TOI-7166 b from a statistical perspective, we estimated their masses, bulk densities, and RV semi-amplitudes by applying the probabilistic mass--radius relationship from the \texttt{SPRIGHT} code \citep{2024MNRAS.527.5693P}. This code utilizes a pre-trained mass--density--radius probability distribution derived from the updated STPM catalogue \citep{2022Sci...377.1211L}, which is applicable to planets between 0.5 and 4 $R_\oplus$ orbiting M dwarfs. 
 
\begin{figure}
    \centering
    \includegraphics[width=\linewidth]{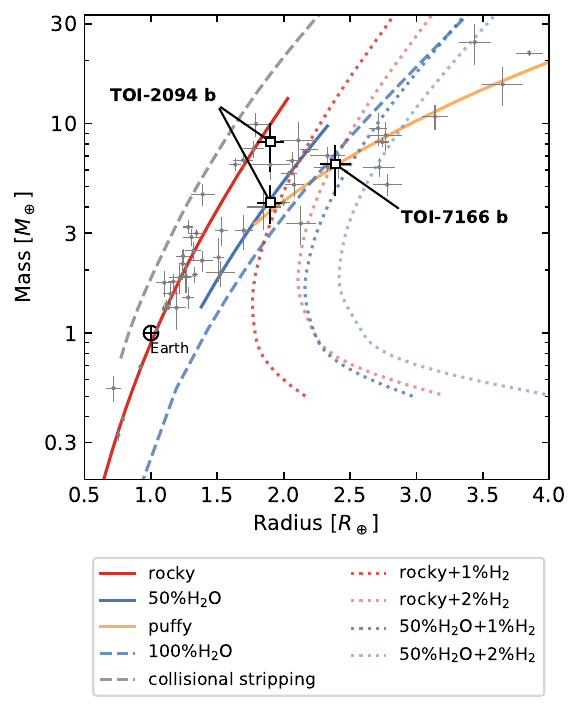}
    \caption{Mass-radius diagram of Earth-to-Neptune-sized exoplanets. The gray error bars represent the planets from the STPM catalogue. The black error bars represent the masses of TOI-2094 b and TOI-7166 b predicted using \texttt{SPRIGHT}. The solid lines represent the empirical relationships defined in \texttt{SPRIGHT}. The dashed and dotted lines represent the composition models from \citet{2019PNAS..116.9723Z}. }
    \label{fig:composition}
\end{figure}

We found that TOI-2094 b exhibits a bimodal mass distribution ($4.2\pm0.9~M_\oplus$ at 76\% possibility; $8.2\pm2.0~M_\oplus$ at 24\% possibility; Fig. \ref{fig:composition}) due to its radius of $1.90 \pm 0.10~R_\oplus$ in the transition region between rocky compositions (higher mass solution) and volatile-rich compositions (lower mass solution). To test the sensitivity of this result to the input planetary catalogue, we generated a new empirical model using the updated PlanetS catalogue\footnote{\url{https://dace.unige.ch/exoplanets/?}} from \cite{2024A&A...688A..59P} in \texttt{SPRIGHT}. We found that the mass predictions were highly consistent between catalogues and thus the results are robust. We note that the bimodal mass prediction of TOI-2094 b does not primarily reflect compositional uncertainties from measurement errors but arises from the prior-driven ambiguity intrinsic to the \texttt{SPRIGHT} model, stemming from the degenerate density-radius relationship of mini-Neptunes. Further mass measurements and atmospheric characterization will therefore be essential to resolve this ambiguity and constrain the planet's interior structure and bulk composition. 

With a larger radius of $2.39 \pm 0.12~R_\oplus$, TOI-7166 b shows a single-peaked mass distribution of $6.4 \pm 2.0~M_\oplus$ and a bulk density of $2.6\pm0.9~{\rm g/cm^3}$ (Fig. \ref{fig:composition}). This suggests TOI-7166 b is likely to be volatile-rich. However, several different compositions can explain the inferred mass and radius of TOI-7166 b, including a gaseous planet, a water world, or a rocky/water-rich core with a H$_2$/He envelope. As shown in Fig. \ref{fig:composition}, this degeneracy persists even with mass and radius uncertainties below 5\%. Atmospheric characterisation through transmission spectroscopy may provide additional constraints to distinguish between these scenarios.

To assess the feasibility of mass determination via RV measurements, we estimated the required telescope time using the \texttt{RVFollowupCalculator} code \citep{2018AJ....156...82C}. We simulated observations with the Gemini/MAROON-X spectrograph, which is one of the few instruments suitable for these two targets given their stellar magnitudes and declinations. Adopting optimistic assumptions, including a stellar RV jitter of 1 m/s and no signal from additional planets, we determined the time needed to achieve a mass precision of $\pm$20\% for precise atmospheric characterisation \citep{2019ApJ...885L..25B}. Our simulations indicate that that $\sim$83 hours (142 RV measurements) are needed for the lower-mass solution of TOI-2094 b. For TOI-7166 b, $\sim$15 hours (25 RV measurements) are needed to achieve a similar precision. Therefore, TOI-7166 b is a more feasible target for RV follow-up with current facilities.

The transmission and emission spectroscopy metric (TSM and ESM) proposed by \cite{2018PASP..130k4401K} are key metrics for prioritizing atmospheric follow-up observations. Since both planets are too cold for emission spectroscopy, we focus on transmission spectroscopy feasibility. The TSM is calculated as
\begin{equation}
    {\rm TSM} = ({\rm scale~factor}) \times \frac{R_{\rm p}^3 T_{\rm eq}}{R_\star^2 M_{\rm p}} \times 10^{-m_J/5},
\end{equation}
where the scale factor is taken from the Table 1 in \cite{2018PASP..130k4401K}, $R_{\rm p}$, $M_{\rm p}$, and $R_\star$ are in units of Earth radii, Earth masses, and solar radii, and $m_J$ is the 2MASS $J$-band magnitude. 

\begin{figure}
    \centering
    \includegraphics[width=\linewidth]{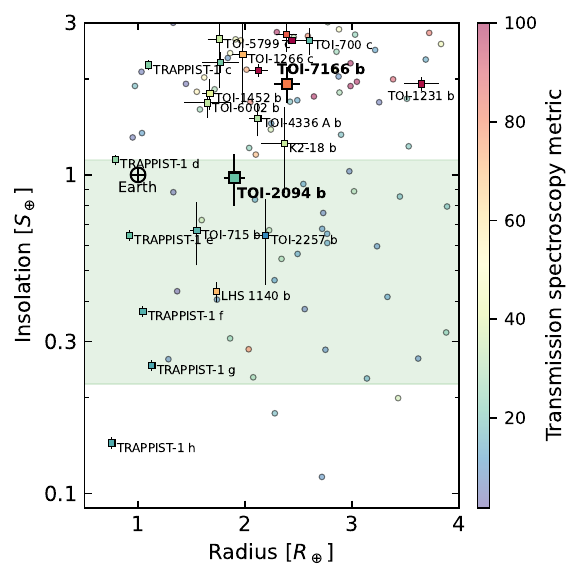}
    \caption{Transiting exoplanets of 0.5--4 Earth radii around habitable zones with TSM greater than 10. The error bars with square markers are confirmed or validated planets. The coloured dots are TESS candidates from the ExoFOP database. The colour of each marker indicates the TSM defined in \citet{2018PASP..130k4401K}. The green shaded area is the habitable-zone boundaries determined by the moist greenhouse (inner edge) and maximum greenhouse (outer edge) for host stars with $T_{\rm eff}$ in the range 3000--7000 K according to \citet{2013ApJ...765..131K}. }
    \label{fig:insolation}
\end{figure}

Using the predicted masses, our calculations suggest that TOI-7166 b has a TSM of $78\pm5$,  ranking it among the most favourable temperate mini-Neptunes for transmission spectroscopy (Fig. \ref{fig:insolation}). In contrast, TOI-2094 b has a lower TSM of $20\pm6$ assuming $M_{\rm p}\approx 4.2~R_\oplus$, or $10\pm2$ assuming $M_{\rm p}\approx 8.2~R_\oplus$. While this presents greater observational challenges, TOI-2094 b remains a compelling target as one of the few validated habitable-zone planets with a potentially observable atmosphere. If future RV measurements confirm the lower mass solution, its TSM would still place it within the range of interest for atmospheric characterisation.

Furthermore,  \cite{2025arXiv250925323D} suggests using transmission spectroscopy as an alternative approach to constraining planetary masses, provided that distinct spectral features are detected in the transmission spectra. Due to the model degeneracy between planetary mass and the atmospheric mean molecular weight, this method is generally applicable for gaseous planets larger than 2 $R_\oplus$ and with a TSM $\ge50$. \cite{2025arXiv250925323D} found that the relative mass uncertainty constrained by JWST transmission spectroscopy of a single transit observation scales linearly with the TSM. For TOI-7166 b with a TSM of 78, one transit observation with JWST is expected to constrain the mass with an uncertainty of $\sim$25\%. This highlights the potential of atmospheric follow-up with JWST to provide complementary mass constraints for TOI-7166 b.

\subsection{Implications for atmospheric characterisation}
 
To further demonstrate their atmospheric characterisation prospects, we simulated JWST transmission spectra for both planets across the 0.85--10 $\upmu$m wavelength range using the NIRISS, NIRSpec, and MIRI instruments. We first estimated the observational precision for each instrument using the \texttt{PandExo} code \citep{2017PASP..129f4501B} with default configurations. Theoretical transmission spectra were then generated using the \texttt{petitRADTRANS} code \citep{2019A&A...627A..67M} under various atmospheric model assumptions, to which we added Gaussian noise based on the transit depth uncertainties derived by \texttt{PandExo}.

The NIRISS simulations were based on the single object slitless spectroscopy \citep[SOSS;][]{2023PASP..135i8001D, 2023PASP..135g5001A} mode with the GR700XD grism, CLEAR filter, SUBSTRIP96 subarray, and NISRAPID readout mode. The NIRSpec simulations were based on the bright object time-series \citep[BOTS;][]{2022A&A...661A..80J, 2023PASP..135c8001B} mode with the G395H grating, SUB2048 subarray, and NRSRAPID readout mode. The MIRI simulations were based on the slitless low-resolution spectroscopy \citep[LRS;][]{2015PASP..127..623K} mode with the SLITLESSPRISM subarray and FASTR1 readout mode. The integration time was optimized for all the instrumental configurations to reach an 80\% saturation level with an out-of-transit baseline equal to the in-transit time.

\begin{figure}
    \includegraphics[width=\linewidth]{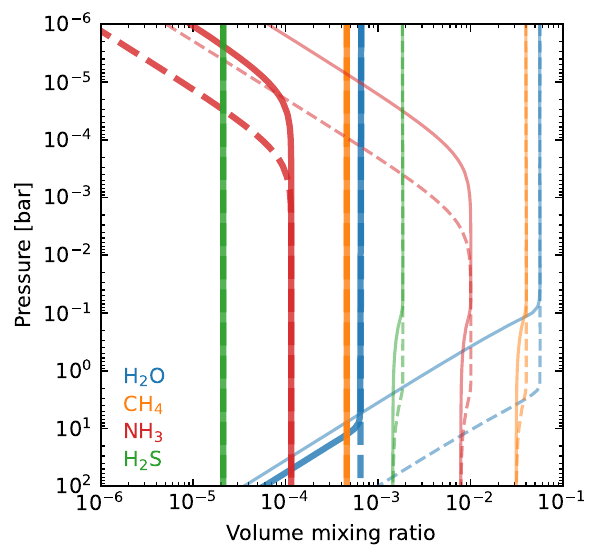}
    \caption{Volume mixing ratios of major chemical species as a function of pressure for TOI-2094 b (solid line) and TOI-7166 b (dashed line). The colours of the lines indicate different species. The thick lines correspond to solar metallicity ($\rm [Fe/H]=0$ and $\rm C/O=0.55$), and the thin lines correspond to enhanced metallicity ($\rm [Fe/H]=2$ and $\rm C/O=0.55$).}
    \label{fig:atm_composition}
\end{figure}

For the transmission spectra modelling, we assumed H$_2$-He dominated, isothermal atmospheres in thermochemical equilibrium. The atmospheric temperatures were approximated to planetary equilibrium temperatures. The continuum opacity sources were contributed by Rayleigh scattering and collisionally-induced absorption from H$_2$ and He. We derived the volume mixing ratios (VMRs) of gaseous species using the pre-calculated chemistry grid provided by \texttt{petitRADTRANS}. Given their similar equilibrium temperatures, TOI-2094 b and TOI-7166 b are expect to possess temperate atmospheres. The equilibrium chemistry model suggests H$_2$O, CH$_4$, NH$_3$, and H$_2$S as the major gaseous species (VMR > $10^{-6}$). As shown in Fig. \ref{fig:atm_composition}, the VMR profiles for these species are generally consistent between the two planets, although NH$_3$ in the upper atmosphere and H$_2$O in the lower atmosphere exhibit greater sensitivity to temperature differences. The corresponding mean molecular weights (MMWs) are 2.33 u and 4.43 u for solar and 100$\times$ solar metallicities, respectively. We note that the composition of such cool atmospheres can be altered by disequilibrium processes like photochemistry and atmospheric dynamics \citep{2019ApJ...883..194M, 2025A&A...699A.306A}. For instance, stellar UV radiation can photodissociate H$_2$S, potentially producing near-infrared absorption signatures from SO and SO$_2$ (e.g. GJ 3470 b, \citealt{2024ApJ...970L..10B}; L 98-59 d, \citealt{2024ApJ...975L..11B}; HAT-P-26 b, \citealt{2025arXiv250916082G}). Meanwhile, chemical quenching from vertical mixing or horizontal transport can also modify the mixing ratios of major species \citep[e.g.][]{2023ApJ...959L..30T}. As demonstrated by \cite{2025A&A...699A.306A} for GJ 1214 b and GJ 436 b, these disequilibrium processes can significantly deplete species such as CH$_4$ and H$_2$S in the upper atmosphere. Consequently, the spectral features of these species are particularly sensitive to such effects. A detailed investigation of disequilibrium chemistry is, however, beyond the scope of this work. 

\begin{figure*}
    \includegraphics[width=\linewidth]{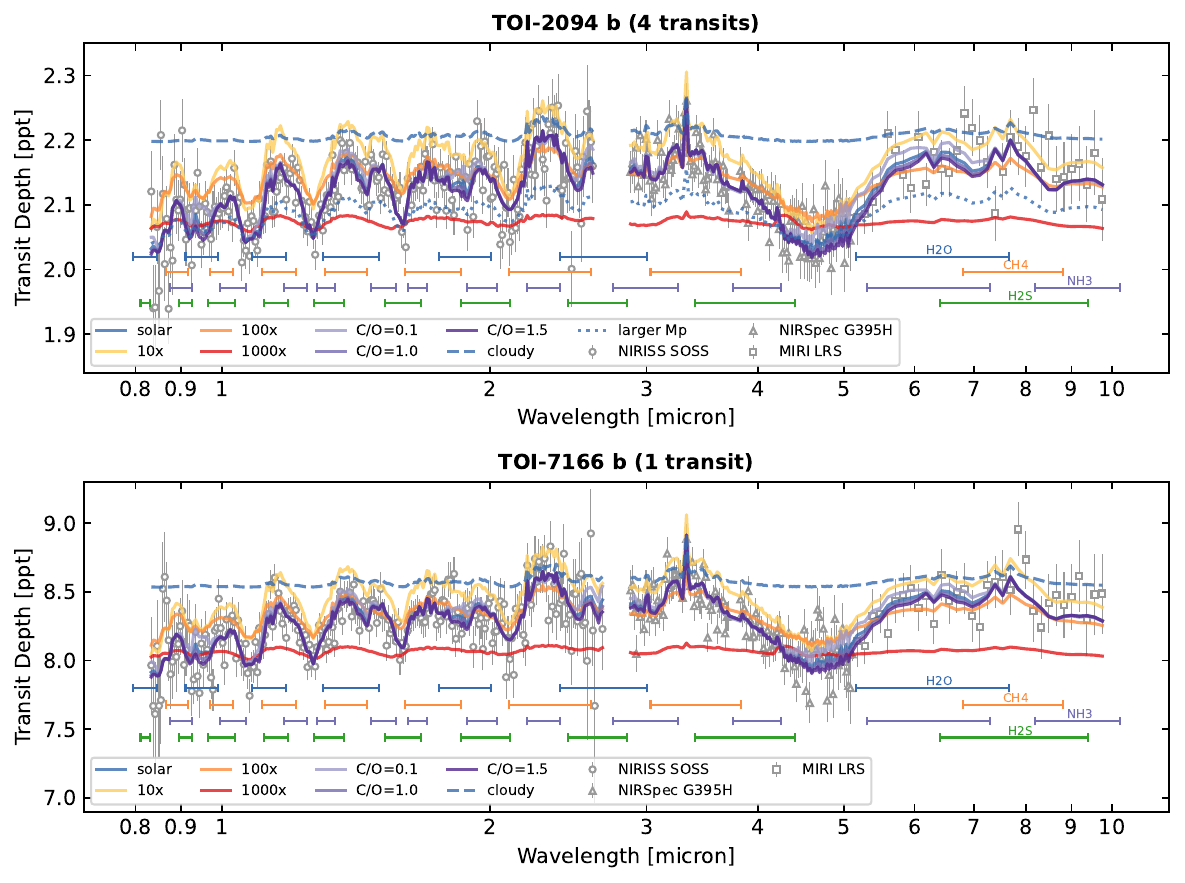}
    \caption{Simulated JWST transmission spectra of TOI-2094 b (upper panel) and TOI-7166 b (lower panel) assuming equilibrium chemistry. The NIRISS and NIRSpec spectra were rebinned to $\lambda/\Delta\lambda=200$, and the MIRI/LRS spectra were rebinned to $\lambda/\Delta\lambda=50$. The error bars correspond to the simulated spectra with solar-abundance chemical compositions (fiducial model), and different markers indicate different instruments. The horizontal line segments indicate the band heads of different molecules. The SNRs for TOI-2094 b were enhanced by four transit observations for each instrument.  
    }
    \label{fig:jwst_simulations}
\end{figure*}

We adopted a solar-abundance atmosphere ($\rm [Fe/H]=0$, $\rm C/O=0.55$) as the fiducial model and compared it with other scenarios including enhanced metallicities (10--1000$\times$), varying carbon-to-oxygen ratios (C/O at 0.1, 1.0, and 1.5), and the presence of a uniform cloud deck at 1 mbar. The model transmission spectra were calculated at a spectral resolution of 1000 using the line lists sampled from the correlated-k method. As shown in Fig.~\ref{fig:jwst_simulations}, the simulated spectra for TOI-2094 b and TOI-7166 b exhibit nearly identical features due to their similar temperatures, but the variation amplitudes are significantly smaller for TOI-2094 b. TOI-7166 b can exhibit prominent absorption features from H$_2$O and CH$_4$ in clear atmosphere scenarios with $\rm [Fe/H] \lesssim 2 $ (MMW $\lesssim 4.4$~u). Using the fiducial model as a benchmark, a featureless flat spectrum can be rejected at significance levels of 17.4~$\sigma$ (NIRISS SOSS), 17.9~$\sigma$ (NIRSpec G395H), and 3.8~$\sigma$ (MIRI LRS), establishing TOI-7166 b as a highly favourable target for atmospheric characterisation with NIRISS and NIRSpec. 

Atmospheric characterisation of TOI-2094 b is more challenging due to its smaller transit depth and scale height\footnote{Pressure scale height $H=kT/(\mu g)$, where $k$ is the Boltzmann constant, $T$ is the temperature, $\mu$ is the mean molecular weight, and $g$ is the gravity.}. Achieving a S/N comparable to TOI-7166 b would require approximately four times the number of transit observations. Furthermore, the higher-mass solution for TOI-2094 b would result in a higher gravity and a smaller scale height. This would mute the spectral features and has model degeneracy with the high MMW (or metallicity) scenarios (upper panel of Fig. \ref{fig:jwst_simulations}). Despite these challenges, a successful detection of spectral features would still provide an upper limit on the planetary mass. With multiple observations, its mass would be constrained to a 25\% uncertainty, similar to the prospects for TOI-7166 b, as discussed in section \ref{sect:statistical}. We show that atmospheric characterisation remains feasible for TOI-2094 b under the most favourable conditions, such as a metal-rich, haze-free atmosphere. Given that it is the fourth habitable-zone mini-Neptunes amenable to atmospheric studies (Fig.~\ref{fig:insolation}), it warrants significant observational resources to explore its atmospheric properties, including potential biosignatures.

\section{Conclusion}
\label{sect:conclusion}

We have validated two TESS candidates, TOI-2094 b and TOI-7166 b, as transiting exoplanets through multicolour transit photometry. TOI-2094 b is a temperate mini-Neptune ($R_{\rm p}=1.90\pm0.10~R_\oplus$, $P=18.793193\pm0.000018$ days, $T_{\rm eq}=276\pm11$ K) located within the habitable zone of its M3V host star. TOI-7166 b is a temperate mini-Neptune ($R_{\rm p}=2.39\pm0.12~R_\oplus$, $P=12.920616\pm0.000061$ days, $T_{\rm eq}=329\pm9$ K) near the inner edge of the habitable zone of its M4.5V host star. Our analysis, combining TESS data with ground-based follow-up observations from GTC/HiPERCAM and other instruments, robustly ruled out false positive scenarios through Bayesian model comparison and provided precise measurements of the planetary transit and physical parameters.

Both planets are promising targets for atmospheric characterisation. TOI-7166 b, with its high TSM of $\sim$78, ranks among the most favourable temperate mini-Neptunes for JWST transmission spectroscopy. TOI-2094 b, despite a lower TSM (10--20 depending on the mass scenario), remains a compelling target for habitability studies due to its location within the habitable zone. Future observations with JWST, Ariel, and ground-based extremely large telescopes could probe the atmospheric properties of these planets and provide insights into the nature of habitable-zone worlds around M dwarfs.

This work demonstrates the power of multicolour validation with 8--10-m class telescopes equipped with high-speed, multi-band imagers such as HiPERCAM. The photometric precision achieved by GTC/HiPERCAM enables the validation of Earth-sized transiting exoplanets and the rejection of false positives mimicking their signals. This method is particularly valuable for validating small planets in the habitable zone, where RV confirmation is observationally expensive. multicolour validation with GTC/HiPERCAM opens a pathway to efficiently confirm a growing sample of small habitable-zone planet candidates from TESS and future missions like PLATO and ET. It will help build a statistically significant sample of terrestrial exoplanets for demographic studies and prioritize targets for detailed atmospheric characterisation, advancing our understanding of planet formation and the potential for life beyond the Solar System.

\section*{Acknowledgements}
We thank the anonymous referee for their careful review and valuable comments on our manuscript.
We acknowledge financial support from the Agencia Estatal de Investigaci\'on of the Ministerio de Ciencia e Innovaci\'on MCIN/AEI/10.13039/501100011033 and the ERDF "A way of making Europe" through projects PID2021-125627OB-C32 and PID2024-158486OB-C32, and from the Centre of Excellence "Severo Ochoa" award to the Instituto de Astrofisica de Canarias.
This work is partly supported by JSPS KAKENHI Grant Numbers JP24H00017, JP24K00689, JP21K13955, JP24K17082, JP24H00248 and JSPS Bilateral Program Number JPJSBP120249910. 
This work is based on observations made with 
(1) the HiPERCAM instrument at the Gran Telescopio Canarias (GTC) operated by the Observatorio del Roque de los Muchachos (La Palma, Spain) under Spanish Comisión de Asignación de Tiempos (CAT): GTC126-23B (PI: J. Orell-Miquel) and Director's Discretionary Time (DDT): GTC07-25ADDT (PI: C. Jiang), 
(2) the MuSCAT2 instrument at the Telescopio Carlos Sánchez operated by the Observatorio del Teide (Tenerife, Spain), supported by JSPS KAKENHI (JP24H00017, JP24K00689) and JSPS Bilateral Program (JPJSBP120249910), 
(3) the MuSCAT3 instrument at the Faulkes Telescope North (Maui, HI), operated by the Las Cumbres Observatory, developed by the Astrobiology Center and supported by JSPS KAKENHI (JP18H05439) and JST PRESTO (JPMJPR1775),
(4) the ALFOSC instrument at the Nordic Optical Telescope under the fast-track service observing program P71-410 (PI: C. Jiang), which is provided by the Instituto de Astrofisica de Andalucia (IAA) under a joint agreement with the University of Copenhagen and NOT,
and (5) the High-resolution imaging instrument Zorro mounted on the Gemini South telescope of the international Gemini Observatory under Gemini LLP Proposal Number: GS-2025A-Q-219, funded by the NASA Exoplanet Exploration Program and built at the NASA Ames Research Center by Steve B. Howell, Nic Scott, Elliott P. Horch, and Emmett Quigley. 
This work used the public data from the following teams and resources: 
(1) the TESS mission and MAST data archive\footnote{\url{https://mast.stsci.edu/portal/Mashup/Clients/Mast/Portal.html}} at the Space Telescope Science Institute (STScI, NASA contract NAS 5-26555),  
(2) the European Space Agency (ESA) space mission {\it Gaia}\footnote{\url{https://www.cosmos.esa.int/gaia}} and the {\it Gaia} Data Processing and Analysis Consortium (DPAC), 
(3) the Exoplanet Follow-up Observation Program (ExoFOP, DOI: 10.26134/ExoFOP5), 
which is operated by the California Institute of Technology, under contract with the National Aeronautics and Space Administration under the Exoplanet Exploration Program,
(4) the Pan-STARRS1 Surveys (PS1) and the PS1 public science archive\footnote{\url{https://catalogs.mast.stsci.edu/panstarrs/}},
(5) the Spanish Virtual Observatory project\footnote{\url{https://svo.cab.inta-csic.es}} funded by MCIN/AEI/10.13039/501100011033/ through grant PID2020-112949GB-I00, 
and (6) the VizieR catalogue service (DOI : 10.26093/cds/vizier) at CDS\footnote{\url{https://cds.unistra.fr/}} (Strasbourg, France). 
This work used the software \texttt{IRAF} distributed by the National Optical Astronomy Observatory.  
This work used the following Python packages:
\texttt{astroARIADNE}, \texttt{astropy},  \texttt{astroquery}, \texttt{celerite}, \texttt{corner}, \texttt{hcam\_finder}, \texttt{hipercam}, \texttt{LDTK}, \texttt{lightkurve}, \texttt{matplotlib}, \texttt{mpi4py}, \texttt{numba}, \texttt{numpy}, \texttt{PandExo}, \texttt{PyAstronomy}, \texttt{PyRAF}, \texttt{PyTransit},  \texttt{PyMultiNest}, \texttt{scipy}, and \texttt{tpfplotter}.

\section*{Data Availability}

Our photometric data observed using GTC/HiPERCAM and spectroscopic data observed using NOT/ALFOSC will be shared on reasonable request to the corresponding author. The TESS data are accessible from the MAST website\footnote{\url{https://mast.stsci.edu/portal/Mashup/Clients/Mast/Portal.html}}. Other ground-based follow-up observations are accessible from the ExoFOP website\footnote{\url{https://exofop.ipac.caltech.edu/tess/}}.



\bibliographystyle{mnras}
\bibliography{references} 

@ARTICLE{2021AJ....161...24G,
       author = {{Giacalone}, Steven and {Dressing}, Courtney D. and {Jensen}, Eric L.~N. and {Collins}, Karen A. and {Ricker}, George R. and {Vanderspek}, Roland and {Seager}, S. and {Winn}, Joshua N. and {Jenkins}, Jon M. and {Barclay}, Thomas and {Barkaoui}, Khalid and {Cadieux}, Charles and {Charbonneau}, David and {Collins}, Kevin I. and {Conti}, Dennis M. and {Doyon}, Ren{\'e} and {Evans}, Phil and {Ghachoui}, Mourad and {Gillon}, Micha{\"e}l and {Guerrero}, Natalia M. and {Hart}, Rhodes and {Jehin}, Emmanu{\"e}l and {Kielkopf}, John F. and {McLean}, Brian and {Murgas}, Felipe and {Palle}, Enric and {Parviainen}, Hannu and {Pozuelos}, Francisco J. and {Relles}, Howard M. and {Shporer}, Avi and {Socia}, Quentin and {Stockdale}, Chris and {Tan}, Thiam-Guan and {Torres}, Guillermo and {Twicken}, Joseph D. and {Waalkes}, William C. and {Waite}, Ian A.},
        title = "{Vetting of 384 TESS Objects of Interest with TRICERATOPS and Statistical Validation of 12 Planet Candidates}",
      journal = {\aj},
         year = 2021,
        month = jan,
       volume = {161},
       number = {1},
          eid = {24},
        pages = {24},
          doi = {10.3847/1538-3881/abc6af},
archivePrefix = {arXiv},
       eprint = {2002.00691},
 primaryClass = {astro-ph.EP},
       adsurl = {https://ui.adsabs.harvard.edu/abs/2021AJ....161...24G}
}

@ARTICLE{2025AJ....170..298S,
       author = {{Schmidt}, Stephen P. and {MacDonald}, Ryan J. and {Tsai}, Shang-Min and {Radica}, Michael and {Wang}, Le-Chris and {Ahrer}, Eva-Maria and {Bell}, Taylor J. and {Fisher}, Chloe and {Thorngren}, Daniel P. and {Wogan}, Nicholas and {May}, Erin M. and {Ferrari}, Piero and {Bennett}, Katherine A. and {Rustamkulov}, Zafar and {L{\'o}pez-Morales}, Mercedes and {Sing}, David K.},
        title = "{A Comprehensive Reanalysis of K2-18 b's JWST NIRISS+NIRSpec Transmission Spectrum}",
      journal = {\aj},
         year = 2025,
        month = dec,
       volume = {170},
       number = {6},
          eid = {298},
        pages = {298},
          doi = {10.3847/1538-3881/ae019a},
archivePrefix = {arXiv},
       eprint = {2501.18477},
 primaryClass = {astro-ph.EP},
       adsurl = {https://ui.adsabs.harvard.edu/abs/2025AJ....170..298S}
}

@ARTICLE{2024arXiv240303325B,
       author = {{Benneke}, Bj{\"o}rn and {Roy}, Pierre-Alexis and {Coulombe}, Louis-Philippe and {Radica}, Michael and {Piaulet}, Caroline and {Ahrer}, Eva-Maria and {Pierrehumbert}, Raymond and {Krissansen-Totton}, Joshua and {Schlichting}, Hilke E. and {Hu}, Renyu and {Yang}, Jeehyun and {Christie}, Duncan and {Thorngren}, Daniel and {Young}, Edward D. and {Pelletier}, Stefan and {Knutson}, Heather A. and {Miguel}, Yamila and {Evans-Soma}, Thomas M. and {Dorn}, Caroline and {Gagnebin}, Anna and {Fortney}, Jonathan J. and {Komacek}, Thaddeus and {MacDonald}, Ryan and {Raul}, Eshan and {Cloutier}, Ryan and {Acuna}, Lorena and {Lafreni{\`e}re}, David and {Cadieux}, Charles and {Doyon}, Ren{\'e} and {Welbanks}, Luis and {Allart}, Romain},
        title = "{JWST Reveals CH$_4$, CO$_2$, and H$_2$O in a Metal-rich Miscible Atmosphere on a Two-Earth-Radius Exoplanet}",
      journal = {arXiv e-prints},
         year = 2024,
        month = mar,
          eid = {arXiv:2403.03325},
        pages = {arXiv:2403.03325},
          doi = {10.48550/arXiv.2403.03325},
archivePrefix = {arXiv},
       eprint = {2403.03325},
 primaryClass = {astro-ph.EP},
       adsurl = {https://ui.adsabs.harvard.edu/abs/2024arXiv240303325B}
}

@ARTICLE{2025arXiv250925323D,
       author = {{de Wit}, Julien and {Seager}, Sara and {Niraula}, Prajwal},
        title = "{Combined Exoplanet Mass and Atmospheric Characterization for Accelerated Exoplanetology}",
      journal = {arXiv e-prints},
         year = 2025,
        month = sep,
          eid = {arXiv:2509.25323},
        pages = {arXiv:2509.25323},
          doi = {10.48550/arXiv.2509.25323},
archivePrefix = {arXiv},
       eprint = {2509.25323},
 primaryClass = {astro-ph.EP},
       adsurl = {https://ui.adsabs.harvard.edu/abs/2025arXiv250925323D}
}

@ARTICLE{2019ApJ...883..194M,
       author = {{Molaverdikhani}, Karan and {Henning}, Thomas and {Molli{\`e}re}, Paul},
        title = "{From Cold to Hot Irradiated Gaseous Exoplanets: Fingerprints of Chemical Disequilibrium in Atmospheric Spectra}",
      journal = {\apj},
         year = 2019,
        month = oct,
       volume = {883},
       number = {2},
          eid = {194},
        pages = {194},
          doi = {10.3847/1538-4357/ab3e30},
archivePrefix = {arXiv},
       eprint = {1908.09847},
 primaryClass = {astro-ph.EP},
       adsurl = {https://ui.adsabs.harvard.edu/abs/2019ApJ...883..194M}
}

@ARTICLE{2023ApJ...959L..30T,
       author = {{Tsai}, Shang-Min and {Moses}, Julianne I. and {Powell}, Diana and {Lee}, Elspeth K.~H.},
        title = "{Day{\textendash}Night Transport-induced Chemistry and Clouds on WASP-39b: Gas-phase Composition}",
      journal = {\apjl},
         year = 2023,
        month = dec,
       volume = {959},
       number = {2},
          eid = {L30},
        pages = {L30},
          doi = {10.3847/2041-8213/ad1405},
archivePrefix = {arXiv},
       eprint = {2305.19403},
 primaryClass = {astro-ph.EP},
       adsurl = {https://ui.adsabs.harvard.edu/abs/2023ApJ...959L..30T}
}

@ARTICLE{2025A&A...699A.306A,
       author = {{Ag{\'u}ndez}, Marcelino},
        title = "{The mutual influence of disequilibrium composition and temperature in exoplanet atmospheres}",
      journal = {\aap},
         year = 2025,
        month = jul,
       volume = {699},
          eid = {A306},
        pages = {A306},
          doi = {10.1051/0004-6361/202554732},
archivePrefix = {arXiv},
       eprint = {2506.11658},
 primaryClass = {astro-ph.EP},
       adsurl = {https://ui.adsabs.harvard.edu/abs/2025A&A...699A.306A}
}

@ARTICLE{2024ChJSS..44..400G,
       author = {{Ge}, Jian and {Chen}, Wen and {Chen}, Yonghe and {Song}, Zongxi and {Wang}, Jian and {Zhang}, Hui and {Li}, Yan and {Zang}, Weicheng and {Zhou}, Dan and {Zhang}, Yongshuai and {Chen}, Kun and {Yang}, Yingquan and {Mao}, Shude and {Huang}, Chelsea and {Yao}, Xinyu and {Li}, Xinglong and {Jiang}, Haijiao and {Yu}, Yong and {Tang}, Zhenghong and {Dong}, Feng and {Gao}, Wei and {Zhang}, Hongfei and {Shen}, Chao and {Wang}, Fengtao and {Wei}, Chuanxin and {Yang}, Baoyu and {Li}, Yudong and {Wen}, Lin and {Zhang}, Pengjun and {Zhang}, Congcong and {Xie}, Jiwei and {Ma}, Bo and {Deng}, Hongping and {Liu}, Huigen and {Duan}, Xuliang and {Wang}, Haoyu and {Huang}, Jiangjiang and {Gao}, Yang and {Wang}, Yifei and {Wang}, Lei and {Qin}, Genjian and {Liu}, Xinyu and {Gao}, Jie},
        title = "{Search for a Second Earth - the Earth 2.0 (ET) Space Mission}",
      journal = {Chinese Journal of Space Science},
         year = 2024,
        month = may,
       volume = {44},
       number = {3},
        pages = {400-424},
          doi = {10.11728/cjss2024.03.yg05},
       adsurl = {https://ui.adsabs.harvard.edu/abs/2024ChJSS..44..400G}
}

@ARTICLE{2021ExA....51..109B,
       author = {{Benz}, W. and {Broeg}, C. and {Fortier}, A. and {Rando}, N. and {Beck}, T. and {Beck}, M. and {Queloz}, D. and {Ehrenreich}, D. and {Maxted}, P.~F.~L. and {Isaak}, K.~G. and {Billot}, N. and {Alibert}, Y. and {Alonso}, R. and {Ant{\'o}nio}, C. and {Asquier}, J. and {Bandy}, T. and {B{\'a}rczy}, T. and {Barrado}, D. and {Barros}, S.~C.~C. and {Baumjohann}, W. and {Bekkelien}, A. and {Bergomi}, M. and {Biondi}, F.},
        title = "{The CHEOPS mission}",
      journal = {Experimental Astronomy},
         year = 2021,
        month = feb,
       volume = {51},
       number = {1},
        pages = {109-151},
          doi = {10.1007/s10686-020-09679-4},
archivePrefix = {arXiv},
       eprint = {2009.11633},
 primaryClass = {astro-ph.IM},
       adsurl = {https://ui.adsabs.harvard.edu/abs/2021ExA....51..109B}
}

@ARTICLE{2025ExA....59...26R,
       author = {{Rauer}, Heike and {Aerts}, Conny and {Cabrera}, Juan and {Deleuil}, Magali and {Erikson}, Anders and {Gizon}, Laurent and {Goupil}, Mariejo and {Heras}, Ana and {Walloschek}, Thomas and {Lorenzo-Alvarez}, Jose and {Marliani}, Filippo and {Martin-Garcia}, C{\'e}sar and {Mas-Hesse}, J. Miguel and {O'Rourke}, Laurence and {Osborn}, Hugh},
        title = "{The PLATO mission}",
      journal = {Experimental Astronomy},
         year = 2025,
        month = jun,
       volume = {59},
       number = {3},
          eid = {26},
        pages = {26},
          doi = {10.1007/s10686-025-09985-9},
archivePrefix = {arXiv},
       eprint = {2406.05447},
 primaryClass = {astro-ph.IM},
       adsurl = {https://ui.adsabs.harvard.edu/abs/2025ExA....59...26R}
}

@ARTICLE{2024ApJ...975L..11B,
       author = {{Banerjee}, Agnibha and {Barstow}, Joanna K. and {Gressier}, Am{\'e}lie and {Espinoza}, N{\'e}stor and {Sing}, David K. and {Allen}, Natalie H. and {Birkmann}, Stephan M. and {Challener}, Ryan C. and {Crouzet}, Nicolas and {Haswell}, Carole A. and {Lewis}, Nikole K. and {Lewis}, Stephen R. and {Yang}, Jingxuan},
        title = "{Atmospheric Retrievals Suggest the Presence of a Secondary Atmosphere and Possible Sulfur Species on L98-59 d from JWST Nirspec G395H Transmission Spectroscopy}",
      journal = {\apjl},
         year = 2024,
        month = nov,
       volume = {975},
       number = {1},
          eid = {L11},
        pages = {L11},
          doi = {10.3847/2041-8213/ad73d0},
archivePrefix = {arXiv},
       eprint = {2408.15707},
 primaryClass = {astro-ph.EP},
       adsurl = {https://ui.adsabs.harvard.edu/abs/2024ApJ...975L..11B}
}

@ARTICLE{2025arXiv250916082G,
       author = {{Gressier}, Am{\'e}lie and {Batalha}, Natasha E. and {Wogan}, Nicholas and {Alderson}, Lili and {Doud}, Dominic and {Espinoza}, N{\'e}stor and {MacDonald}, Ryan J. and {Wakeford}, Hannah R. and {Valenti}, Jeff A. and {Lewis}, Nikole K. and {Seager}, Sara and {Stevenson}, Kevin B. and {Allen}, Natalie H. and {Ca{\~n}as}, Caleb I. and {Challener}, Ryan C. and {Glidden}, Ana and {Huang}, Jingcheng and {Lin}, Zifan and {Louie}, Dana R. and {Maguire}, Cathal and {Mullens}, Elijah and {Sotzen}, Kristin and {Valentine}, Daniel and {Clampin}, Mark and {Pueyo}, Laurent and {van der Marel}, Roeland P. and {Mountain}, C. Matt},
        title = "{JWST-TST DREAMS: Sulfur dioxide in the atmosphere of the Neptune-mass planet HAT-P-26 b from NIRSpec G395H transmission spectroscopy}",
      journal = {arXiv e-prints},
         year = 2025,
        month = sep,
          eid = {arXiv:2509.16082},
        pages = {arXiv:2509.16082},
          doi = {10.48550/arXiv.2509.16082},
archivePrefix = {arXiv},
       eprint = {2509.16082},
 primaryClass = {astro-ph.EP},
       adsurl = {https://ui.adsabs.harvard.edu/abs/2025arXiv250916082G}
}

@ARTICLE{2024ApJ...970L..10B,
       author = {{Beatty}, Thomas G. and {Welbanks}, Luis and {Schlawin}, Everett and {Bell}, Taylor J. and {Line}, Michael R. and {Murphy}, Matthew and {Edelman}, Isaac and {Greene}, Thomas P. and {Fortney}, Jonathan J. and {Henry}, Gregory W. and {Mukherjee}, Sagnick and {Ohno}, Kazumasa and {Parmentier}, Vivien and {Rauscher}, Emily and {Wiser}, Lindsey S. and {Arnold}, Kenneth E.},
        title = "{Sulfur Dioxide and Other Molecular Species in the Atmosphere of the Sub-Neptune GJ 3470 b}",
      journal = {\apjl},
         year = 2024,
        month = jul,
       volume = {970},
       number = {1},
          eid = {L10},
        pages = {L10},
          doi = {10.3847/2041-8213/ad55e9},
archivePrefix = {arXiv},
       eprint = {2406.04450},
 primaryClass = {astro-ph.EP},
       adsurl = {https://ui.adsabs.harvard.edu/abs/2024ApJ...970L..10B}
}

@ARTICLE{2023PASP..135g5001A,
       author = {{Albert}, Lo{\"\i}c and {Lafreni{\`e}re}, David and {Ren{\'e}},, Doyon and {Artigau}, {\'E}tienne and {Volk}, Kevin and {Goudfrooij}, Paul and {Martel}, Andr{\'e} R. and {Radica}, Michael and {Rowe}, Jason and {Espinoza}, N{\'e}stor and {Roy}, Arpita and {Filippazzo}, Joseph C. and {Darveau-Bernier}, Antoine and {Talens}, Geert Jan and {Sivaramakrishnan}, Anand and {Willott}, Chris J. and {Fullerton}, Alexander W. and {LaMassa}, Stephanie and {Hutchings}, John B. and {Rowlands}, Neil and {Vila}, M. Bego{\~n}a and {Zhou}, Julia and {Aldridge}, David and {Maszkiewicz}, Michael and {Beaulieu}, Mathilde and {Cook}, Neil J. and {Piaulet}, Caroline and {Roy}, Pierre-Alexis and {Lamontagne}, Pierrot and {Morel}, Kim and {Frost}, William and {Salhi}, Salma and {Coulombe}, Louis-Philippe and {Benneke}, Bj{\"o}rn and {MacDonald}, Ryan J. and {Johnstone}, Doug and {Turner}, Jake D. and {Fournier-Tondreau}, Marylou and {Allart}, Romain and {Kaltenegger}, Lisa},
        title = "{The Near Infrared Imager and Slitless Spectrograph for the James Webb Space Telescope. III. Single Object Slitless Spectroscopy}",
      journal = {\pasp},
         year = 2023,
        month = jul,
       volume = {135},
       number = {1049},
          eid = {075001},
        pages = {075001},
          doi = {10.1088/1538-3873/acd7a3},
archivePrefix = {arXiv},
       eprint = {2306.04572},
 primaryClass = {astro-ph.IM},
       adsurl = {https://ui.adsabs.harvard.edu/abs/2023PASP..135g5001A}
}

@ARTICLE{2023PASP..135i8001D,
       author = {{Doyon}, Ren{\'e} and {Willott}, Chris J. and {Hutchings}, John B. and {Sivaramakrishnan}, Anand and {Albert}, Lo{\"\i}c and {Lafreni{\`e}re}, David and {Rowlands}, Neil and {Bego{\~n}a Vila}, M. and {Martel}, Andr{\'e} R. and {LaMassa}, Stephanie and {Aldridge}, David and {Artigau}, {\'E}tienne and {Cameron}, Peter and {Chayer}, Pierre and {Cook}, Neil J. and {Cooper}, Rachel A. and {Darveau-Bernier}, Antoine and {Dupuis}, Jean and {Earnshaw}, Colin and {Espinoza}, N{\'e}stor and {Filippazzo}, Joseph C. and {Fullerton}, Alexander W. and {Gaudreau}, Daniel and {Gawlik}, Roman and {Goudfrooij}, Paul and {Haley}, Craig and {Kammerer}, Jens and {Kendall}, David and {Lambros}, Scott D. and {Ignat}, Luminita Ilinca and {Maszkiewicz}, Michael and {McColgan}, Ashley and {Morishita}, Takahiro and {Ouellette}, Nathalie N. -Q. and {Pacifici}, Camilla and {Philippi}, Natasha and {Radica}, Michael and {Ravindranath}, Swara and {Rowe}, Jason and {Roy}, Arpita and {Roy}, Niladri and {Saad}, Karl and {Sohn}, Sangmo Tony and {Talens}, Geert Jan and {Touahri}, Driss and {Thatte}, Deepashri and {Taylor}, Joanna M. and {Vandal}, Thomas and {Volk}, Kevin and {Wander}, Michel and {Warner}, Gerald and {Zheng}, Sheng-Hai and {Zhou}, Julia and {Abraham}, Roberto and {Beaulieu}, Mathilde and {Benneke}, Bj{\"o}rn and {Ferrarese}, Laura and {Jayawardhana}, Ray and {Johnstone}, Doug and {Kaltenegger}, Lisa and {Meyer}, Michael R. and {Pipher}, Judy L. and {Rameau}, Julien and {Rieke}, Marcia and {Salhi}, Salma and {Sawicki}, Marcin},
        title = "{The Near Infrared Imager and Slitless Spectrograph for the James Webb Space Telescope. I. Instrument Overview and In-flight Performance}",
      journal = {\pasp},
         year = 2023,
        month = sep,
       volume = {135},
       number = {1051},
          eid = {098001},
        pages = {098001},
          doi = {10.1088/1538-3873/acd41b},
archivePrefix = {arXiv},
       eprint = {2306.03277},
 primaryClass = {astro-ph.IM},
       adsurl = {https://ui.adsabs.harvard.edu/abs/2023PASP..135i8001D}
}

@ARTICLE{2023PASP..135c8001B,
       author = {{B{\"o}ker}, T. and {Beck}, T.~L. and {Birkmann}, S.~M. and {Giardino}, G. and {Keyes}, C. and {Kumari}, N. and {Muzerolle}, J. and {Rawle}, T. and {Zeidler}, P. and {Abul-Huda}, Y. and {Alves de Oliveira}, C. and {Arribas}, S. and {Bechtold}, K. and {Bhatawdekar}, R. and {Bonaventura}, N. and {Bunker}, A.~J. and {Cameron}, A.~J. and {Carniani}, S. and {Charlot}, S. and {Curti}, M. and {Espinoza}, N. and {Ferruit}, P. and {Franx}, M. and {Jakobsen}, P. and {Karakla}, D. and {L{\'o}pez-Caniego}, M. and {L{\"u}tzgendorf}, N. and {Maiolino}, R. and {Manjavacas}, E. and {Marston}, A.~P. and {Moseley}, S.~H. and {Ogle}, P. and {Perna}, M. and {Pe{\~n}a-Guerrero}, M. and {Pirzkal}, N. and {Plesha}, R. and {Proffitt}, C.~R. and {Rauscher}, B.~J. and {Rix}, H. -W. and {Rodr{\'\i}guez del Pino}, B. and {Rustamkulov}, Z. and {Sabbi}, E. and {Sing}, D.~K. and {Sirianni}, M. and {te Plate}, M. and {{\'U}beda}, L. and {Wahlgren}, G.~M. and {Wislowski}, E. and {Wu}, R. and {Willott}, Chris J.},
        title = "{In-orbit Performance of the Near-infrared Spectrograph NIRSpec on the James Webb Space Telescope}",
      journal = {\pasp},
         year = 2023,
        month = mar,
       volume = {135},
       number = {1045},
          eid = {038001},
        pages = {038001},
          doi = {10.1088/1538-3873/acb846},
archivePrefix = {arXiv},
       eprint = {2301.13766},
 primaryClass = {astro-ph.IM},
       adsurl = {https://ui.adsabs.harvard.edu/abs/2023PASP..135c8001B}
}

@ARTICLE{2022A&A...661A..80J,
       author = {{Jakobsen}, P. and {Ferruit}, P. and {Alves de Oliveira}, C. and {Arribas}, S. and {Bagnasco}, G. and {Barho}, R. and {Beck}, T.~L. and {Birkmann}, S. and {B{\"o}ker}, T. and {Bunker}, A.~J. and {Charlot}, S. and {de Jong}, P. and {de Marchi}, G. and {Ehrenwinkler}, R. and {Falcolini}, M. and {Fels}, R. and {Franx}, M. and {Franz}, D. and {Funke}, M. and {Giardino}, G. and {Gnata}, X. and {Holota}, W. and {Honnen}, K. and {Jensen}, P.~L. and {Jentsch}, M. and {Johnson}, T. and {Jollet}, D. and {Karl}, H. and {Kling}, G. and {K{\"o}hler}, J. and {Kolm}, M. -G. and {Kumari}, N. and {Lander}, M.~E. and {Lemke}, R. and {L{\'o}pez-Caniego}, M. and {L{\"u}tzgendorf}, N. and {Maiolino}, R. and {Manjavacas}, E. and {Marston}, A. and {Maschmann}, M. and {Maurer}, R. and {Messerschmidt}, B. and {Moseley}, S.~H. and {Mosner}, P. and {Mott}, D.~B. and {Muzerolle}, J. and {Pirzkal}, N. and {Pittet}, J. -F. and {Plitzke}, A. and {Posselt}, W. and {Rapp}, B. and {Rauscher}, B.~J. and {Rawle}, T. and {Rix}, H. -W. and {R{\"o}del}, A. and {Rumler}, P. and {Sabbi}, E. and {Salvignol}, J. -C. and {Schmid}, T. and {Sirianni}, M. and {Smith}, C. and {Strada}, P. and {te Plate}, M. and {Valenti}, J. and {Wettemann}, T. and {Wiehe}, T. and {Wiesmayer}, M. and {Willott}, C.~J. and {Wright}, R. and {Zeidler}, P. and {Zincke}, C.},
        title = "{The Near-Infrared Spectrograph (NIRSpec) on the James Webb Space Telescope. I. Overview of the instrument and its capabilities}",
      journal = {\aap},
         year = 2022,
        month = may,
       volume = {661},
          eid = {A80},
        pages = {A80},
          doi = {10.1051/0004-6361/202142663},
archivePrefix = {arXiv},
       eprint = {2202.03305},
 primaryClass = {astro-ph.IM},
       adsurl = {https://ui.adsabs.harvard.edu/abs/2022A&A...661A..80J}
}

@ARTICLE{2015PASP..127..623K,
       author = {{Kendrew}, Sarah and {Scheithauer}, Silvia and {Bouchet}, Patrice and {Amiaux}, Jerome and {Azzollini}, Ruym{\'a}n and {Bouwman}, Jeroen and {Chen}, C.~H. and {Dubreuil}, D. and {Fischer}, Sebastian and {Glasse}, Alistair and {Greene}, T.~P. and {Lagage}, P. -O. and {Lahuis}, Fred and {Ronayette}, Samuel and {Wright}, David and {Wright}, G.~S.},
        title = "{The Mid-Infrared Instrument for the James Webb Space Telescope, IV: The Low-Resolution Spectrometer}",
      journal = {\pasp},
         year = 2015,
        month = jul,
       volume = {127},
       number = {953},
        pages = {623},
          doi = {10.1086/682255},
archivePrefix = {arXiv},
       eprint = {1512.03000},
 primaryClass = {astro-ph.IM},
       adsurl = {https://ui.adsabs.harvard.edu/abs/2015PASP..127..623K}
}

@ARTICLE{2017PASP..129f4501B,
       author = {{Batalha}, Natasha E. and {Mandell}, Avi and {Pontoppidan}, Klaus and {Stevenson}, Kevin B. and {Lewis}, Nikole K. and {Kalirai}, Jason and {Earl}, Nick and {Greene}, Thomas and {Albert}, Lo{\"\i}c and {Nielsen}, Louise D.},
        title = "{PandExo: A Community Tool for Transiting Exoplanet Science with JWST \& HST}",
      journal = {\pasp},
         year = 2017,
        month = jun,
       volume = {129},
       number = {976},
        pages = {064501},
          doi = {10.1088/1538-3873/aa65b0},
archivePrefix = {arXiv},
       eprint = {1702.01820},
 primaryClass = {astro-ph.IM},
       adsurl = {https://ui.adsabs.harvard.edu/abs/2017PASP..129f4501B}
}

@ARTICLE{2019ApJ...885L..25B,
       author = {{Batalha}, Natasha E. and {Lewis}, Taylor and {Fortney}, Jonathan J. and {Batalha}, Natalie M. and {Kempton}, Eliza and {Lewis}, Nikole K. and {Line}, Michael R.},
        title = "{The Precision of Mass Measurements Required for Robust Atmospheric Characterization of Transiting Exoplanets}",
      journal = {\apjl},
         year = 2019,
        month = nov,
       volume = {885},
       number = {1},
          eid = {L25},
        pages = {L25},
          doi = {10.3847/2041-8213/ab4909},
archivePrefix = {arXiv},
       eprint = {1910.00076},
 primaryClass = {astro-ph.EP},
       adsurl = {https://ui.adsabs.harvard.edu/abs/2019ApJ...885L..25B}
}

@ARTICLE{2024A&A...688A..59P,
       author = {{Parc}, L{\'e}na and {Bouchy}, Fran{\c{c}}ois and {Venturini}, Julia and {Dorn}, Caroline and {Helled}, Ravit},
        title = "{From super-Earths to sub-Neptunes: Observational constraints and connections to theoretical models}",
      journal = {\aap},
         year = 2024,
        month = aug,
       volume = {688},
          eid = {A59},
        pages = {A59},
          doi = {10.1051/0004-6361/202449911},
archivePrefix = {arXiv},
       eprint = {2406.04311},
 primaryClass = {astro-ph.EP},
       adsurl = {https://ui.adsabs.harvard.edu/abs/2024A&A...688A..59P}
}

@ARTICLE{2019PNAS..116.9723Z,
       author = {{Zeng}, Li and {Jacobsen}, Stein B. and {Sasselov}, Dimitar D. and {Petaev}, Michail I. and {Vanderburg}, Andrew and {Lopez-Morales}, Mercedes and {Perez-Mercader}, Juan and {Mattsson}, Thomas R. and {Li}, Gongjie and {Heising}, Matthew Z. and {Bonomo}, Aldo S. and {Damasso}, Mario and {Berger}, Travis A. and {Cao}, Hao and {Levi}, Amit and {Wordsworth}, Robin D.},
        title = "{Growth model interpretation of planet size distribution}",
      journal = {Proceedings of the National Academy of Science},
         year = 2019,
        month = may,
       volume = {116},
       number = {20},
        pages = {9723-9728},
          doi = {10.1073/pnas.1812905116},
archivePrefix = {arXiv},
       eprint = {1906.04253},
 primaryClass = {astro-ph.EP},
       adsurl = {https://ui.adsabs.harvard.edu/abs/2019PNAS..116.9723Z}
}

@ARTICLE{2020A&A...634A..43O,
       author = {{Otegi}, J.~F. and {Bouchy}, F. and {Helled}, R.},
        title = "{Revisited mass-radius relations for exoplanets below 120 M$_{{\ensuremath{\oplus}}}$}",
      journal = {\aap},
         year = 2020,
        month = feb,
       volume = {634},
          eid = {A43},
        pages = {A43},
          doi = {10.1051/0004-6361/201936482},
archivePrefix = {arXiv},
       eprint = {1911.04745},
 primaryClass = {astro-ph.EP},
       adsurl = {https://ui.adsabs.harvard.edu/abs/2020A&A...634A..43O}
}

@ARTICLE{2014ApJ...783L...6W,
       author = {{Weiss}, Lauren M. and {Marcy}, Geoffrey W.},
        title = "{The Mass-Radius Relation for 65 Exoplanets Smaller than 4 Earth Radii}",
      journal = {\apjl},
         year = 2014,
        month = mar,
       volume = {783},
       number = {1},
          eid = {L6},
        pages = {L6},
          doi = {10.1088/2041-8205/783/1/L6},
archivePrefix = {arXiv},
       eprint = {1312.0936},
 primaryClass = {astro-ph.EP},
       adsurl = {https://ui.adsabs.harvard.edu/abs/2014ApJ...783L...6W}
}

@INPROCEEDINGS{2010ASSP...14..211D,
       author = {{Djupvik}, Anlaug Amanda and {Andersen}, Johannes},
        title = "{The Nordic Optical Telescope}",
    booktitle = {Highlights of Spanish Astrophysics V},
         year = 2010,
       editor = {{Diego}, Jose M. and {Goicoechea}, Luis J. and {Gonz{\'a}lez-Serrano}, J. Ignacio and {Gorgas}, Javier},
       series = {Astrophysics and Space Science Proceedings},
       volume = {14},
        month = jan,
        pages = {211},
          doi = {10.1007/978-3-642-11250-8_21},
archivePrefix = {arXiv},
       eprint = {0901.4015},
 primaryClass = {astro-ph.IM},
       adsurl = {https://ui.adsabs.harvard.edu/abs/2010ASSP...14..211D}
}

@ARTICLE{2018AJ....156..102S,
       author = {{Stassun}, Keivan G. and {Oelkers}, Ryan J. and {Pepper}, Joshua and {Paegert}, Martin and {De Lee}, Nathan and {Torres}, Guillermo and {Latham}, David W. and {Charpinet}, St{\'e}phane and {Dressing}, Courtney D. and {Huber}, Daniel and {Kane}, Stephen R. and {L{\'e}pine}, S{\'e}bastien and {Mann}, Andrew and {Muirhead}, Philip S. and {Rojas-Ayala}, B{\'a}rbara and {Silvotti}, Roberto and {Fleming}, Scott W. and {Levine}, Al and {Plavchan}, Peter},
        title = "{The TESS Input Catalog and Candidate Target List}",
      journal = {\aj},
         year = 2018,
        month = sep,
       volume = {156},
       number = {3},
          eid = {102},
        pages = {102},
          doi = {10.3847/1538-3881/aad050},
archivePrefix = {arXiv},
       eprint = {1706.00495},
 primaryClass = {astro-ph.EP},
       adsurl = {https://ui.adsabs.harvard.edu/abs/2018AJ....156..102S}
}

@ARTICLE{2019PASP..131b4506L,
       author = {{Li}, Jie and {Tenenbaum}, Peter and {Twicken}, Joseph D. and {Burke}, Christopher J. and {Jenkins}, Jon M. and {Quintana}, Elisa V. and {Rowe}, Jason F. and {Seader}, Shawn E.},
        title = "{Kepler Data Validation II-Transit Model Fitting and Multiple-planet Search}",
      journal = {\pasp},
         year = 2019,
        month = feb,
       volume = {131},
       number = {996},
        pages = {024506},
          doi = {10.1088/1538-3873/aaf44d},
archivePrefix = {arXiv},
       eprint = {1812.00103},
 primaryClass = {astro-ph.IM},
       adsurl = {https://ui.adsabs.harvard.edu/abs/2019PASP..131b4506L}
}

@ARTICLE{2021ApJS..254...39G,
       author = {{Guerrero}, Natalia M. and {Seager}, S. and {Huang}, Chelsea X. and {Vanderburg}, Andrew and {Garcia Soto}, Aylin and {Mireles}, Ismael and {Hesse}, Katharine and {Fong}, William and {Glidden}, Ana and {Shporer}, Avi and {Latham}, David W. and {Collins}, Karen A. and {Quinn}, Samuel N. and {Burt}, Jennifer and {Dragomir}, Diana and {Crossfield}, Ian and {Vanderspek}, Roland and {Fausnaugh}, Michael and {Burke}, Christopher J. and {Ricker}, George and {Daylan}, Tansu and {Essack}, Zahra and {G{\"u}nther}, Maximilian N. and {Osborn}, Hugh P. and {Pepper}, Joshua and {Rowden}, Pamela and {Sha}, Lizhou and {Villanueva}, Jr., Steven and {Yahalomi}, Daniel A. and {Yu}, Liang and {Ballard}, Sarah and {Batalha}, Natalie M. and {Berardo}, David and {Chontos}, Ashley and {Dittmann}, Jason A. and {Esquerdo}, Gilbert A. and {Mikal-Evans}, Thomas and {Jayaraman}, Rahul and {Krishnamurthy}, Akshata and {Louie}, Dana R. and {Mehrle}, Nicholas and {Niraula}, Prajwal and {Rackham}, Benjamin V. and {Rodriguez}, Joseph E. and {Rowden}, Stephen J.~L. and {Sousa-Silva}, Clara and {Watanabe}, David and {Wong}, Ian and {Zhan}, Zhuchang and {Zivanovic}, Goran and {Christiansen}, Jessie L. and {Ciardi}, David R. and {Swain}, Melanie A. and {Lund}, Michael B. and {Mullally}, Susan E. and {Fleming}, Scott W. and {Rodriguez}, David R. and {Boyd}, Patricia T. and {Quintana}, Elisa V. and {Barclay}, Thomas and {Col{\'o}n}, Knicole D. and {Rinehart}, S.~A. and {Schlieder}, Joshua E. and {Clampin}, Mark and {Jenkins}, Jon M. and {Twicken}, Joseph D. and {Caldwell}, Douglas A. and {Coughlin}, Jeffrey L. and {Henze}, Chris and {Lissauer}, Jack J. and {Morris}, Robert L. and {Rose}, Mark E. and {Smith}, Jeffrey C. and {Tenenbaum}, Peter and {Ting}, Eric B. and {Wohler}, Bill and {Bakos}, G. {\'A}. and {Bean}, Jacob L. and {Berta-Thompson}, Zachory K. and {Bieryla}, Allyson and {Bouma}, Luke G. and {Buchhave}, Lars A. and {Butler}, Nathaniel and {Charbonneau}, David and {Doty}, John P. and {Ge}, Jian and {Holman}, Matthew J. and {Howard}, Andrew W. and {Kaltenegger}, Lisa and {Kane}, Stephen R. and {Kjeldsen}, Hans and {Kreidberg}, Laura and {Lin}, Douglas N.~C. and {Minsky}, Charlotte and {Narita}, Norio and {Paegert}, Martin and {P{\'a}l}, Andr{\'a}s and {Palle}, Enric and {Sasselov}, Dimitar D. and {Spencer}, Alton and {Sozzetti}, Alessandro and {Stassun}, Keivan G. and {Torres}, Guillermo and {Udry}, Stephane and {Winn}, Joshua N.},
        title = "{The TESS Objects of Interest Catalog from the TESS Prime Mission}",
      journal = {\apjs},
         year = 2021,
        month = jun,
       volume = {254},
       number = {2},
          eid = {39},
        pages = {39},
          doi = {10.3847/1538-4365/abefe1},
archivePrefix = {arXiv},
       eprint = {2103.12538},
 primaryClass = {astro-ph.EP},
       adsurl = {https://ui.adsabs.harvard.edu/abs/2021ApJS..254...39G}
}

@ARTICLE{2018PASP..130f4502T,
       author = {{Twicken}, Joseph D. and {Catanzarite}, Joseph H. and {Clarke}, Bruce D. and {Girouard}, Forrest and {Jenkins}, Jon M. and {Klaus}, Todd C. and {Li}, Jie and {McCauliff}, Sean D. and {Seader}, Shawn E. and {Tenenbaum}, Peter and {Wohler}, Bill and {Bryson}, Stephen T. and {Burke}, Christopher J. and {Caldwell}, Douglas A. and {Haas}, Michael R. and {Henze}, Christopher E. and {Sanderfer}, Dwight T.},
        title = "{Kepler Data Validation I{\textemdash}Architecture, Diagnostic Tests, and Data Products for Vetting Transiting Planet Candidates}",
      journal = {\pasp},
         year = 2018,
        month = jun,
       volume = {130},
       number = {988},
        pages = {064502},
          doi = {10.1088/1538-3873/aab694},
archivePrefix = {arXiv},
       eprint = {1803.04526},
 primaryClass = {astro-ph.EP},
       adsurl = {https://ui.adsabs.harvard.edu/abs/2018PASP..130f4502T}
}

@MISC{2020ksci.rept....9J,
       author = {{Jenkins}, Jon M. and {Tenenbaum}, Peter and {Seader}, Shawn and {Burke}, Christopher J. and {McCauliff}, Sean D. and {Smith}, Jeffrey C. and {Twicken}, Joseph D. and {Chandrasekaran}, Hema},
        title = "{Kepler Data Processing Handbook: Transiting Planet Search}",
 howpublished = {Kepler Science Document KSCI-19081-003, id. 9. Edited by Jon M. Jenkins.},
         year = 2020,
        month = mar,
          eid = {9},
        pages = {9},
       adsurl = {https://ui.adsabs.harvard.edu/abs/2020ksci.rept....9J}
}

@INPROCEEDINGS{2010SPIE.7740E..0DJ,
       author = {{Jenkins}, Jon M. and {Chandrasekaran}, Hema and {McCauliff}, Sean D. and {Caldwell}, Douglas A. and {Tenenbaum}, Peter and {Li}, Jie and {Klaus}, Todd C. and {Cote}, Miles T. and {Middour}, Christopher},
        title = "{Transiting planet search in the Kepler pipeline}",
    booktitle = {Software and Cyberinfrastructure for Astronomy},
         year = 2010,
       editor = {{Radziwill}, Nicole M. and {Bridger}, Alan},
       series = {Society of Photo-Optical Instrumentation Engineers (SPIE) Conference Series},
       volume = {7740},
        month = jul,
          eid = {77400D},
        pages = {77400D},
          doi = {10.1117/12.856764},
       adsurl = {https://ui.adsabs.harvard.edu/abs/2010SPIE.7740E..0DJ}
}

@ARTICLE{2002ApJ...575..493J,
       author = {{Jenkins}, Jon M.},
        title = "{The Impact of Solar-like Variability on the Detectability of Transiting Terrestrial Planets}",
      journal = {\apj},
         year = 2002,
        month = aug,
       volume = {575},
       number = {1},
        pages = {493-505},
          doi = {10.1086/341136},
       adsurl = {https://ui.adsabs.harvard.edu/abs/2002ApJ...575..493J}
}

@ARTICLE{2015JATIS...1a4003R,
       author = {{Ricker}, George R. and {Winn}, Joshua N. and {Vanderspek}, Roland and {Latham}, David W. and {Bakos}, G{\'a}sp{\'a}r {\'A}. and {Bean}, Jacob L. and {Berta-Thompson}, Zachory K. and {Brown}, Timothy M. and {Buchhave}, Lars and {Butler}, Nathaniel R. and {Butler}, R. Paul and {Chaplin}, William J. and {Charbonneau}, David and {Christensen-Dalsgaard}, J{\o}rgen and {Clampin}, Mark and {Deming}, Drake and {Doty}, John and {De Lee}, Nathan and {Dressing}, Courtney and {Dunham}, Edward W. and {Endl}, Michael and {Fressin}, Francois and {Ge}, Jian and {Henning}, Thomas and {Holman}, Matthew J. and {Howard}, Andrew W. and {Ida}, Shigeru and {Jenkins}, Jon M. and {Jernigan}, Garrett and {Johnson}, John Asher and {Kaltenegger}, Lisa and {Kawai}, Nobuyuki and {Kjeldsen}, Hans and {Laughlin}, Gregory and {Levine}, Alan M. and {Lin}, Douglas and {Lissauer}, Jack J. and {MacQueen}, Phillip and {Marcy}, Geoffrey and {McCullough}, Peter R. and {Morton}, Timothy D. and {Narita}, Norio and {Paegert}, Martin and {Palle}, Enric and {Pepe}, Francesco and {Pepper}, Joshua and {Quirrenbach}, Andreas and {Rinehart}, Stephen A. and {Sasselov}, Dimitar and {Sato}, Bun'ei and {Seager}, Sara and {Sozzetti}, Alessandro and {Stassun}, Keivan G. and {Sullivan}, Peter and {Szentgyorgyi}, Andrew and {Torres}, Guillermo and {Udry}, Stephane and {Villasenor}, Joel},
        title = "{Transiting Exoplanet Survey Satellite (TESS)}",
      journal = {Journal of Astronomical Telescopes, Instruments, and Systems},
         year = 2015,
        month = jan,
       volume = {1},
          eid = {014003},
        pages = {014003},
          doi = {10.1117/1.JATIS.1.1.014003},
       adsurl = {https://ui.adsabs.harvard.edu/abs/2015JATIS...1a4003R}
}

@ARTICLE{2021FrASS...8..138S,
       author = {{Scott}, Nicholas J. and {Howell}, Steve B. and {Gnilka}, Crystal L. and {Stephens}, Andrew W. and {Salinas}, Ricardo and {Matson}, Rachel A. and {Furlan}, Elise and {Horch}, Elliott P. and {Everett}, Mark E. and {Ciardi}, David R. and {Mills}, Dave and {Quigley}, Emmett A.},
        title = "{Twin High-resolution, High-speed Imagers for the Gemini Telescopes: Instrument description and science verification results}",
      journal = {Frontiers in Astronomy and Space Sciences},
         year = 2021,
        month = sep,
       volume = {8},
          eid = {138},
        pages = {138},
          doi = {10.3389/fspas.2021.716560},
       adsurl = {https://ui.adsabs.harvard.edu/abs/2021FrASS...8..138S}
}

@INPROCEEDINGS{2014SPIE.9148E..3AM,
       author = {{McGurk}, Rosalie and {Rockosi}, Constance and {Gavel}, Donald and {Kupke}, Renate and {Peck}, Michael and {Pfister}, Terry and {Ward}, Jim and {Deich}, William and {Gates}, John and {Gates}, Elinor and {Alcott}, Barry and {Cowley}, David and {Dillon}, Daren and {Lanclos}, Kyle and {Sandford}, Dale and {Saylor}, Mike and {Srinath}, Srikar and {Weiss}, Jason and {Norton}, Andrew},
        title = "{Commissioning ShARCS: the Shane adaptive optics infrared camera-spectrograph for the Lick Observatory Shane 3-m telescope}",
    booktitle = {Adaptive Optics Systems IV},
         year = 2014,
       editor = {{Marchetti}, Enrico and {Close}, Laird M. and {Vran}, Jean-Pierre},
       series = {Society of Photo-Optical Instrumentation Engineers (SPIE) Conference Series},
       volume = {9148},
        month = jul,
          eid = {91483A},
        pages = {91483A},
          doi = {10.1117/12.2057027},
archivePrefix = {arXiv},
       eprint = {1407.8205},
 primaryClass = {astro-ph.IM},
       adsurl = {https://ui.adsabs.harvard.edu/abs/2014SPIE.9148E..3AM}
}

@INPROCEEDINGS{2014SPIE.9148E..05G,
       author = {{Gavel}, Donald and {Kupke}, Renate and {Dillon}, Daren and {Norton}, Andrew and {Ratliff}, Chris and {Cabak}, Jerry and {Phillips}, Andrew and {Rockosi}, Connie and {McGurk}, Rosalie and {Srinath}, Srikar and {Peck}, Michael and {Deich}, William and {Lanclos}, Kyle and {Gates}, John and {Saylor}, Michael and {Ward}, Jim and {Pfister}, Terry},
        title = "{ShaneAO: wide science spectrum adaptive optics system for the Lick Observatory}",
    booktitle = {Adaptive Optics Systems IV},
         year = 2014,
       editor = {{Marchetti}, Enrico and {Close}, Laird M. and {Vran}, Jean-Pierre},
       series = {Society of Photo-Optical Instrumentation Engineers (SPIE) Conference Series},
       volume = {9148},
        month = jul,
          eid = {914805},
        pages = {914805},
          doi = {10.1117/12.2055256},
archivePrefix = {arXiv},
       eprint = {1407.8207},
 primaryClass = {astro-ph.IM},
       adsurl = {https://ui.adsabs.harvard.edu/abs/2014SPIE.9148E..05G}
}

@ARTICLE{2021MNRAS.507..350D,
       author = {{Dhillon}, V.~S. and {Bezawada}, N. and {Black}, M. and {Dixon}, S.~D. and {Gamble}, T. and {Gao}, X. and {Henry}, D.~M. and {Kerry}, P. and {Littlefair}, S.~P. and {Lunney}, D.~W. and {Marsh}, T.~R. and {Miller}, C. and {Parsons}, S.~G. and {Ashley}, R.~P. and {Breedt}, E. and {Brown}, A. and {Dyer}, M.~J. and {Green}, M.~J. and {Pelisoli}, I. and {Sahman}, D.~I. and {Wild}, J. and {Ives}, D.~J. and {Mehrgan}, L. and {Stegmeier}, J. and {Dubbeldam}, C.~M. and {Morris}, T.~J. and {Osborn}, J. and {Wilson}, R.~W. and {Casares}, J. and {Mu{\~n}oz-Darias}, T. and {Pall{\'e}}, E. and {Rodr{\'\i}guez-Gil}, P. and {Shahbaz}, T. and {Torres}, M.~A.~P. and {de Ugarte Postigo}, A. and {Cabrera-Lavers}, A. and {Corradi}, R.~L.~M. and {Dom{\'\i}nguez}, R.~D. and {Garc{\'\i}a-Alvarez}, D.},
        title = "{HiPERCAM: a quintuple-beam, high-speed optical imager on the 10.4-m Gran Telescopio Canarias}",
      journal = {\mnras},
         year = 2021,
        month = oct,
       volume = {507},
       number = {1},
        pages = {350-366},
          doi = {10.1093/mnras/stab2130},
archivePrefix = {arXiv},
       eprint = {2107.10124},
 primaryClass = {astro-ph.IM},
       adsurl = {https://ui.adsabs.harvard.edu/abs/2021MNRAS.507..350D}
}

@ARTICLE{2013ApJ...765..131K,
       author = {{Kopparapu}, Ravi Kumar and {Ramirez}, Ramses and {Kasting}, James F. and {Eymet}, Vincent and {Robinson}, Tyler D. and {Mahadevan}, Suvrath and {Terrien}, Ryan C. and {Domagal-Goldman}, Shawn and {Meadows}, Victoria and {Deshpande}, Rohit},
        title = "{Habitable Zones around Main-sequence Stars: New Estimates}",
      journal = {\apj},
         year = 2013,
        month = mar,
       volume = {765},
       number = {2},
          eid = {131},
        pages = {131},
          doi = {10.1088/0004-637X/765/2/131},
archivePrefix = {arXiv},
       eprint = {1301.6674},
 primaryClass = {astro-ph.EP},
       adsurl = {https://ui.adsabs.harvard.edu/abs/2013ApJ...765..131K}
}

@INPROCEEDINGS{2011ASPC..448...91A,
       author = {{Allard}, F. and {Homeier}, D. and {Freytag}, B.},
        title = "{Model Atmospheres From Very Low Mass Stars to Brown Dwarfs}",
    booktitle = {16th Cambridge Workshop on Cool Stars, Stellar Systems, and the Sun},
         year = 2011,
       editor = {{Johns-Krull}, Christopher and {Browning}, Matthew K. and {West}, Andrew A.},
       series = {Astronomical Society of the Pacific Conference Series},
       volume = {448},
        month = dec,
        pages = {91},
          doi = {10.48550/arXiv.1011.5405},
archivePrefix = {arXiv},
       eprint = {1011.5405},
 primaryClass = {astro-ph.SR},
       adsurl = {https://ui.adsabs.harvard.edu/abs/2011ASPC..448...91A}
}

@ARTICLE{2018PASP..130k4401K,
       author = {{Kempton}, Eliza M. -R. and {Bean}, Jacob L. and {Louie}, Dana R. and {Deming}, Drake and {Koll}, Daniel D.~B. and {Mansfield}, Megan and {Christiansen}, Jessie L. and {L{\'o}pez-Morales}, Mercedes and {Swain}, Mark R. and {Zellem}, Robert T. and {Ballard}, Sarah and {Barclay}, Thomas and {Barstow}, Joanna K. and {Batalha}, Natasha E. and {Beatty}, Thomas G. and {Berta-Thompson}, Zach and {Birkby}, Jayne and {Buchhave}, Lars A. and {Charbonneau}, David and {Cowan}, Nicolas B. and {Crossfield}, Ian and {de Val-Borro}, Miguel and {Doyon}, Ren{\'e} and {Dragomir}, Diana and {Gaidos}, Eric and {Heng}, Kevin and {Hu}, Renyu and {Kane}, Stephen R. and {Kreidberg}, Laura and {Mallonn}, Matthias and {Morley}, Caroline V. and {Narita}, Norio and {Nascimbeni}, Valerio and {Pall{\'e}}, Enric and {Quintana}, Elisa V. and {Rauscher}, Emily and {Seager}, Sara and {Shkolnik}, Evgenya L. and {Sing}, David K. and {Sozzetti}, Alessandro and {Stassun}, Keivan G. and {Valenti}, Jeff A. and {von Essen}, Carolina},
        title = "{A Framework for Prioritizing the TESS Planetary Candidates Most Amenable to Atmospheric Characterization}",
      journal = {\pasp},
         year = 2018,
        month = nov,
       volume = {130},
       number = {993},
        pages = {114401},
          doi = {10.1088/1538-3873/aadf6f},
archivePrefix = {arXiv},
       eprint = {1805.03671},
 primaryClass = {astro-ph.EP},
       adsurl = {https://ui.adsabs.harvard.edu/abs/2018PASP..130k4401K}
}

@ARTICLE{2022AJ....163..152S,
       author = {{Sprague}, Dani and {Culhane}, Connor and {Kounkel}, Marina and {Olney}, Richard and {Covey}, K.~R. and {Hutchinson}, Brian and {Lingg}, Ryan and {Stassun}, Keivan G. and {Rom{\'a}n-Z{\'u}{\~n}iga}, Carlos G. and {Roman-Lopes}, Alexandre and {Nidever}, David and {Beaton}, Rachael L. and {Borissova}, Jura and {Stutz}, Amelia and {Stringfellow}, Guy S. and {Ram{\'\i}rez}, Karla Pe{\~n}a and {Ram{\'\i}rez-Preciado}, Valeria and {Hern{\'a}ndez}, Jes{\'u}s and {Kim}, Jinyoung Serena and {Lane}, Richard R.},
        title = "{APOGEE Net: An Expanded Spectral Model of Both Low-mass and High-mass Stars}",
      journal = {\aj},
         year = 2022,
        month = apr,
       volume = {163},
       number = {4},
          eid = {152},
        pages = {152},
          doi = {10.3847/1538-3881/ac4de7},
archivePrefix = {arXiv},
       eprint = {2201.03661},
 primaryClass = {astro-ph.GA},
       adsurl = {https://ui.adsabs.harvard.edu/abs/2022AJ....163..152S}
}

@MISC{2013wise.rept....1C,
       author = {{Cutri}, R.~M. and {Wright}, E.~L. and {Conrow}, T. and {Fowler}, J.~W. and {Eisenhardt}, P.~R.~M. and {Grillmair}, C. and {Kirkpatrick}, J.~D. and {Masci}, F. and {McCallon}, H.~L. and {Wheelock}, S.~L. and {Fajardo-Acosta}, S. and {Yan}, L. and {Benford}, D. and {Harbut}, M. and {Jarrett}, T. and {Lake}, S. and {Leisawitz}, D. and {Ressler}, M.~E. and {Stanford}, S.~A. and {Tsai}, C.~W. and {Liu}, F. and {Helou}, G. and {Mainzer}, A. and {Gettings}, D. and {Gonzalez}, A. and {Hoffman}, D. and {Marsh}, K.~A. and {Padgett}, D. and {Skrutskie}, M.~F. and {Beck}, R.~P. and {Papin}, M. and {Wittman}, M.},
        title = "{Explanatory Supplement to the AllWISE Data Release Products}",
 howpublished = {Explanatory Supplement to the AllWISE Data Release Products, by R. M. Cutri et al.},
         year = 2013,
        month = nov,
        pages = {1},
       adsurl = {https://ui.adsabs.harvard.edu/abs/2013wise.rept....1C}
}

@ARTICLE{2020ApJS..251....7F,
       author = {{Flewelling}, H.~A. and {Magnier}, E.~A. and {Chambers}, K.~C. and {Heasley}, J.~N. and {Holmberg}, C. and {Huber}, M.~E. and {Sweeney}, W. and {Waters}, C.~Z. },
        title = "{The Pan-STARRS1 Database and Data Products}",
      journal = {\apjs},
         year = 2020,
        month = nov,
       volume = {251},
       number = {1},
          eid = {7},
        pages = {7},
          doi = {10.3847/1538-4365/abb82d},
archivePrefix = {arXiv},
       eprint = {1612.05243},
 primaryClass = {astro-ph.IM},
       adsurl = {https://ui.adsabs.harvard.edu/abs/2020ApJS..251....7F}
}

@ARTICLE{2016arXiv161205560C,
       author = {{Chambers}, K.~C. and {Magnier}, E.~A. and {Metcalfe}, N. and {Flewelling}, H.~A. and {Huber}, M.~E. and {Waters}, C.~Z. and {Denneau}, L. },
        title = "{The Pan-STARRS1 Surveys}",
      journal = {arXiv e-prints},
         year = 2016,
        month = dec,
          eid = {arXiv:1612.05560},
        pages = {arXiv:1612.05560},
          doi = {10.48550/arXiv.1612.05560},
archivePrefix = {arXiv},
       eprint = {1612.05560},
 primaryClass = {astro-ph.IM},
       adsurl = {https://ui.adsabs.harvard.edu/abs/2016arXiv161205560C}
}

@ARTICLE{2022ApJ...927...31T,
       author = {{Tayar}, Jamie and {Claytor}, Zachary R. and {Huber}, Daniel and {van Saders}, Jennifer},
        title = "{A Guide to Realistic Uncertainties on the Fundamental Properties of Solar-type Exoplanet Host Stars}",
      journal = {\apj},
         year = 2022,
        month = mar,
       volume = {927},
       number = {1},
          eid = {31},
        pages = {31},
          doi = {10.3847/1538-4357/ac4bbc},
archivePrefix = {arXiv},
       eprint = {2012.07957},
 primaryClass = {astro-ph.EP},
       adsurl = {https://ui.adsabs.harvard.edu/abs/2022ApJ...927...31T}
}

@ARTICLE{2012RSPTA.370.2765A,
       author = {{Allard}, F. and {Homeier}, D. and {Freytag}, B.},
        title = "{Models of very-low-mass stars, brown dwarfs and exoplanets}",
      journal = {Philosophical Transactions of the Royal Society of London Series A},
         year = 2012,
        month = jun,
       volume = {370},
       number = {1968},
        pages = {2765-2777},
          doi = {10.1098/rsta.2011.0269},
archivePrefix = {arXiv},
       eprint = {1112.3591},
 primaryClass = {astro-ph.SR},
       adsurl = {https://ui.adsabs.harvard.edu/abs/2012RSPTA.370.2765A}
}

@ARTICLE{2022MNRAS.513.2719V,
       author = {{Vines}, Jose I. and {Jenkins}, James S.},
        title = "{ARIADNE: measuring accurate and precise stellar parameters through SED fitting}",
      journal = {\mnras},
         year = 2022,
        month = jun,
       volume = {513},
       number = {2},
        pages = {2719-2731},
          doi = {10.1093/mnras/stac956},
archivePrefix = {arXiv},
       eprint = {2204.03769},
 primaryClass = {astro-ph.SR},
       adsurl = {https://ui.adsabs.harvard.edu/abs/2022MNRAS.513.2719V}
}

@INPROCEEDINGS{1986SPIE..627..733T,
       author = {{Tody}, Doug},
        title = "{The IRAF Data Reduction and Analysis System}",
    booktitle = {Instrumentation in astronomy VI},
         year = 1986,
       editor = {{Crawford}, David L.},
       series = {Society of Photo-Optical Instrumentation Engineers (SPIE) Conference Series},
       volume = {627},
        month = jan,
        pages = {733},
          doi = {10.1117/12.968154},
       adsurl = {https://ui.adsabs.harvard.edu/abs/1986SPIE..627..733T}
}

@ARTICLE{2022Sci...377.1211L,
       author = {{Luque}, Rafael and {Pall{\'e}}, Enric},
        title = "{Density, not radius, separates rocky and water-rich small planets orbiting M dwarf stars}",
      journal = {Science},
         year = 2022,
        month = sep,
       volume = {377},
       number = {6611},
        pages = {1211-1214},
          doi = {10.1126/science.abl7164},
archivePrefix = {arXiv},
       eprint = {2209.03871},
 primaryClass = {astro-ph.EP},
       adsurl = {https://ui.adsabs.harvard.edu/abs/2022Sci...377.1211L}
}

@ARTICLE{2019AJ....158..138S,
       author = {{Stassun}, Keivan G. and {Oelkers}, Ryan J. and {Paegert}, Martin and {Torres}, Guillermo and {Pepper}, Joshua and {De Lee}, Nathan },
        title = "{The Revised TESS Input Catalog and Candidate Target List}",
      journal = {\aj},
         year = 2019,
        month = oct,
       volume = {158},
       number = {4},
          eid = {138},
        pages = {138},
          doi = {10.3847/1538-3881/ab3467},
archivePrefix = {arXiv},
       eprint = {1905.10694},
 primaryClass = {astro-ph.SR},
       adsurl = {https://ui.adsabs.harvard.edu/abs/2019AJ....158..138S}
}

@ARTICLE{2024MNRAS.527.5693P,
       author = {{Parviainen}, Hannu and {Luque}, Rafael and {Palle}, Enric},
        title = "{SPRIGHT: a probabilistic mass-density-radius relation for small planets}",
      journal = {\mnras},
         year = 2024,
        month = jan,
       volume = {527},
       number = {3},
        pages = {5693-5716},
          doi = {10.1093/mnras/stad3504},
archivePrefix = {arXiv},
       eprint = {2311.07255},
 primaryClass = {astro-ph.EP},
       adsurl = {https://ui.adsabs.harvard.edu/abs/2024MNRAS.527.5693P}
}

@ARTICLE{2017AJ....153...77C,
       author = {{Collins}, Karen A. and {Kielkopf}, John F. and {Stassun}, Keivan G. and {Hessman}, Frederic V.},
        title = "{AstroImageJ: Image Processing and Photometric Extraction for Ultra-precise Astronomical Light Curves}",
      journal = {\aj},
         year = 2017,
        month = feb,
       volume = {153},
       number = {2},
          eid = {77},
        pages = {77},
          doi = {10.3847/1538-3881/153/2/77},
archivePrefix = {arXiv},
       eprint = {1701.04817},
 primaryClass = {astro-ph.IM},
       adsurl = {https://ui.adsabs.harvard.edu/abs/2017AJ....153...77C}
}

@ARTICLE{2020MNRAS.499.1633P,
       author = {{Parviainen}, H.},
        title = "{RoadRunner: a fast and flexible exoplanet transit model}",
      journal = {\mnras},
         year = 2020,
        month = dec,
       volume = {499},
       number = {2},
        pages = {1633-1639},
          doi = {10.1093/mnras/staa2901},
archivePrefix = {arXiv},
       eprint = {2009.08500},
 primaryClass = {astro-ph.EP},
       adsurl = {https://ui.adsabs.harvard.edu/abs/2020MNRAS.499.1633P}
}

@ARTICLE{2008ConPh..49...71T,
       author = {{Trotta}, Roberto},
        title = "{Bayes in the sky: Bayesian inference and model selection in cosmology}",
      journal = {Contemporary Physics},
         year = 2008,
        month = mar,
       volume = {49},
       number = {2},
        pages = {71-104},
          doi = {10.1080/00107510802066753},
archivePrefix = {arXiv},
       eprint = {0803.4089},
 primaryClass = {astro-ph},
       adsurl = {https://ui.adsabs.harvard.edu/abs/2008ConPh..49...71T}
}

@ARTICLE{2024A&A...690A..62P,
       author = {{Pel{\'a}ez-Torres}, A. and {Esparza-Borges}, E. and {Pall{\'e}}, E. and {Parviainen}, H. and {Murgas}, F. and {Morello}, G. and {Zapatero-Osorio}, M.~R. },
        title = "{Validation of up to seven TESS planet candidates through multi-colour transit photometry using MuSCAT2 data}",
      journal = {\aap},
         year = 2024,
        month = oct,
       volume = {690},
          eid = {A62},
        pages = {A62},
          doi = {10.1051/0004-6361/202347251},
archivePrefix = {arXiv},
       eprint = {2409.07400},
 primaryClass = {astro-ph.EP},
       adsurl = {https://ui.adsabs.harvard.edu/abs/2024A&A...690A..62P}
}

@ARTICLE{2024A&A...690A.263G,
       author = {{Ghachoui}, M. and {Rackham}, B.~V. and {D{\'e}vora-Pajares}, M. and {Chouqar}, J. and {Timmermans}, M. and {Kaltenegger}, L. and {Sebastian}, D. },
        title = "{TESS discovery of two super-Earths orbiting the M-dwarf stars TOI-6002 and TOI-5713 near the radius valley}",
      journal = {\aap},
         year = 2024,
        month = oct,
       volume = {690},
          eid = {A263},
        pages = {A263},
          doi = {10.1051/0004-6361/202451120},
archivePrefix = {arXiv},
       eprint = {2408.00709},
 primaryClass = {astro-ph.EP},
       adsurl = {https://ui.adsabs.harvard.edu/abs/2024A&A...690A.263G}
}

@ARTICLE{2024A&A...683A.170P,
       author = {{Parviainen}, H. and {Murgas}, F. and {Esparza-Borges}, E. and {Pel{\'a}ez-Torres}, A. and {Palle}, E. and {Luque}, R. and {Zapatero-Osorio}, M.~R. },
        title = "{TOI-2266 b: A keystone super-Earth at the edge of the M dwarf radius valley}",
      journal = {\aap},
         year = 2024,
        month = mar,
       volume = {683},
          eid = {A170},
        pages = {A170},
          doi = {10.1051/0004-6361/202347431},
archivePrefix = {arXiv},
       eprint = {2401.11879},
 primaryClass = {astro-ph.EP},
       adsurl = {https://ui.adsabs.harvard.edu/abs/2024A&A...683A.170P}
}

@ARTICLE{2022A&A...666A..10E,
       author = {{Esparza-Borges}, E. and {Parviainen}, H. and {Murgas}, F. and {Pall{\'e}}, E. and {Maas}, A. and {Morello}, G. and {Zapatero-Osorio}, M.~R. and {Barkaoui}, K.  },
        title = "{A hot sub-Neptune in the desert and a temperate super-Earth around faint M dwarfs. Color validation of TOI-4479b and TOI-2081b}",
      journal = {\aap},
         year = 2022,
        month = oct,
       volume = {666},
          eid = {A10},
        pages = {A10},
          doi = {10.1051/0004-6361/202243731},
archivePrefix = {arXiv},
       eprint = {2206.10643},
 primaryClass = {astro-ph.EP},
       adsurl = {https://ui.adsabs.harvard.edu/abs/2022A&A...666A..10E}
}

@ARTICLE{2021AJ....162...54H,
       author = {{Hedges}, Christina and {Hughes}, Alex and {Zhou}, George and {David}, Trevor J. and {Becker}, Juliette and {Giacalone}, Steven and {Vanderburg}, Andrew },
        title = "{TOI-2076 and TOI-1807: Two Young, Comoving Planetary Systems within 50 pc Identified by TESS that are Ideal Candidates for Further Follow Up}",
      journal = {\aj},
         year = 2021,
        month = aug,
       volume = {162},
       number = {2},
          eid = {54},
        pages = {54},
          doi = {10.3847/1538-3881/ac06cd},
archivePrefix = {arXiv},
       eprint = {2111.01311},
 primaryClass = {astro-ph.EP},
       adsurl = {https://ui.adsabs.harvard.edu/abs/2021AJ....162...54H}
}

@ARTICLE{2019A&A...630A..89P,
       author = {{Parviainen}, H. and {Tingley}, B. and {Deeg}, H.~J. and {Palle}, E. and {Alonso}, R. and {Montanes Rodriguez}, P. and {Murgas}, F. and {Narita}, N. },
        title = "{Multicolour photometry for exoplanet candidate validation}",
      journal = {\aap},
         year = 2019,
        month = oct,
       volume = {630},
          eid = {A89},
        pages = {A89},
          doi = {10.1051/0004-6361/201935709},
archivePrefix = {arXiv},
       eprint = {1907.09776},
 primaryClass = {astro-ph.EP},
       adsurl = {https://ui.adsabs.harvard.edu/abs/2019A&A...630A..89P}
}

@ARTICLE{2020A&A...633A..28P,
       author = {{Parviainen}, H. and {Palle}, E. and {Zapatero-Osorio}, M.~R. and {Montanes Rodriguez}, P. and {Murgas}, F. and {Narita}, N. and {Hidalgo Soto}, D.},
        title = "{MuSCAT2 multicolour validation of TESS candidates: an ultra-short-period substellar object around an M dwarf}",
      journal = {\aap},
         year = 2020,
        month = jan,
       volume = {633},
          eid = {A28},
        pages = {A28},
          doi = {10.1051/0004-6361/201935958},
archivePrefix = {arXiv},
       eprint = {1911.04366},
 primaryClass = {astro-ph.EP},
       adsurl = {https://ui.adsabs.harvard.edu/abs/2020A&A...633A..28P}
}

@ARTICLE{2017Natur.544..333D,
       author = {{Dittmann}, Jason A. and {Irwin}, Jonathan M. and {Charbonneau}, David and {Bonfils}, Xavier and {Astudillo-Defru}, Nicola and {Haywood}, Rapha{\"e}lle D. },
        title = "{A temperate rocky super-Earth transiting a nearby cool star}",
      journal = {\nat},
         year = 2017,
        month = apr,
       volume = {544},
       number = {7650},
        pages = {333-336},
          doi = {10.1038/nature22055},
archivePrefix = {arXiv},
       eprint = {1704.05556},
 primaryClass = {astro-ph.EP},
       adsurl = {https://ui.adsabs.harvard.edu/abs/2017Natur.544..333D}
}

@ARTICLE{2024MNRAS.527...35D,
       author = {{Dransfield}, Georgina and {Timmermans}, Mathilde and {Triaud}, Amaury H.~M.~J. and {D{\'e}vora-Pajares}, Mart{\'\i}n and {Aganze}, Christian },
        title = "{A 1.55 R$_{{\ensuremath{\oplus}}}$ habitable-zone planet hosted by TOI-715, an M4 star near the ecliptic South Pole}",
      journal = {\mnras},
         year = 2024,
        month = jan,
       volume = {527},
       number = {1},
        pages = {35-52},
          doi = {10.1093/mnras/stad1439},
archivePrefix = {arXiv},
       eprint = {2305.06206},
 primaryClass = {astro-ph.EP},
       adsurl = {https://ui.adsabs.harvard.edu/abs/2024MNRAS.527...35D}
}

@ARTICLE{2020AJ....160..117R,
       author = {{Rodriguez}, Joseph E. and {Vanderburg}, Andrew and {Zieba}, Sebastian and {Kreidberg}, Laura and {Morley}, Caroline V. and {Eastman}, Jason D. },
        title = "{The First Habitable-zone Earth-sized Planet from TESS. II. Spitzer Confirms TOI-700 d}",
      journal = {\aj},
         year = 2020,
        month = sep,
       volume = {160},
       number = {3},
          eid = {117},
        pages = {117},
          doi = {10.3847/1538-3881/aba4b3},
archivePrefix = {arXiv},
       eprint = {2001.00954},
 primaryClass = {astro-ph.EP},
       adsurl = {https://ui.adsabs.harvard.edu/abs/2020AJ....160..117R}
}

@ARTICLE{2020AJ....160..116G,
       author = {{Gilbert}, Emily A. and {Barclay}, Thomas and {Schlieder}, Joshua E. and {Quintana}, Elisa V. and {Hord}, Benjamin J. and {Kostov}, Veselin B. },
        title = "{The First Habitable-zone Earth-sized Planet from TESS. I. Validation of the TOI-700 System}",
      journal = {\aj},
         year = 2020,
        month = sep,
       volume = {160},
       number = {3},
          eid = {116},
        pages = {116},
          doi = {10.3847/1538-3881/aba4b2},
archivePrefix = {arXiv},
       eprint = {2001.00952},
 primaryClass = {astro-ph.EP},
       adsurl = {https://ui.adsabs.harvard.edu/abs/2020AJ....160..116G}
}

@ARTICLE{2018AJ....156...82C,
    author = {{Cloutier}, Ryan and {Doyon}, Ren{\'e} and {Bouchy}, Francois and {H{\'e}brard}, Guillaume},
    title = "{Quantifying the Observational Effort Required for the Radial Velocity Characterization of TESS Planets}",
    journal = {\aj},
    year = 2018,
    month = Aug,
    volume = {156},
    eid = {82},
    pages = {82},
    doi = {10.3847/1538-3881/aacea9},
    archivePrefix = {arXiv},
    eprint = {1807.01263},
    primaryClass = {astro-ph.EP},
    adsurl = {https://ui.adsabs.harvard.edu/\#abs/2018AJ....156...82C}
}

@INPROCEEDINGS{2018SPIE10707E..0KM,
       author = {{McCully}, Curtis and {Volgenau}, Nikolaus H. and {Harbeck}, Daniel-Rolf and {Lister}, Tim A. and {Saunders}, Eric S. and {Turner}, Monica L. and {Siiverd}, Robert J. and {Bowman}, Mark},
        title = "{Real-time processing of the imaging data from the network of Las Cumbres Observatory Telescopes using BANZAI}",
    booktitle = {Software and Cyberinfrastructure for Astronomy V},
         year = 2018,
       editor = {{Guzman}, Juan C. and {Ibsen}, Jorge},
       series = {SPIE Conference Series},
       volume = {10707},
        month = jul,
          eid = {107070K},
        pages = {107070K},
          doi = {10.1117/12.2314340},
archivePrefix = {arXiv},
       eprint = {1811.04163},
 primaryClass = {astro-ph.IM},
       adsurl = {https://ui.adsabs.harvard.edu/abs/2018SPIE10707E..0KM}
}

@INPROCEEDINGS{2020SPIE11447E..5KN,
       author = {{Narita}, Norio and {Fukui}, Akihiko and {Yamamuro}, Tomoyasu and {Harbeck}, Daniel and {Bowman}, Mark and {Elphick}, Mark and {Nation}, Jon },
        title = "{MuSCAT3: a 4-color simultaneous camera for the 2m Faulkes Telescope North}",
    booktitle = {Ground-based and Airborne Instrumentation for Astronomy VIII},
         year = 2020,
       editor = {{Evans}, Christopher J. and {Bryant}, Julia J. and {Motohara}, Kentaro},
       series = {SPIE Conference Series},
       volume = {11447},
        month = dec,
          eid = {114475K},
        pages = {114475K},
          doi = {10.1117/12.2559947},
       adsurl = {https://ui.adsabs.harvard.edu/abs/2020SPIE11447E..5KN}
}

@ARTICLE{2012PASP..124.1000S,
       author = {{Smith}, Jeffrey C. and {Stumpe}, Martin C. and {Van Cleve}, Jeffrey E. and {Jenkins}, Jon M. and {Barclay}, Thomas S. and {Fanelli}, Michael N. and {Girouard}, Forrest R. and {Kolodziejczak}, Jeffery J. and {McCauliff}, Sean D. and {Morris}, Robert L. and {Twicken}, Joseph D.},
        title = "{Kepler Presearch Data Conditioning II - A Bayesian Approach to Systematic Error Correction}",
      journal = {\pasp},
         year = 2012,
        month = sep,
       volume = {124},
       number = {919},
        pages = {1000},
          doi = {10.1086/667697},
archivePrefix = {arXiv},
       eprint = {1203.1383},
 primaryClass = {astro-ph.IM},
       adsurl = {https://ui.adsabs.harvard.edu/abs/2012PASP..124.1000S}
}

@ARTICLE{2014PASP..126..100S,
       author = {{Stumpe}, Martin C. and {Smith}, Jeffrey C. and {Catanzarite}, Joseph H. and {Van Cleve}, Jeffrey E. and {Jenkins}, Jon M. and {Twicken}, Joseph D. and {Girouard}, Forrest R.},
        title = "{Multiscale Systematic Error Correction via Wavelet-Based Bandsplitting in Kepler Data}",
      journal = {\pasp},
         year = 2014,
        month = jan,
       volume = {126},
       number = {935},
        pages = {100},
          doi = {10.1086/674989},
       adsurl = {https://ui.adsabs.harvard.edu/abs/2014PASP..126..100S}
}

@ARTICLE{2012PASP..124..985S,
       author = {{Stumpe}, Martin C. and {Smith}, Jeffrey C. and {Van Cleve}, Jeffrey E. and {Twicken}, Joseph D. and {Barclay}, Thomas S. and {Fanelli}, Michael N. and {Girouard}, Forrest R. and {Jenkins}, Jon M. and {Kolodziejczak}, Jeffery J. and {McCauliff}, Sean D. and {Morris}, Robert L.},
        title = "{Kepler Presearch Data Conditioning I{\textemdash}Architecture and Algorithms for Error Correction in Kepler Light Curves}",
      journal = {\pasp},
         year = 2012,
        month = sep,
       volume = {124},
       number = {919},
        pages = {985},
          doi = {10.1086/667698},
archivePrefix = {arXiv},
       eprint = {1203.1382},
 primaryClass = {astro-ph.IM},
       adsurl = {https://ui.adsabs.harvard.edu/abs/2012PASP..124..985S}
}

@INPROCEEDINGS{2016SPIE.9913E..3EJ,
       author = {{Jenkins}, Jon M. and {Twicken}, Joseph D. and {McCauliff}, Sean and {Campbell}, Jennifer},
        title = "{The TESS science processing operations center}",
    booktitle = {Software and Cyberinfrastructure for Astronomy IV},
         year = 2016,
       editor = {{Chiozzi}, Gianluca and {Guzman}, Juan C.},
       series = {SPIE Conference Series},
       volume = {9913},
        month = aug,
          eid = {99133E},
        pages = {99133E},
          doi = {10.1117/12.2233418},
       adsurl = {https://ui.adsabs.harvard.edu/abs/2016SPIE.9913E..3EJ}
}

@ARTICLE{2020A&A...635A.128A,
       author = {{Aller}, A. and {Lillo-Box}, J. and {Jones}, D. and {Miranda}, L.~F. and {Barcel{\'o} Forteza}, S.},
        title = "{Planetary nebulae seen with TESS: Discovery of new binary central star candidates from Cycle 1}",
      journal = {\aap},
         year = 2020,
        month = mar,
       volume = {635},
          eid = {A128},
        pages = {A128},
          doi = {10.1051/0004-6361/201937118},
archivePrefix = {arXiv},
       eprint = {1911.09991},
 primaryClass = {astro-ph.SR}
}

@ARTICLE{2017AJ....154..220F,
       author = {{Foreman-Mackey}, Daniel and {Agol}, Eric and {Ambikasaran}, Sivaram and {Angus}, Ruth},
        title = "{Fast and Scalable Gaussian Process Modeling with Applications to Astronomical Time Series}",
      journal = {\aj},
         year = 2017,
        month = dec,
       volume = {154},
       number = {6},
          eid = {220},
        pages = {220},
          doi = {10.3847/1538-3881/aa9332},
archivePrefix = {arXiv},
       eprint = {1703.09710},
 primaryClass = {astro-ph.IM},
       adsurl = {https://ui.adsabs.harvard.edu/abs/2017AJ....154..220F}
}

@ARTICLE{2013A&A...553A...6H,
       author = {{Husser}, T. -O. and {Wende-von Berg}, S. and {Dreizler}, S. and {Homeier}, D. and {Reiners}, A. and {Barman}, T. and {Hauschildt}, P.~H.},
        title = "{A new extensive library of PHOENIX stellar atmospheres and synthetic spectra}",
      journal = {\aap},
         year = 2013,
        month = may,
       volume = {553},
          eid = {A6},
        pages = {A6},
          doi = {10.1051/0004-6361/201219058},
archivePrefix = {arXiv},
       eprint = {1303.5632},
 primaryClass = {astro-ph.SR},
       adsurl = {https://ui.adsabs.harvard.edu/abs/2013A&A...553A...6H}
}

@ARTICLE{2023A&A...674A...1G,
       author = {{Gaia Collaboration} and {Vallenari}, A. and {Brown}, A.~G.~A. and {Prusti}, T. and {de Bruijne}, J.~H.~J. },
        title = "{Gaia Data Release 3. Summary of the content and survey properties}",
      journal = {\aap},
         year = 2023,
        month = jun,
       volume = {674},
          eid = {A1},
        pages = {A1},
          doi = {10.1051/0004-6361/202243940},
archivePrefix = {arXiv},
       eprint = {2208.00211},
 primaryClass = {astro-ph.GA},
       adsurl = {https://ui.adsabs.harvard.edu/abs/2023A&A...674A...1G}
}

@ARTICLE{2016A&A...595A...1G,
       author = {{Gaia Collaboration} and {Prusti}, T. and {de Bruijne}, J.~H.~J. and {Brown}, A.~G.~A. and {Vallenari}, A. and {Babusiaux}, C. and {Bailer-Jones}, C.~A.~L. and {Bastian}, U. and {Biermann}, M. and {Evans}, D.~W. and {Eyer}, L. and {Jansen}, F. and {Jordi}, C. and {Klioner}, S.~A. },
        title = "{The Gaia mission}",
      journal = {\aap},
         year = 2016,
        month = nov,
       volume = {595},
          eid = {A1},
        pages = {A1},
          doi = {10.1051/0004-6361/201629272},
archivePrefix = {arXiv},
       eprint = {1609.04153},
 primaryClass = {astro-ph.IM},
       adsurl = {https://ui.adsabs.harvard.edu/abs/2016A&A...595A...1G}
}

@ARTICLE{2013ApJ...778..153B,
       author = {{Benneke}, Bj{\"o}rn and {Seager}, Sara},
        title = "{How to Distinguish between Cloudy Mini-Neptunes and Water/Volatile-dominated Super-Earths}",
      journal = {\apj},
         year = 2013,
        month = dec,
       volume = {778},
       number = {2},
          eid = {153},
        pages = {153},
          doi = {10.1088/0004-637X/778/2/153},
archivePrefix = {arXiv},
       eprint = {1306.6325},
 primaryClass = {astro-ph.EP},
       adsurl = {https://ui.adsabs.harvard.edu/abs/2013ApJ...778..153B}
}

@ARTICLE{2019A&A...627A..67M,
       author = {{Molli{\`e}re}, P. and {Wardenier}, J.~P. and {van Boekel}, R. and {Henning}, Th. and {Molaverdikhani}, K. and {Snellen}, I.~A.~G.},
        title = "{petitRADTRANS. A Python radiative transfer package for exoplanet characterization and retrieval}",
      journal = {\aap},
         year = 2019,
        month = jul,
       volume = {627},
          eid = {A67},
        pages = {A67},
          doi = {10.1051/0004-6361/201935470},
archivePrefix = {arXiv},
       eprint = {1904.11504},
 primaryClass = {astro-ph.EP},
       adsurl = {https://ui.adsabs.harvard.edu/abs/2019A&A...627A..67M}
}

@ARTICLE{2009MNRAS.398.1601F,
       author = {{Feroz}, F. and {Hobson}, M.~P. and {Bridges}, M.},
        title = "{MULTINEST: an efficient and robust Bayesian inference tool for cosmology and particle physics}",
      journal = {\mnras},
         year = 2009,
        month = oct,
       volume = {398},
       number = {4},
        pages = {1601-1614},
          doi = {10.1111/j.1365-2966.2009.14548.x},
archivePrefix = {arXiv},
       eprint = {0809.3437},
 primaryClass = {astro-ph},
       adsurl = {https://ui.adsabs.harvard.edu/abs/2009MNRAS.398.1601F}
}

@ARTICLE{2014A&A...564A.125B,
       author = {{Buchner}, J. and {Georgakakis}, A. and {Nandra}, K. and {Hsu}, L. and {Rangel}, C. and {Brightman}, M. and {Merloni}, A. and {Salvato}, M. and {Donley}, J. and {Kocevski}, D.},
        title = "{X-ray spectral modelling of the AGN obscuring region in the CDFS: Bayesian model selection and catalogue}",
      journal = {\aap},
         year = 2014,
        month = apr,
       volume = {564},
          eid = {A125},
        pages = {A125},
          doi = {10.1051/0004-6361/201322971},
archivePrefix = {arXiv},
       eprint = {1402.0004},
 primaryClass = {astro-ph.HE},
       adsurl = {https://ui.adsabs.harvard.edu/abs/2014A&A...564A.125B}
}

@BOOK{2003tmc..book.....C,
       author = {{Cutri}, R.~M. and {Skrutskie}, M.~F. and {van Dyk}, S. and {Beichman}, C.~A. and {Carpenter}, J.~M. and {Chester}, T. and {Cambresy}, L. and {Evans}, T. and {Fowler}, J. and {Gizis}, J. and {Howard}, E. and {Huchra}, J. and {Jarrett}, T. and {Kopan}, E.~L. and {Kirkpatrick}, J.~D. and {Light}, R.~M. and {Marsh}, K.~A. and {McCallon}, H. and {Schneider}, S. and {Stiening}, R. and {Sykes}, M. and {Weinberg}, M. and {Wheaton}, W.~A. and {Wheelock}, S. and {Zacarias}, N.},
        title = "{2MASS All Sky Catalog of point sources.}",
         year = 2003,
       adsurl = {https://ui.adsabs.harvard.edu/abs/2003tmc..book.....C}
}

@ARTICLE{2015MNRAS.450.3233P,
       author = {{Parviainen}, Hannu},
        title = "{PYTRANSIT: fast and easy exoplanet transit modelling in PYTHON}",
      journal = {\mnras},
         year = 2015,
        month = jul,
       volume = {450},
       number = {3},
        pages = {3233-3238},
          doi = {10.1093/mnras/stv894},
archivePrefix = {arXiv},
       eprint = {1504.07433},
 primaryClass = {astro-ph.EP},
       adsurl = {https://ui.adsabs.harvard.edu/abs/2015MNRAS.450.3233P}
}

@ARTICLE{2015MNRAS.453.3821P,
       author = {{Parviainen}, H. and {Aigrain}, S.},
        title = "{LDTK: Limb Darkening Toolkit}",
      journal = {\mnras},
         year = 2015,
        month = nov,
       volume = {453},
       number = {4},
        pages = {3821-3826},
          doi = {10.1093/mnras/stv1857},
archivePrefix = {arXiv},
       eprint = {1508.02634},
 primaryClass = {astro-ph.EP},
       adsurl = {https://ui.adsabs.harvard.edu/abs/2015MNRAS.453.3821P}
}

@ARTICLE{Narita2019,
       author = {{Narita}, Norio and {Fukui}, Akihiko and {Kusakabe}, Nobuhiko and {Watanabe}, Noriharu and {Palle}, Enric and {Parviainen}, Hannu and {Monta{\~n}{\'e}s-Rodr{\'\i}guez}, Pilar and {Murgas}, Felipe and {Monelli}, Matteo and {Aguiar}, Marta and {Perez Prieto}, Jorge Andres and {Oscoz}, {\'A}lex and {de Leon}, Jerome and {Mori}, Mayuko and {Tamura}, Motohide and {Yamamuro}, Tomoyasu and {B{\'e}jar}, Victor J.~S. and {Crouzet}, Nicolas and {Hidalgo}, Diego and {Klagyivik}, Peter and {Luque}, Rafael and {Nishiumi}, Taku},
        title = "{MuSCAT2: four-color simultaneous camera for the 1.52-m Telescopio Carlos S{\'a}nchez}",
      journal = {Journal of Astronomical Telescopes, Instruments, and Systems},
         year = 2019,
        month = jan,
       volume = {5},
          eid = {015001},
        pages = {015001},
          doi = {10.1117/1.JATIS.5.1.015001},
archivePrefix = {arXiv},
       eprint = {1807.01908},
 primaryClass = {astro-ph.IM},
       adsurl = {https://ui.adsabs.harvard.edu/abs/2019JATIS...5a5001N}
}

@ARTICLE{Parviainen2019,
       author = {{Parviainen}, H. and {Tingley}, B. and {Deeg}, H.~J. and {Palle}, E. and {Alonso}, R. and {Montanes Rodriguez}, P. and {Murgas}, F. and {Narita}, N. and {Fukui}, A. and {Watanabe}, N. and {Kusakabe}, N. and {Tamura}, M. and {Nishiumi}, T. and {Prieto-Arranz}, J. and {Klagyivik}, P. and {B{\'e}jar}, V.~J.~S. and {Crouzet}, N. and {Mori}, M. and {Hidalgo Soto}, D. and {Casasayas Barris}, N. and {Luque}, R.},
        title = "{Multicolour photometry for exoplanet candidate validation}",
      journal = {\aap},
         year = 2019,
        month = oct,
       volume = {630},
          eid = {A89},
        pages = {A89},
          doi = {10.1051/0004-6361/201935709},
archivePrefix = {arXiv},
       eprint = {1907.09776},
 primaryClass = {astro-ph.EP},
       adsurl = {https://ui.adsabs.harvard.edu/abs/2019A&A...630A..89P}
}


\bsp	
\label{lastpage}

\appendix
\section{Supplementary figures and tables}

\begin{table*} 
    \caption{Posterior parameter estimates for the false positive hypotheses $\mathcal{H}_2$ and $\mathcal{H}_3$.}
    \begin{tabular}{llcc}
    \hline
        Parameter & Prior & TOI-2094 b & TOI-7166 b \\
    \hline
    $\mathcal{H}_2$: \textit{an eclipsing white dwarf} \\
    \hline
        Radius ratio $R_{\rm p}/R_\star$ & $\mathcal{U}(0, 0.2)$ & $0.0246^{+0.0017}_{-0.0017}$ & $0.0662^{+0.0059}_{-0.0055}$\\
        Transit epoch $T_0$ [$\rm BJD_{TDB}$] & $\mathcal{U}(-0.1, 0.1)+{\rm Const.}$ $^a$ & $ 2\,460\,232.360959^{+0.00024}_{-0.00024}$ & $2\,460\,886.50479^{+0.00040}_{-0.00039}$\\
        Orbital period $P$ [d] & $\mathcal{U}(-0.1, 0.1)+{\rm Const.}$ $^b$ & $18.793192^{+0.000018}_{-0.000020}$ & $12.920585^{+0.000067}_{-0.000070}$\\
        Scaled semimajor axis $a/R_\star$ & $\mathcal{U}(0.3, 1)\times {\rm Const.}$ $^c$  & $51.1^{+2.6}_{-2.6}$ & $33.2^{+1.8}_{-1.6}$\\
        Orbital impact parameter $b$ & $\mathcal{U}(0, 1+R_{\rm p}/R_\star)$ & $0.9269^{+0.0075}_{-0.0078}$ & $0.866^{+0.015}_{-0.019}$\\
        Stellar temperature $T_{\rm eff}$ [K] & $\mathcal{N}(\mu,\sigma)$ from Table \ref{tab:stellar_parameters} & $3397^{+71}_{-67}$ & $3100^{+60}_{-63}$ \\
        Stellar gravity $\lg g$ [c.g.s] & $\mathcal{N}(\mu,\sigma)$ from Table \ref{tab:stellar_parameters} & $4.89^{+0.04}_{-0.04}$ & $5.01^{+0.06}_{-0.06}$\\
        Stellar metallicity [M/H] & $\mathcal{N}(\mu,\sigma)$ from Table \ref{tab:stellar_parameters} & $-0.06^{+0.02}_{-0.02}$ & $0.24^{+0.04}_{-0.04}$\\
        White dwarf temperature $T_{\rm eff,WD}$ [K] & $\mathcal{U}(2000, 20\,000)$ & $4145^{+154}_{-142}$ & $3369^{+115}_{-111}$ \\ 
    \hline
    $\mathcal{H}_3$: \textit{a brown dwarf transiting an unresolved star} \\
    \hline
        Radius ratio $R_{\rm p}/R_\star$ & $\mathcal{U}(0.1, 1)$ & $0.224^{+0.042}_{-0.017}$ & $0.136^{+0.010}_{-0.007}$\\
        Transit epoch $T_0$ [$\rm BJD_{TDB}$] & $\mathcal{U}(-0.1, 0.1)+{\rm Const.}$ $^a$ & $2\,460\,232.36092^{+0.00039}_{-0.00059}$ & $2\,460\,886.50468^{+0.00042}_{-0.00041}$\\
        Orbital period $P$ [d] & $\mathcal{U}(-0.1, 0.1)+{\rm Const.}$ $^b$ & $18.793176^{+0.000024}_{-0.000025}$ & $12.920737^{+0.000067}_{-0.000070}$\\
        Scaled semimajor axis $a/R_\star$ & $\mathcal{U}(20, 200)$  & $129.8^{+2.4}_{-2.6}$ & $54.6^{+3.3}_{-2.3}$\\
        Orbital impact parameter $b$ & $\mathcal{U}(0, 1+R_{\rm p}/R_\star)$ & $0.094^{+0.092}_{-0.063}$ & $0.48^{+0.07}_{-0.12}$\\
        Stellar temperature $T_{\rm eff}$ [K] & $\mathcal{N}(\mu,\sigma)$ from Table \ref{tab:stellar_parameters} & $3430^{+61}_{-69}$ & $2993^{+73}_{-68}$ \\
        Contaminant temperature $T_{\rm eff,EB}$ [K] & $\mathcal{U}(2300, 7000)$ & $4403^{+132}_{-409}$ & $6830^{+122}_{-268}$ \\
        Contaminant gravity $\lg g$ [c.g.s] & $\mathcal{U}(4.0, 5.5)$ & $4.73^{+0.52}_{-0.47}$ & $4.72^{+0.16}_{-0.18}$\\
        Contaminant metallicity [M/H] & $\mathcal{U}(-1, 0.5)$ & $-0.46^{+0.56}_{-0.38}$ & $-0.13^{+0.42}_{-0.49}$ \\
        Rescaling factor $\lg \gamma$ & $\mathcal{U}(1, 6)$ & $1.011^{+0.016}_{-0.008}$ & $1.007^{+0.009}_{-0.005}$ \\
    \hline
    \end{tabular}

    \medskip
    \begin{minipage}{\linewidth}
    \textbf{Notes.}
    $^a$ Constants are 2\,460\,232.35970 for TOI-2094 b and 2\,460\,886.50520 for TOI-7166 b, corresponding to the mid-transit epoch predicted by the ephemeris from SPOC DV for the latest HiPERCAM observation. Specially for $\mathcal{H}_2$, the value $T_0$ indicates the mid-eclipse epoch.
    $^b$ Constants are 18.793175 for TOI-2094 b and 12.920670 for TOI-7166 b.
    $^c$ Using Kepler's third law, the constants are calculated to be 120 for TOI-2094 b and 100 for TOI-7166 b, given the prior knowledge of the stellar bulk density and planetary orbital period. 
    \end{minipage}

    \label{tab:planet_params_other_scenarios}
\end{table*}

 \begin{table*} 
    \caption{Posterior estimates of the planetary parameters of TOI-2094 b and TOI-7166 b assuming eccentric orbits.}
    \begin{tabular}{llcc}
    \hline
        Parameter & Prior & TOI-2094 b & TOI-7166 b \\
    \hline 
        Radius ratio $R_{\rm p}/R_\star$ & $\mathcal{U}(0, 0.2)$ & $0.0455^{+0.0013}_{-0.0014}$ & $0.0896^{+0.0025}_{-0.0027}$\\
        Transit epoch $T_0$ [$\rm BJD_{TDB}$] & $\mathcal{U}(-0.1, 0.1)+{\rm Const.}$ $^a$ & $2\,460\,232.36093^{+0.00033}_{-0.00031}$ & $2\,460\,886.50492^{+0.00037}_{-0.00036}$\\
        Orbital period $P$ [d] & $\mathcal{U}(-0.1, 0.1)+{\rm Const.}$ $^b$ & $18.793194^{+0.000017}_{-0.000018}$ & $12.920616^{+0.000056}_{-0.000058}$\\
        Scaled semimajor axis $a/R_\star$ & $\mathcal{U}(0.3, 1)\times {\rm Const.}$ $^c$  & $79.4^{+16.9}_{-9.3}$ & $44.3^{+10.2}_{-6.6}$\\
        Orbital impact parameter $b$ & $\mathcal{U}(0, 1+R_{\rm p}/R_\star)$ & $0.81^{+0.10}_{-0.11}$ & $0.71^{+0.14}_{-0.11}$\\
        Orbital eccentricity $e$ & $\mathcal{U}(0, 1)$  & $0.18^{+0.19}_{-0.12}$ & $0.20^{+0.16}_{-0.14}$\\
        Argument of periastron $\omega$ [deg] & $\mathcal{U}(0, 360)$ & $216^{+100}_{-141}$ & $191^{+111}_{-130}$\\
        Stellar temperature $T_{\rm eff}$ [K] & $\mathcal{N}(\mu,\sigma)$ from Table \ref{tab:stellar_parameters} & $3435^{+55}_{-62}$ & $3092^{+59}_{-59}$ \\
        Stellar gravity $\lg g$ [c.g.s] & $\mathcal{N}(\mu,\sigma)$ from Table \ref{tab:stellar_parameters} & $4.89^{+0.04}_{-0.04}$ & $5.01^{+0.06}_{-0.06}$\\
        Stellar metallicity [Fe/H] & $\mathcal{N}(\mu,\sigma)$ from Table \ref{tab:stellar_parameters} & $-0.06^{+0.02}_{-0.02}$ & $0.24^{+0.04}_{-0.04}$\\ 
    \hline
    \end{tabular}
    
    \medskip
    \begin{minipage}{\linewidth}
    \textbf{Notes.}
    $^a$ Constants are 2\,460\,232.35970 for TOI-2094 b and 2\,460\,886.50520 for TOI-7166 b, corresponding to the mid-transit epoch predicted by the ephemeris from SPOC DV for the latest HiPERCAM observation.
    $^b$ Constants are 18.793175 for TOI-2094 b and 12.920670 for TOI-7166 b.
    $^c$ Using Kepler's third law, the constants are calculated to be 120 for TOI-2094 b and 100 for TOI-7166 b, given the prior knowledge of the stellar bulk density and planetary orbital period.  
    \end{minipage}
    \label{tab:planet_parameters_eccentric}
\end{table*}

\begin{figure*}
    \centering
    \includegraphics[width=0.49\linewidth]{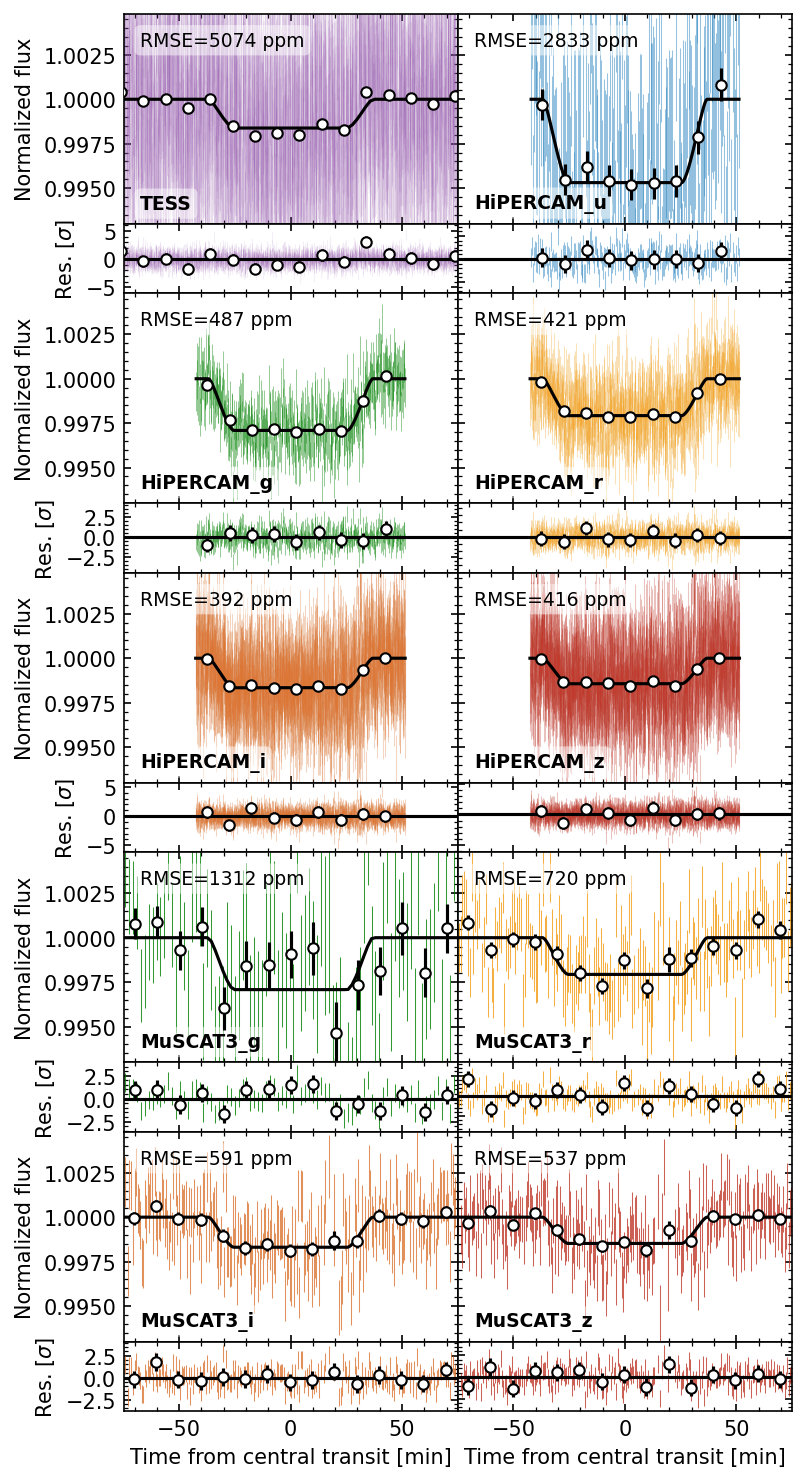}
    \includegraphics[width=0.49\linewidth]{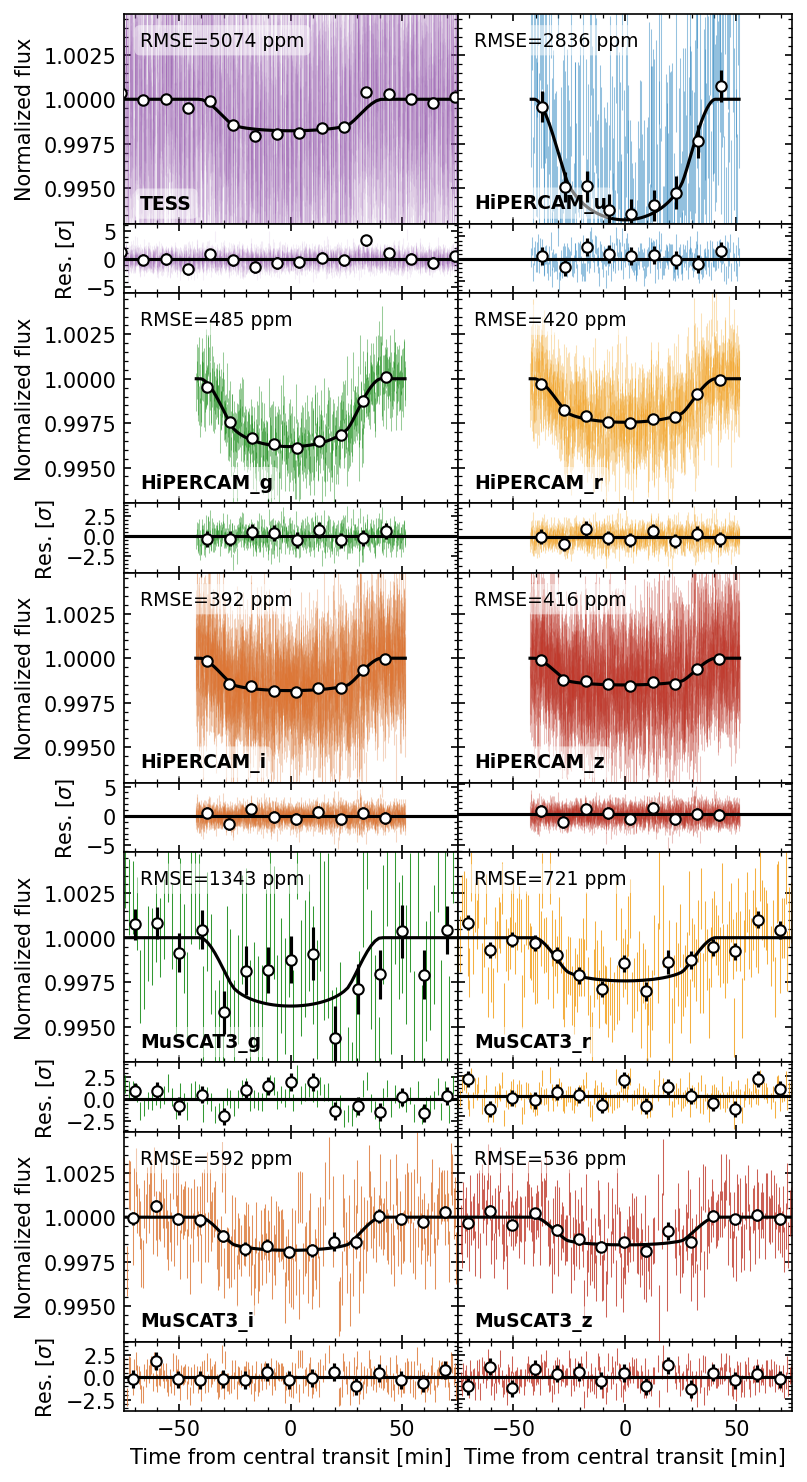}
    \caption{Best-fit light curves of TOI-2094 b assuming an eclipsing white dwarf ($\mathcal{H}_2$, \textit{left}) and assuming a brown dwarf transiting an unresolved background star ($\mathcal{H}_3$, \textit{right}). The coloured error bars are the normalized and detrended fluxes, while the black error bars are the fluxes after 10-min binning. The TESS light curves have been phase-folded before binning. The black solid lines are the best-fit model. The small panels below the light curves are the residuals normalized by flux uncertainties. The RMSE in each panel has been normalized to one-minute integration time for comparison.}
    \label{fig:lc-2094-s23}
\end{figure*}

\begin{figure*}
    \centering
    \includegraphics[width=0.49\linewidth]{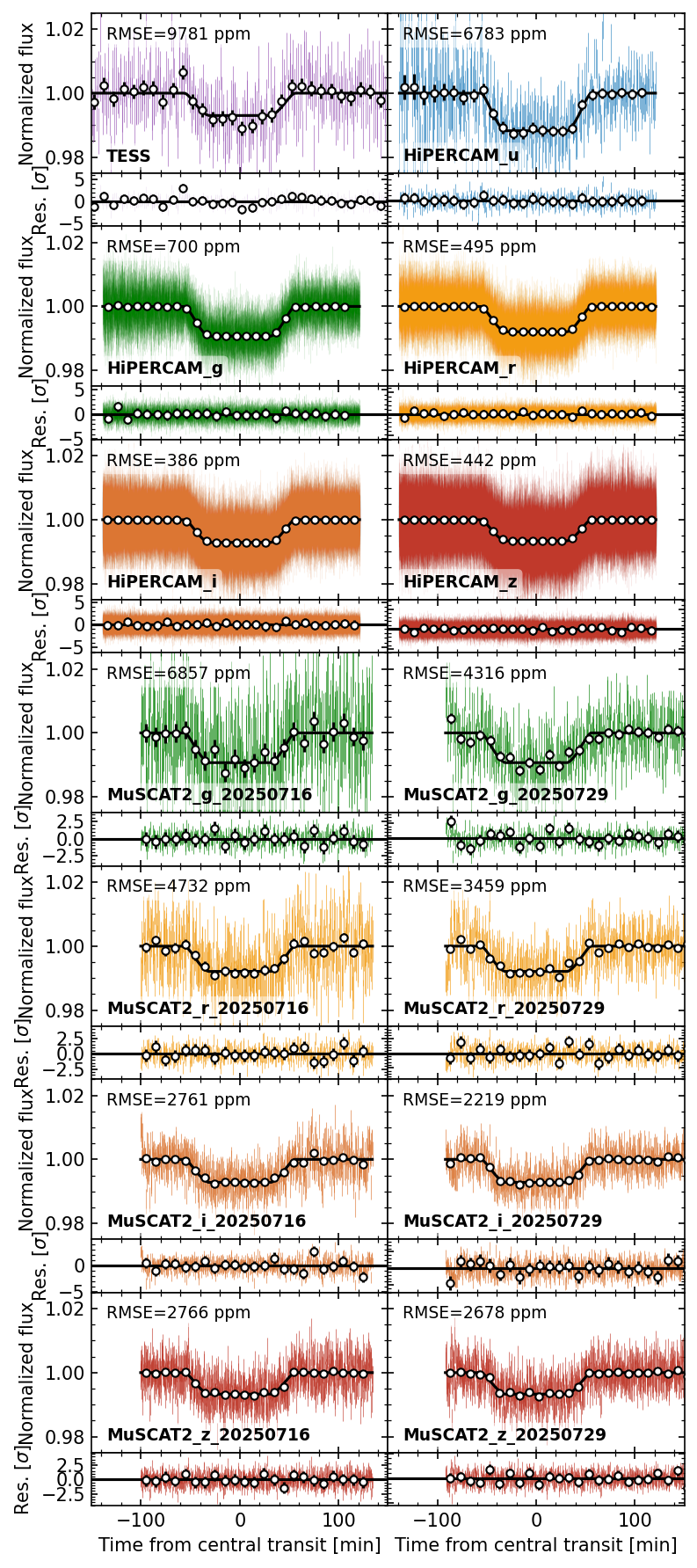}
    \includegraphics[width=0.49\linewidth]{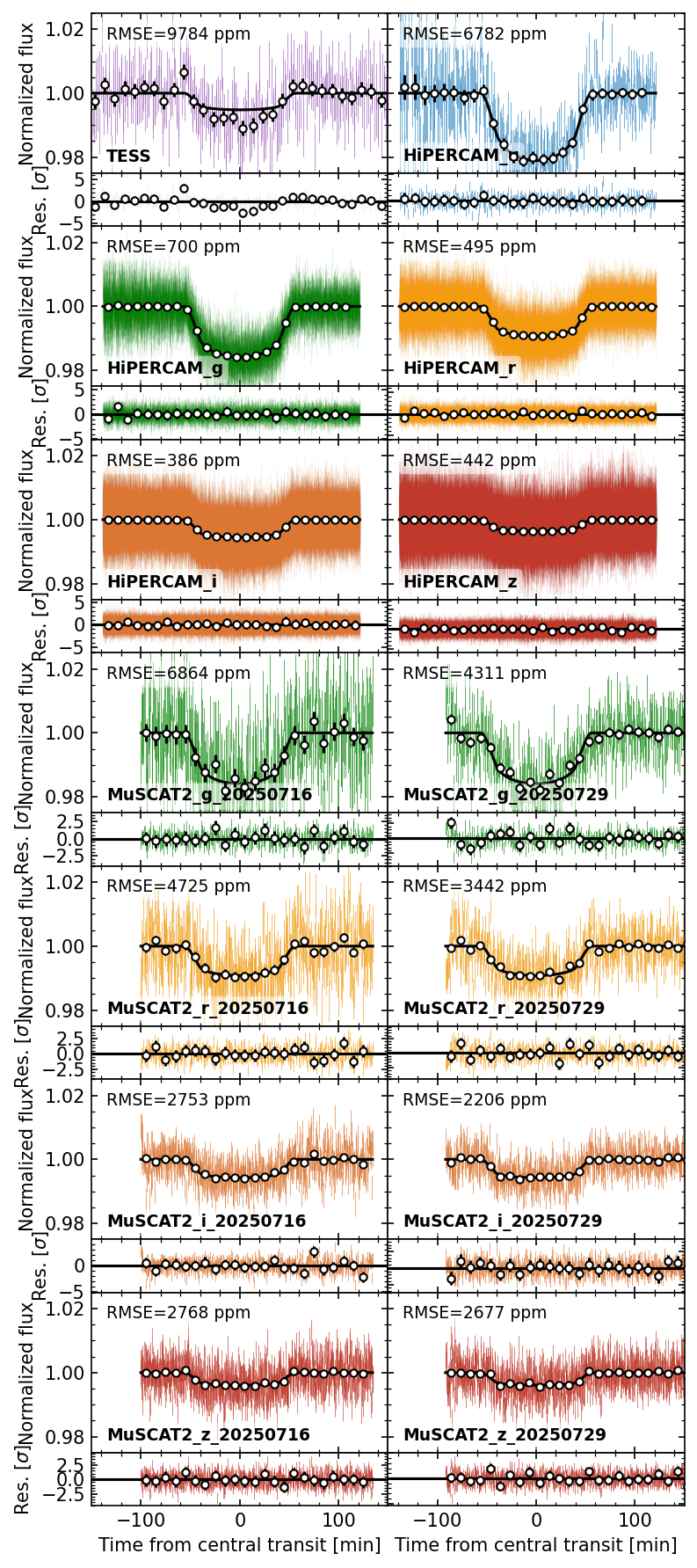}
    \caption{Best-fit light curves of TOI-7166 b assuming an eclipsing white dwarf ($\mathcal{H}_2$, \textit{left}) and assuming a brown dwarf transiting an unresolved background star ($\mathcal{H}_3$, \textit{right}). The coloured error bars are the normalized and detrended fluxes, while the black error bars are the fluxes after 10-min binning. The black solid lines are the best-fit model. The small panels below the light curves are the residuals normalized by flux uncertainties. The TESS light curves have been phase-folded before binning. The RMSE in each panel has been normalized to one-minute integration time for comparison.}
    \label{fig:lc-7166-s23}
\end{figure*}

\end{document}